\documentclass[lettersize,journal]{IEEEtran}
\usepackage{amsmath,amsfonts}
\usepackage{algorithm,algpseudocode}
\usepackage{marvosym}

\usepackage{array}
\usepackage[caption=false,font=normalsize,labelfont=sf,textfont=sf]{subfig}
\usepackage{textcomp}
\usepackage{stfloats}
\usepackage{url}
\usepackage{circledsteps}   
\usepackage{pifont}
\usepackage{verbatim}
\usepackage{graphicx}
\usepackage[normalem]{ulem}
\usepackage{makecell}   
\usepackage{graphicx}   
\usepackage{amssymb}
\usepackage{cite}
\usepackage{graphicx}
\usepackage{amssymb}
\usepackage{booktabs}
\usepackage{multirow}
\usepackage{xcolor}
\usepackage{colortbl}
\usepackage[capitalize]{cleveref}
\Crefname{section}{Section}{Sections}
\Crefname{table}{Table}{Tables}
\Crefname{figure}{Figure}{Figures}
\definecolor{my_color}{HTML}{e8eef1}

\renewcommand{\baselinestretch}{0.95}

\newcommand{\lad}[1]{\textcolor{black}{#1}}

\ifCLASSINFOpdf
\else
\fi
\begin{document}

\title{Scalable Neural Vocoder from Range-Null Space Decomposition}

\author{Andong~Li, \IEEEmembership{Member,~IEEE},
	Tong Lei,
	Zhihang Sun,
	Rilin Chen,
	Xiaodong Li,
	Dong~Yu, \IEEEmembership{Fellow,~IEEE},
	Chengshi~Zheng, \IEEEmembership{Senior Member,~IEEE},
	\IEEEcompsocitemizethanks{
		\IEEEcompsocthanksitem Andong Li, Xiaodong Li, and Chengshi Zheng are with the Key Laboratory of Noise and Vibration Research, Institute of Acoustics, Chinese Academy of Sciences, Beijing, 100190, China, and also with University of Chinese Academy of Sciences, Beijing, 100049, China. (Email: liandong@mail.ioa.ac.cn, lxd@mail.ioa.ac.cn, cszheng@mail.ioa.ac.cn)
		\IEEEcompsocthanksitem Zhihang Sun is with School of Communications and Information Engineering, Chongqing University of Posts and Telecommunications, Chongqing 400065, China.
		\IEEEcompsocthanksitem Tong Lei, Rilin Chen and Dong Yu are with Tencent AI Lab. 
		\IEEEcompsocthanksitem Corresponding author: Chengshi Zheng.
}}
\maketitle
\IEEEpeerreviewmaketitle

\begin{abstract}
Although deep neural networks have facilitated significant progress of neural vocoders in recent years, they usually suffer from intrinsic challenges like opaque modeling, inflexible retraining under different input configurations, and parameter-performance trade-off. These inherent hurdles can heavily impede the development of this field. To resolve these problems, in this paper, we propose a novel neural vocoder in the time-frequency (T-F) domain. Specifically, we bridge the connection between the classical range-null decomposition (RND) theory and the vocoder task, where the reconstruction of the target spectrogram is formulated into the superimposition between range-space and null-space. The former aims to project the representation in the original mel-domain into the target linear-scale domain, and the latter can be instantiated via neural networks to further infill the spectral details. To fully leverage the spectrum prior, an elaborate dual-path framework is devised, where the spectrum is hierarchically encoded and decoded, and the cross- and narrow-band modules are leveraged for effectively modeling along sub-band and time dimensions. To enable inference under various configurations, we propose a simple yet effective strategy, which transforms the multi-condition adaption in the inference stage into the data augmentation in the training stage. Comprehensive experiments are conducted on various benchmarks. Quantitative and qualitative results show that while enjoying lightweight network structure and scalable inference paradigm, the proposed framework achieves state-of-the-art performance among existing advanced methods. Code is available at https://github.com/Andong-Li-speech/RNDVoC.
\end{abstract}


\section{Introduction}\label{sec:introduction}
\lad{Vocoders are designed to reconstruct audible time-domain audio waveforms using electronic and computational methods.} As a well-established technique in the signal processing community, \lad{vocoders play a crucial role in various audio generation and processing tasks, \emph{e.g.}, text-to-speech (TTS)~{\cite{wang2017tacotron,shen2018natural,ren2019fastspeech,ren2020fastspeech,tan2024naturalspeech}}, text-to-audio (TTA)~{\cite{kreukaudiogen,liu2023audioldm,ghosal2023text}}, singing voice synthesis (SVS)~{\cite{liu2022diffsinger,hwang2025hiddensinger}}, and speech enhancement~{\cite{liu22x_interspeech,zhou2024mel,su2021hifi}}).} Conventional digital signal processing (DSP)-based vocoders, exemplified by STRAIGHT~{\cite{kawahara2006straight}} and WORLD~{\cite{morise2016world}}, typically involve procedures for estimating various acoustic parameters (\emph{e.g.}, fundamental frequency and vocal tract resonance) to control the vocoding process via manually-designed algorithms. Despite their effectiveness, they often suffer from substantial information loss in spectral details, leading to unsatisfactory synthesis quality. \lad{With the rapid development of deep neural networks (DNNs), particularly generative models~{\cite{chen2022snis,liao2023famm,fu2024waverecovery,li2024screen}}, neural vocoders have advanced rapidly.} Compared with traditional DSP-based methods, neural vocoders exhibit superior performance in generation naturalness and quality, which constitutes the focus in this paper.


\begin{figure}
	\centering
	\includegraphics[width=0.87\linewidth]{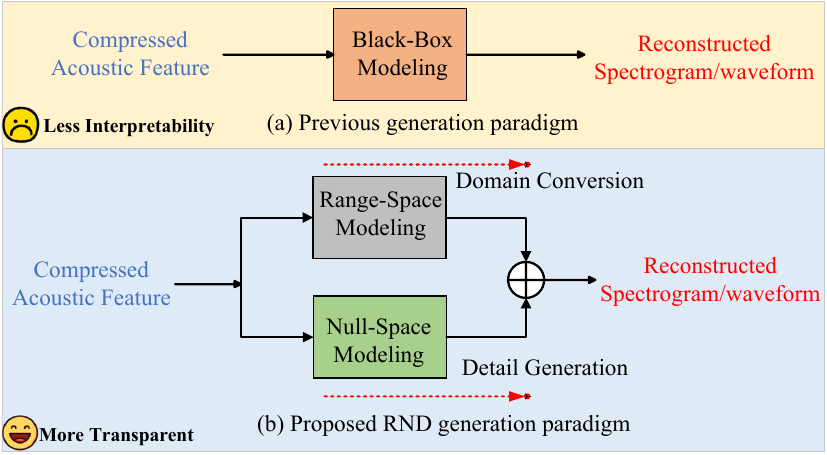}
	\vspace{-2pt}
	\caption{Illustrations of the previous and the proposed neural vocoder generation pipelines with range-null decomposition (RND) theory. (a) In previous T-F domain neural vocoders, the mapping relation from mel-spectrogram to the target spectrogram/waveform was designed in a black-box manner. (b) We exploit the linear degradation prior to develop a more transparent generation pipeline, where the range-space module is to transform the acoustic feature in the original mel-scale domain into the target linear-scale domain, and the null-space module is responsible for generating fine-grained spectral details.}
	\label{fig:black_box_illu}
	\vspace{-0.3cm}
\end{figure}

When evaluating neural vocoders, reconstruction quality and processing efficiency are two key factors for practical applications. In the initial stage, autoregressive-based vocoders, such as WaveNet~{\cite{van2016wavenet}} and WaveRNN~{\cite{kalchbrenner2018efficient}}, have been proposed and effectively utilized in neural vocoders. Although achieving significant breakthrough in synthesized speech quality compared to traditional ones, they often encounter considerably slow inference speed owing to the autoregressive generation nature at the sample level. To mitigate the deficiency, several other schemes have been subsequently introduced, including knowledge-distillation-based methods~{\cite{oord2018parallel,ping2018clarinet}}, flow-based methods~{\cite{ping2020waveflow,prenger2019waveglow}}, glottis-based methods~{\cite{valin2019lpcnet,juvela2019glotnet}}. Although improved inference speed is attained, the computational complexity is still relatively high, significantly impeding deployment in practical scenarios. Besides, diffusion-based methods~{\cite{kongdiffwave,chenwavegrad,leepriorgrad}} are also proposed in recent years for improved performance. Recently, generative adversarial network (GAN)-based vocoders have garnered increasing attention due to their non-autoregressive structure and high-fidelity generation quality~{\cite{donahueadversarial,kumar2019melgan,kong2020hifi,kaneko2022istftnet}}. In these vocoders, high inference speeds are achieved through paralleled convolution layers, and high quality is ensured by the adversarial training paradigm. By doing so, GAN-based vocoders can achieve a favorable trade-off between inference speed and generation quality.

\lad{In the early GAN literature, generators typically estimate raw target waveforms by stacking residual blocks interleaved with upsampling layers. However, multiple consecutive upsampling operations along the time-axis are required to reach the target temporal length, which can slow down inference.} In contrast to end-to-end (E2E) waveform generation, the Fourier transform is regarded as an effective representation tool for audio generation, as many informative acoustic features are well represented in the time-frequency (T-F) domain, such as the fundamental frequency F0~{\cite{kawahara1999restructuring}}. In~{\cite{kaneko2022istftnet}}, an early-exit mechanism is utilized to expedite generation, where the low T-F resolution spectral magnitude and phase are estimated in the intermediate layer of HiFiGAN. In~{\cite{siuzdakvocos,ai2023apnet}}, stacked ConvNext blocks~{\cite{liu2022convnet}} are used to serve as the backbone to directly model the target complex spectrogram in a frame-wise manner. Since no upsampling operation is involved, it exhibits a remarkable inference advantage as well as competitive performance compared to prevailing time-domain methods.

Despite the success of existing neural vocoders, several inherent challenges remain, impeding further improvements:

(1) Black-box-style target modeling may lead to the distortion of acoustic features within the original feature space. \lad{As shown in Fig.~{\ref{fig:black_box_illu}}(a), previous works~{\cite{siuzdakvocos,ai2023apnet}} use mel-spectrograms as input acoustic features and employ a NN-based backbone as a black box to map mel-spectrograms to target spectrograms/waveforms. Due to the high nonlinearity of neural networks, acoustic features embedded in target signals tend to be distorted during this mapping process, and thus degrading reconstruction quality.}

(2) Existing approaches lack scalability to multiple inference configurations. Given a specific training configuration, \emph{e.g.,} number of mel-bands and maximum frequency for mel-filter coefficient calculation, retraining is often required when a different set of mel configurations is used for inference. This is highly labor-intensive and energy-consuming as the model must be trained separately for each mel configuration. For instance, in the open-source project of BigVGAN, several pretrained model checkpoints are provided for different configurations{\footnote{https://github.com/NVIDIA/BigVGAN.}}. \lad{Nevertheless, only a limited number of configurations are supported, forcing researchers to retrain models for unsupported configurations. Thus, it is imperative to develop a flexible neural vocoder that supports multiple inference configurations after a single training.}

(3) Existing T-F domain-based vocoder methods lag behind mainstream time-domain methods. While BigVGAN achieves remarkable performance by scaling the trainable parameters to 112 M~{\cite{leebigvgan}}, existing T-F domain methods do not exhibit notable performance superiority. We attribute the main reason to under-utilization of spectral information. Specifically, the short-time Fourier transform (STFT) operation explicitly decouples information across different frequency bands, although not perfectly, it is essential to model each sub-band distribution separately, which has been widely validated in front-end speech tasks~{\cite{yu2022dbt,luo2023music,li2022embedding}}. However, existing T-F domain neural vocoders typically incorporate full-band modules (\emph{e.g.,} ResNet~{\cite{he2016deep}} and ConvNext blocks) for spectrum reconstruction, \lad{leading to suboptimal generation quality}. Besides, as phase spectrum exhibits a periodic structure and causes the wrapping around the principal value range $(-\pi, +\pi]$, it remains a challenging issue on how to estimate the phase more effectively~{\cite{pobloth1999phase,gerkmann2015phase,williamson2015complex}}.

\begin{figure}
	\centering
	\vspace{0pt}
	\includegraphics[width=0.87\columnwidth]{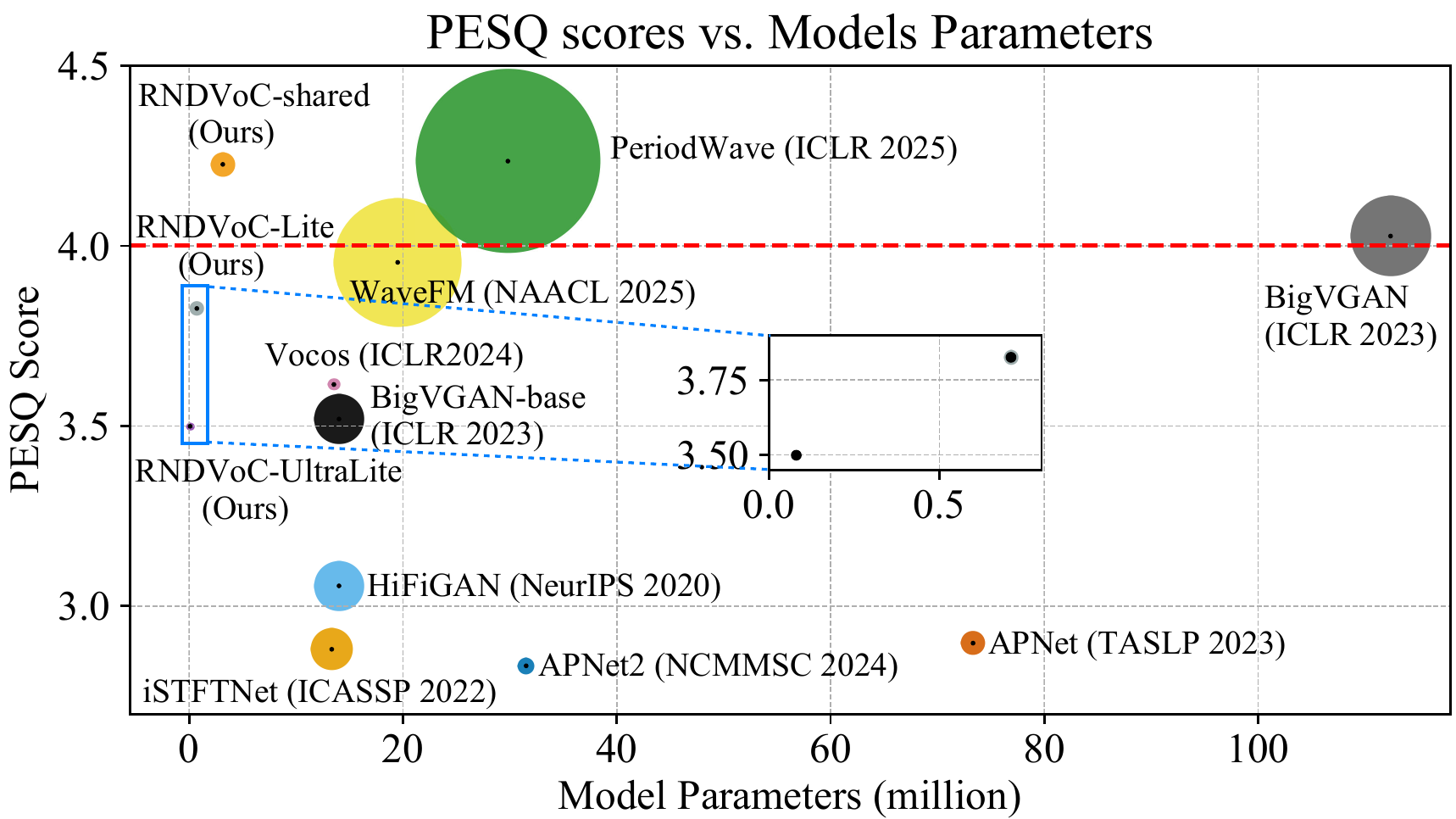}
	\vspace{-4pt}
	\caption{\lad{Comparison of model parameters and PESQ scores between the proposed RNDVoC and other mainstream vocoders on the LibriTTS benchmark. Larger bubbles indicate higher computational complexity. All generators are updated for 1M steps.}}
	\label{fig:bubble_example}
	\vspace{-0.5cm}
\end{figure}

This work aims to resolve the above-mentioned challenges by developing an effective and efficient scalable neural vocoder in the T-F domain. Our contributions are three-fold:

\vspace{3pt}
\noindent \ding{113}~(1) \lad{We introduce the range-null decomposition (RND) theory to neural vocoders. Since the mel-spectrogram is a linear degradation of the linear-scale spectrogram via mel-filtering{\footnote{Strictly speaking, mel-spectrograms are derived via linear mel-filtering of spectral magnitudes, with phase components discarded.}}, we leverage the RND theory~{\cite{wangzero}} to decompose the target spectrogram reconstruction process into two orthogonal subspaces: range-space modeling (RSM) and null-space modeling (NSM), as illustrated in Fig.~{\ref{fig:black_box_illu}}(b). RSM converts the mel-spectrogram to the target linear-scale domain via the pseudo-inverse of the linear degradation operation, while  
NSM acts as a generator to further ``infill'' remaining spectral details.} Unlike previous black-box neural vocoders, our approach comprehensively exploits linear degradation priors to model dual orthogonal signal subspaces, facilitating a more interpretable generation process. To the best of our knowledge, we are the first to introduce RND theory into neural vocoder.

\noindent \ding{113}~(2) We propose a simple yet effective multiple-condition-as-data-augmentation (MCDA) strategy to enable scalable inference under varying mel conditions. \lad{Differences across mel configurations stem from variations in mel filters. The RND theory allows projecting diverse mel-spectrograms into a common linear-scale domain via a simple pseudo-inverse matrix operation. This converts the inference-stage multi-condition adaptation issue into a simple augmentation tactic in the training stage. During training, we randomly sample different mel-configurations as data augmentation, and the model can effectively adapt to various mel conditions during inference. To the best of our knowledge, we are the first to support multi-condition mel-spectrogram inference within a single model{\footnote{While some works~{\cite{song2023dspgan,song2022robust}} claim universal capability, none of them can handle scenarios where mel configurations change, \emph{e.g.}, number of mel bands and $f_{\text{max}}$.}.}}

\noindent \ding{113}~(3) We propose a novel sub-band-based network structure for effective spectrum reconstruction. Inspired by recent advances in music separation~{\cite{luo2023music}} and speech restoration~{\cite{li2022embedding,xu2023zoneformer,yu23b_interspeech}}, we develop a dual-path T-F domain framework dubbed RNDVoC for spectrum reconstruction. \lad{Leveraging prior knowledge of spectral distribution, we incorporate band-split and merge strategies to hierarchically model sub-bands, thereby reducing computational complexity.} Meanwhile, interleaved cross-band and narrow-band modules are employed to excavate inter- and intra-band correlations. Moreover, a novel omnidirectional phase loss that models relations between the center T-F bin and several adjacent bins, is proposed to further improve reconstruction quality. 

\lad{Extensive experiments on the LJSpeech and LibriTTS benchmarks demonstrate the state-of-the-art (SoTA) performance of our method in both quantitative and qualitative comparisons. Concretely, our method outperforms BigVGAN-112 M version with only 2.8\% of its parameters and 8.17\% of its computational complexity, and achieves comparable performance over PeriodWave~{\cite{leeperiodwave}}, a most recently proposed flow-matching (FM)-based method with over 99\% reduction in computational cost, demonstrating the superiority of the proposed framework. Fig.~{\ref{fig:bubble_example}} showcases a comparison in terms of PESQ score among different methods as an example.}

\lad{Our previous work was published at IJCAI 2025~{\cite{li2025learning}}, and this paper extends it with significant advancements. First, we propose a simple yet effective multiple-condition-as-data-augmentation (MCDA) strategy to support scalable inference with a single model. Second, we investigate the band-scaling characteristics in the vocoder task. Besides, we provide more detailed experimental evaluations, including ablation studies and comparisons with diffusion methods, to further validate the effectiveness of the proposed approach. The remainder of this paper is organized as follows. Sec.~{\ref{sec:related-works}} reviews related work. Sec.~{\ref{sec:proposed-methodology}} presents the proposed method. In Sec.~{\ref{sec:experimental-setup}}, datasets and experimental setups are described. Extensive results and analysis are reported in Sec.~{\ref{sec:results-and-analysis}}. In Sec.~{\ref{sec:conclusion}}, we summarize this paper, and draw some valuable conclusions.}   
\vspace{-0.25cm}
\section{Related Work}
\label{sec:related-works}
\vspace{-6pt}
\subsection{Vocoder Methods}
\label{sec:vocoder-methods}
DSP-based Methods: Conventional DSP-based vocoders synthesize speech via statistical parameter estimation. Kawahara \emph{et al.}~{\cite{kawahara2006straight}} introduced the STRAIGHT vocoder, where the excitation and resonant parameters are separately estimated for real-time speech manipulation and synthesis. Morise \emph{et al.}~{\cite{morise2016world}} proposed WORLD, a high-quality speech synthesis method. In this method, the F0, spectral envelope, and aperiodic parameters are determined using analysis algorithms, and a synthesis method is designed for waveform generation. Despite their simplicity, the synthesized speech quality is often unsatisfactory and may contain buzzing artifacts.

\noindent Neural Vocoder: The proliferation of DNNs has driven the rapid development of neural vocoders.

\noindent\textit{Autoregressive Methods}: WaveNet~{\cite{van2016wavenet,oord2018parallel}} and WaveRNN~{\cite{kalchbrenner2018efficient}} are pioneering works, where waveform points were generated autoregressively. In LPCNet~{\cite{valin2019lpcnet}}, the linear prediction coefficients (LPC) are estimated to predict the next sample and a lightweight RNN is adopted for residual calculation. Despite the improvements, these autoregressive methods often suffer from rather slow inference speed.\\
\textit{Flow-based Methods}: As a classical generation paradigm, normalizing flow-based~{\cite{rezende2015variational}} vocoder methods have enabled faster generation speed and improved performance. Typical works include WaveGlow~{\cite{prenger2019waveglow}}, FlowWaveNet~{\cite{kim2019flowavenet}} and RealNVP~{\cite{dinh2016density}}, where a bijective mapping is established between a standard normal distribution and the target data distribution via stacked invertible modules.\\
\textit{GAN-based Methods}: GAN-based neural vocoders have garnered significant attention due to their adversarial training. Plenty of studies have been proposed in recent years, with differences lying in the choice of generators, discriminators and adversarial losses. MelGAN~{\cite{kumar2019melgan}} adopts a non-autoregressive fully-convolutional generator and a multi-scale discriminator for training. HiFiGAN~{\cite{kong2020hifi}} leverages a sinusoidal speech model alongside multi-periodic and multi-scale discriminators for improved speech generation. In BigVGAN~{\cite{leebigvgan}}, the periodic activation function and anti-aliased representation are incorporated into the generator, and the parameters are scaled up to 112 M to achieve impressive performance. More recently, T-F domain based methods begin to gain increasing attention due to their superior inference efficiency. For example, Vocos~{\cite{siuzdakvocos}} attempts to generate target magnitude and phase by ConvNext blocks. In APNet~{\cite{ai2023apnet}}, separate modeling for magnitude and phase are conducted by ResNet blocks. Despite the prominent advantage in inference efficiency they achieved, no notably superiority is observed when compared to existing mainstream time-domain-based schemes.\\
\textit{Diffusion-based Methods}: Diffusion models, successful in text-to-image and text-to-video tasks \cite{galimage,sun2024sora}, have led to diffusion-based vocoders like DiffWave \cite{kongdiffwave}, WaveGrad \cite{chenwavegrad}, and PriorGrad \cite{leepriorgrad}. Despite good performance, they can be slow due to numerous iteration steps, and fast-sampling strategies~\cite{huang2022prodiff,lu2022dpm} are usually needed for further inference cost reduction. \lad{More recently, flow-matching (FM) has attracted increasing attention due to its simple physical formulation and efficient sampling~{\cite{lipman2022flow}}. In~{\cite{leeperiodwave}}, optimal-transport FM (OT-FM) was proposed to model the velocity and multiple tactics were proposed to improve the generation quality. In~{\cite{luo2025wavefm}}, Luo \emph{et al.} incorporate FM and consistency distillation to facilitate single-step generation, notably improving the inference efficiency.}
\vspace{-0.35cm}
\subsection{\lad{Feature Decomposition and Synthesis}}
\label{sec:range-null-decomposition}
\lad{Effective feature decomposition and synthesis are crucial for audio generation and processing tasks. For instance, Luo \emph{et al.}~{\cite{luo2019conv}} proposed an analysis-separation-synthesis pipeline, where waveform samples are embedded via over-complete representations to facilitate discriminative feature learning. In~{\cite{li2024tabe}}, a Taylor-unfolding mapping framework was proposed to mimic the decomposition-synthesis process of Taylor series for audio modeling.} 

More recently, as a classical signal decomposition tool in various signal processing application, range-null decomposition (RND) has begun to be applied in various learning-based inverse problems and generative models. \lad{Mathematically, the target signal is decomposed into orthogonal components in its range space and null space.} Wang~\emph{et al.}~\cite{wang2023gan} proposed a null-space learning strategy for consistent face super-resolution. Later, they incorporated the RND strategy into the diffusion backward process to achieve a better generation-consistency trade-off~\cite{wangzero}. Most recently, in~\cite{caiphocolens}, a two-step lensless imaging algorithm was proposed, which involved the reconstruction of range-space and null-space. 

\lad{To our best knowledge, no prior work has addressed this in the audio domain.}  In this paper, we recast the problem formulation of the vocoder task and propose a novel framework that effectively integrates the insights from the RND theory to improve target spectrum reconstruction quality.
\vspace{-0.45cm}
\section{Proposed Methodology}
\label{sec:proposed-methodology}
\vspace{-0pt}
In this section, we first introduce the RND theory and formulate the problem of the speech vocoder task, and we then demonstrate their interconnection. Subsequently, the proposed learning framework is elucidated, including detailed network module design and training objectives.
\begin{figure}
	\centering
	\vspace{0pt}
	\includegraphics[width=0.43\textwidth]{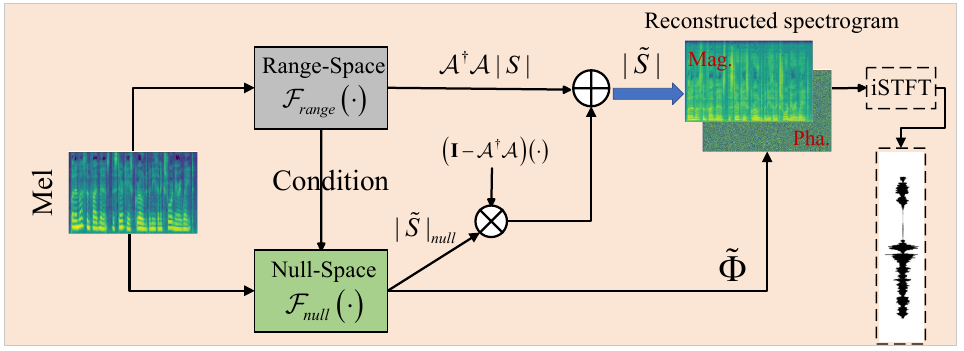}
	\vspace{-2pt}
	\caption{Illustrations of the proposed RNDVoC.} 
	\label{fig:rnd_learn_illu}
	\vspace{-10pt}
\end{figure}
\begin{table}[t]
    \Huge
    \centering
    \caption{\lad{Notation tables between classical signal model and mel degradation formulation.}}
    \resizebox{0.99\columnwidth}{!}{
    \begin{tabular}{c|cc}
    \toprule
    \lad{Category} &\lad{Signal-Level} &\lad{Spectral-Level (Vocoder)} \\
    \hline
    \lad{Compression Matrix} &\lad{$\mathbf{A}$} &\lad{Mel-Filter $\mathcal{A}$}\\
    \lad{Observation} &\lad{$\mathbf{y}$} &\lad{Input Mel-Spectrogram $\mathbf{Y}$, Eq.~$\left(\ref{eqn:5}\right)$} \\
    \lad{Target Signal} &\lad{$\mathbf{x}$} &\lad{Linear-Scale Spectrum $\mathbf{S}$}\\
    \lad{Range-Space} &\lad{$\mathbf{A}^{\dagger}\mathbf{y}$} &\lad{Range-Space Spectrum} \lad{$\left|\tilde{\mathbf{S}}\right|_{range}$, Eq.~$\left(\ref{eqn:9}\right)$} \\ 
    \lad{Null-Space} &\lad{$\left(\mathbf{I} - \mathbf{A}^{\dagger}\mathbf{A}\right)\hat{\mathbf{x}}$} &\lad{Null-Space Spectrum$\left(\mathbf{I} - \mathcal{A}^{\dagger}\mathcal{A}\right)\left|\tilde{\mathbf{S}}\right|_{null}$, Eq.~$\left(\ref{eqn:10}\right)$} \\
    \lad{Estimation} &\lad{$\mathbf{\tilde{x}}$} &\lad{Superimposed Spectrum $\mathbf{\tilde{S}},$, Eq.~$\left(\ref{eqn:11}\right)$} \\
    \hline
    \end{tabular}}
    \label{tab:mapping-notation}
    \vspace{-0.35cm}
\end{table}
\vspace{-0.2cm}
\subsection{Review of Range-Null Space Decomposition}
\label{sec:range-null-space-decomposition}
Without loss of generality, we commence with a classical signal compression physical model:
\begin{align}
	\label{eqn:1}
	\mathbf{y} = \mathbf{A}\mathbf{x} + \mathbf{n},
\end{align} 
where $\mathbf{x}\in\mathbb{R}^{D}, \left\{\mathbf{y}, \mathbf{n}\right\}\in\mathbb{R}^{d}$ denote the target, observed, and noise signals, respectively, and $\mathbf{A}\in\mathbb{R}^{d\times D}$ represents the compression matrix which satisfies $d\ll D$. In the noise-free scenario, Eq.~{(\ref{eqn:1})} can be further simplified into $\mathbf{y} = \mathbf{A}\mathbf{x}$. If we define the pseudo-inverse of the linear matrix $\mathbf{A}$ as $\mathbf{A}^{\dagger}\in\mathbb{R}^{D\times d}$, which satisfies $\mathbf{A}\mathbf{A}^{\dagger}\mathbf{A}\equiv\mathbf{A}$, then the signal $\mathbf{x}$ can be decomposed into two orthogonal subspaces: one residing in the range-space of $\mathbf{A}$, and the other in the null-space:
\begin{align}
	\label{eqn:2}
	\mathbf{x} \equiv \mathbf{A}^{\dagger}\mathbf{A}\mathbf{x} + \left(\mathbf{I} - \mathbf{A}^{\dagger}\mathbf{A}\right)\mathbf{x},
\end{align}
where $\mathbf{A}^{\dagger}\mathbf{A}\mathbf{x}$ defines the range-space component and $\left(\mathbf{I} - \mathbf{A}^{\dagger}\mathbf{A}\right)\mathbf{x}$ corresponds to the remaining null-space part. For the above-mentioned compression model, the solution of $\mathbf{x}$, \emph{i.e.}, $\tilde{\mathbf{x}}$, should meet two constraints: (1) Degradation consistency: $\mathbf{A}\tilde{\mathbf{x}}\equiv\mathbf{y}$, (2) Distribution consistency: $\tilde{\mathbf{x}}\sim p\left(\mathbf{x}\right)$, where the first constraint attempts to maintain the original information embedded in the compressed signal space unaltered after reconstruction, and the second condition requires that the reconstructed signal $\tilde{\mathbf{x}}$ should exhibit a distribution consistent with the target signal. Based on the RND theory, such a solution can be expressed as:
\begin{align}
	\label{eqn:3}
	\tilde{\mathbf{x}} = \mathbf{A}^{\dagger}\mathbf{y} + \left(\mathbf{I} - \mathbf{A}^{\dagger}\mathbf{A}\right)\hat{\mathbf{x}}.
\end{align}

Since the range and null subspaces are orthogonal, $\tilde{\mathbf{x}}$ naturally satisfies the degradation consistency as $\mathbf{A}\tilde{\mathbf{x}}\equiv\mathbf{A}\mathbf{A}^{\dagger}\mathbf{A}\mathbf{x} + \mathbf{A}\left(\mathbf{I} - \mathbf{A}^{\dagger}\mathbf{A}\right)\hat{\mathbf{x}}\equiv\mathbf{A}\mathbf{x}+\mathbf{0}\equiv\mathbf{y}$. Thus we need to consider the estimation of $\hat{\textbf{x}}$ to fulfill the second constraint, which can be achieved through different neural network architectures~{\cite{wangzero}}.
\vspace{-0.25cm}
\subsection{Vocoder Problem Formulation}
\label{sec:problem-formulation}
The neural vocoder aims to recover the target waveform from acoustic features. In this study, we primarily consider the mel-spectrogram as the acoustic feature due to its widespread use and simplicity. The forward degradation process of the mel-spectrogram can be formulated as:
\begin{align}
	\label{eqn:4}
	\mathbf{X}^{mel} = \log\left(\mathcal{A}\left|\mathbf{S}\right|\right),
\end{align}
where $\mathbf{S}\in\mathbb{C}^{F\times T}$ and $\mathbf{X}^{mel}\in\mathbb{R}^{F_{m}\times T}$ denote the target spectrogram and the resultant log-scale mel-spectrogram, respectively, and $\mathcal{A}\in\mathbb{R}^{F_{m}\times F}$ represents the mel-filter matrix. $\left\{F, F_{m}, T\right\}$ denote the frequency size in the linear-, mel-scale, and the number of frames, respectively. $\log\left(\cdot\right)$ and $\left|\cdot\right|$ are the logarithm and modulus operators, respectively. If the logarithm operation is absorbed into the left of Eq.~{(\ref{eqn:4})}, we obtain:
\begin{align}
	\label{eqn:5}
	\mathbf{Y} = \mathcal{A}\left|\mathbf{S}\right|,
\end{align}
where $\mathbf{Y} = \exp\left(\mathbf{X}^{mel}\right)$. Eq.~{(\ref{eqn:5})} involves two simple degradation operations: $\Circled{\footnotesize{1}}$ Discarding the phase information, $\Circled{\footnotesize{2}}$ Magnitude compression with a linear operation. Accordingly, the inverse process of Eq.~{(\ref{eqn:5})} can be illustrated as follows:
\vspace{-1.5pt}
\begin{itemize}
	\item Phase information recovery: It aims to retrieve the phase spectrum from the input compression spectrum information, \emph{i.e.}, establish a mapping relation $f_{p}(\mathbf{Y}) \mapsto \mathbf{\Phi}$, where $\mathbf{\Phi}$ denotes the target phase spectrum. 
	\item Magnitude information recovery: It aims to recover the target magnitude spectrum from the compressed observation, \emph{i.e.}, establish a mapping relation $f_{m}(\mathbf{Y}) \mapsto \left|\mathbf{S}\right|$. 
\end{itemize}
\vspace{-1.5pt}

Regarding the first point, the phase retrieval problem has been a long-standing task in the audio processing field. Given the magnitude spectrum, many conventional methods have been proposed, such as the Griffin-Lim~{\cite{griffin1984signal}} and PGHI~{\cite{pruuvsa2017noniterative}} algorithms. Recently, some NN-based methods have been proposed for phase estimation~{\cite{peer2023diffphase}}, and the difference is that the magnitude spectrum used here undergoes linear compression and may incur information loss. For the second point, since the compression matrix $\mathcal{A}$ is usually known in advance, an ``oracle'' inverse solution $\mathcal{A}^{*}\in\mathbb{R}^{F\times F_{m}}$ is expected to exist in recovering the magnitude perfectly. However, it is often infeasible as $F_{m}\ll F$ in practical settings{\footnote{For example, in LJSpeech speech vocoder benchmark, a common mel setting is $\left\{F=513, F_{m}=80\right\}$.}}, and only its pseudo-inverse counterpart $\mathcal{A}^{\dagger}$ exists{\footnote{In this paper, we use the Moore-Penrose inverse.}}. With $\mathcal{A}^{\dagger}$, we can project the input from the mel-scale domain back to the linear-scale domain, and the overall reverse process can thus be formulated as:
\begin{align}
	\label{eqn:6}
	\left\{|\tilde{\mathbf{S}}|, \tilde{\mathbf{\Phi}}\right\} = \mathcal{F}\left(\mathcal{A}^{\dagger}\mathbf{Y}\right) = \mathcal{F}\left(\mathcal{A}^{\dagger}\mathcal{A}\left|\mathbf{S}\right|\right),
\end{align}
\begin{align}
	\label{eqn:7}
	\tilde{\mathbf{s}} = \text{iSTFT}\left(|\tilde{\mathbf{S}}|e^{j\tilde{\mathbf{\Phi}}}\right),
\end{align}
where $\mathcal{F}\left(\cdot\right)$ indicates the inverse mapping defined in the T-F domain, and $\tilde{\mathbf{s}}\in\mathbb{R}^{L}$ represents the reconstructed waveforms.
\vspace{-0.25cm}
\subsection{Connection between RND and Vocoder Task}
\label{sec:connection-between-rnd-and-vocoder}
\vspace{-0.1cm}
It is noteworthy that Eq.~{(\ref{eqn:5})} exhibits a similar signal formulation as Eq.~{(\ref{eqn:1})} in the noise-free case, thus inspiring us to establish a bridge between RND and the vocoder task, \emph{i.e.}, how to reconstruct the target spectrum from its compressed observation. Since $\mathcal{A}^{\dagger}\mathbf{Y}\equiv\mathcal{A}^{\dagger}\mathcal{A}|\mathbf{S}|\in\mathbb{R}^{F\times T}$ is actually the range-space component of $\left|\mathbf{S}\right|$, we rewrite Eq.~{(\ref{eqn:6})} as:
\begin{align}
	\label{eqn:8}
	\left\{|\tilde{\mathbf{S}}|, \tilde{\mathbf{\Phi}}\right\} = \mathcal{F}\left(\mathcal{F}_{range}\left(\mathbf{Y}\right)\right),
\end{align}
where the range-space function $\mathcal{F}_{range}\left(\mathbf{Y}\right) = \mathcal{A}^{\dagger}\mathbf{Y}$ obtains the range-space component by projecting the compressed mel-spectrogram into the linear-scale domain. We notice that most recently in~{\cite{lv24_interspeech}}, a similar operation was adopted, albeit without a comprehensive explanation. In contrast, in this work, we interpret the pseudo-inverse operation as the projection in the range-space, an insight that has been rarely explored in the context of the neural vocoder task.

Similar to Eq.~{(\ref{eqn:3})}, we rewrite the abstract inverse mapping process in Eq.~{(\ref{eqn:8})} as an explicit superimposition of the range-space and null-space components, given by:
\begin{align}
	\label{eqn:9}
	|\tilde{\mathbf{S}}|_{range} = \mathcal{F}_{range}\left(\mathbf{Y}\right) = \mathcal{A}^{\dagger}\mathbf{Y},
\end{align}
\begin{align}
	\label{eqn:10}
	\left\{|\tilde{\mathbf{S}}|_{null}, \tilde{\mathbf{\Phi}}\right\} = \mathcal{F}_{null}\left(|\tilde{\mathbf{S}}|_{range}\right),
\end{align}
\begin{align}
	\label{eqn:11}
	\lad{\tilde{\mathbf{S}}} &\lad{= \left(|\tilde{\mathbf{S}}|_{range} + \left(\mathbf{I} - \mathcal{A}^{\dagger}\mathcal{A}\right)|\tilde{\mathbf{S}}|_{null}\nonumber\right)e^{j\mathbf{\tilde{\Phi}}}}\\
	&\lad{= \mathcal{A}^{\dagger}\mathcal{A}\left|\mathbf{S}\right| + \left(\mathbf{I} - \mathcal{A}^{\dagger}\mathcal{A}\right)|\tilde{\mathbf{S}}|_{null}e^{j\mathbf{\tilde{\Phi}}}},
\end{align}
where $\mathcal{F}\left(\cdot\right)$ is decomposed into two operations in Eqs.~{(\ref{eqn:10})}-{(\ref{eqn:11})}. Specifically, $|\tilde{\mathbf{S}}|_{range}$ is obtained via pseudo-inverse matrix operation, and the null-space module $\mathcal{F}_{null}\left(\cdot\right)$ is employed to generate the null-space component estimation. These two components are fused to yield the final spectrogram estimation. Note that the phase spectrum is also estimated in the null-space learning stage as a multi-task option{\footnote{\lad{We compare the multi-task and cascaded-task under a close computational complexity in Sec.~III of the Appendix.}}}. The overall forward process is illustrated in Fig.~{\ref{fig:rnd_learn_illu}, and the framework is abbreviated as RNDVoC. \lad{
To clarify the correspondence between general inverse problem formulation and our vocoder-specific implementation, Table {\ref{tab:mapping-notation}} summarizes the key notation mappings}.

\lad{\textbf{Remarks}: We would like to highlight two key benefits of integrating the RND theory into neural vocoding. First, as a classical linear signal processing framework with decades of development in inverse problem, RND endows the vocoder task with better interpretability and model robustness. Second, the range-space component preserves mel-spectrogram acoustic information losslessly via pseudo-inverse transformation, avoiding spectral artifacts induced by end-to-end nonlinear mapping in conventional neural vocoders.} 

        \vspace{-3pt}
\begin{figure*}
	\centering
	\vspace{0pt}
	\includegraphics[width=0.83\textwidth]{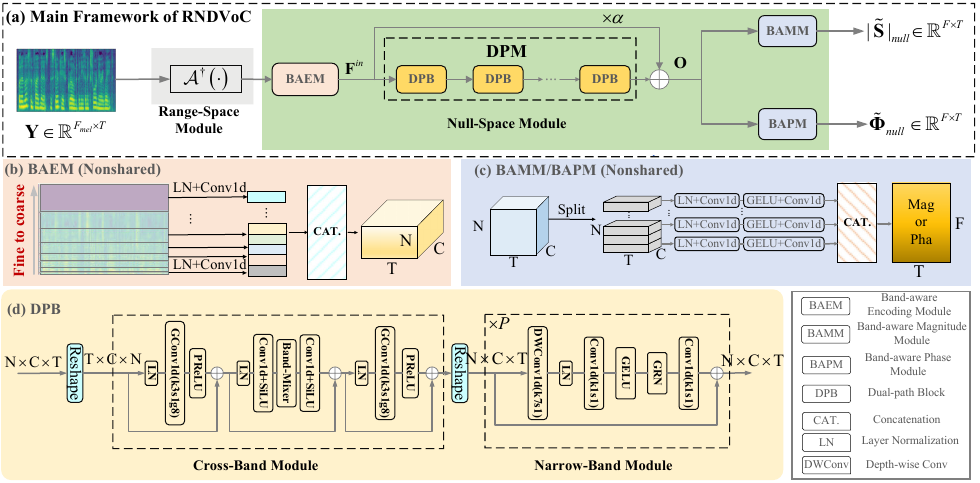}
	\vspace{-2pt}
	\caption{Framework diagram of the proposed RNDVoC, where the range-space module only involves the pseudo-inverse matrix operation and its output will serve as the input of the null-space module. (a) Main framework diagram of the null-space module. (b) Detailed structure of the band-aware spectral encoding module (BAEM). \lad{(c) Detailed structure of the band-aware spectral magnitude/phase module (BAMM/BAPM), which shares a similar network structure. (d) Detailed structure of the dual-path block (DPB).} Note that the nonshared scheme is adopted in (b)-(c).}
	\label{fig:framework}
	\vspace{-0.4cm}
\end{figure*}
			\vspace{-0.25cm}
			\subsection{Architecture Design of RNDVoC}
			\label{sec:architecture-design-of-rndvoc}
			\lad{The overall framework diagram of the proposed RNDVoC is depicted in Fig.~{\ref{fig:framework}}(a). The input mel-spectrogram is first projected to its range-space component via the pseudo-inverse matrix $\mathcal{A}^{\dagger}$ and then fed as input to the null-space module (NSM). NSM primarily consists of three components: Band-aware Encoding Module (BAEM) for hierarchical spectral feature encoding, Dual-Path Module (DPM) for efficient cross- and narrow-band feature modeling, and Band-aware Magnitude/Phase Module (BAMM/BAPM) for spectral magnitude and phase decoding.}
			\subsubsection{Band-aware Encoding/Decoding Module} 
			\label{sec:band-aware-encoding-module}
			Previous vocoder literature has predominantly focused on full-band modeling~{\cite{siuzdakvocos,ai2023apnet}}, overlooking the hierarchical characteristics of spectrograms in the T-F domain. For instance, speech spectra exhibit fine-grained harmonic structures in low- and mid-frequency regions, and the value of fundamental frequency (F0) typically resides below 500 Hz. This motivates us to adopt a hierarchical encoding strategy for spectral features. Specifically, \lad{as shown in Fig.~{\ref{fig:framework}}(b)}, we employ a frequency band-split strategy to partition the frequency dimension into $N$ sub-bands, \lad{\emph{i.e.}, $F = \sum_{n}^{N}F_{n}$}, where $F_{n}$ denotes the frequency size of the $n$-th sub-band. The division follows a ``from-fine-to-coarse'' principle, \emph{i.e.}, we gradually increase $F_{n}$, to preserve the fine-grained spectral information in the relatively low frequency regions while reducing the overall computational cost by compressing more bands in the high frequency regions. Since no phase is available, we regard $|\mathbf{\tilde{S}}|_{range}$ as a special spectrogram with a constant zero-valued imaginary part and construct the input spectrogram as  
			\begin{equation}
				\label{eqn:14}
				\mathbf{\tilde{S}}_{range} = \text{Cat}\left(|\mathbf{\tilde{S}}|_{range}|, \mathbf{0}\right)\in\mathbb{R}^{F\times T\times2},
			\end{equation}
			where $\text{Cat}\left(\cdot\right)$ denotes the concatenation operation, and $\mathbf{0}\in\mathbb{R}^{F\times T}$ is an all-zero tensor with the same size of $|\mathbf{\tilde{S}}|_{range}$. Following~{\cite{luo2024gull}}, for $n$-th subband $\mathbf{\tilde{S}}_{range,n}\in\mathbb{R}^{F_{n}\times T\times 2}$, we extract a gain-shape representation $\mathbf{G}_{n}\in\mathbb{R}^{\left(2F_{n} + 1\right)\times T}$:
			\begin{equation}\small
				\label{eqn:15}
				\mathbf{G}_{n} = \text{Cat}\left(\frac{\mathcal{R}\left(\mathbf{\tilde{S}}_{range,n}\right)}{\left\|\mathbf{\tilde{S}}_{range,n}\right\|_{2}}, \frac{\mathcal{I}\left(\mathbf{\tilde{S}}_{range,n}\right)}{\left\|\mathbf{\tilde{S}}_{range,n}\right\|_{2}}, \log\left(\left\|\mathbf{\tilde{S}}_{range,n}\right\|_{2}\right)\right),
			\end{equation}
			where $\mathcal{R}\left(\cdot\right)$ and $\mathcal{I}\left(\cdot\right)$ denote the real and imaginary operations, respectively, $\left\|\mathbf{\tilde{S}}_{range,n}\right\|_{2}$ is the $L_{2}$ norm of $\mathbf{\tilde{S}}_{range,n}$. With layer normalization (LN)~{\cite{lei2016layer}} and Conv1d, the encoded feature of the $n$-th subband can be represented as:
			\begin{equation}
				\label{eqn:16}
				\mathbf{F}^{in}_{n} = \text{Conv1d}\left(\text{LN}\left(\mathbf{G}_{n}\right)\right)\in\mathbb{R}^{C\times T},
			\end{equation}
			where $C$ is the channel dimension. After BAEM, we obtain the encoded feature $\mathbf{F}_{in} = \text{Cat}\left(\mathbf{F}^{in}_{1},\cdots,\mathbf{F}^{in}_{N}\right)\in\mathbb{R}^{N\times C\times T}$.
			
			Similarly, in the decoding module, we reconstruct the magnitude and phase components hierarchically, \lad{as shown in Fig.~{\ref{fig:framework}}(c)}. Given the output tensor $\textbf{O}\in\mathbb{R}^{N\times C\times T}$, for the $n$-th subband, it passes through LN, Conv1d, GELU activation~{\cite{hendrycks2016gaussian}}, and another Conv1d to obtain the magnitude and phase components, given by:
			\begin{equation}
				\label{eqn:17}
				|\mathbf{\tilde{S}}_{null,n}| = \text{Exp}(\text{Conv1d}\left(\text{GELU}\left(\left(\text{Conv1d}\left(\text{LN}\left(\mathbf{O}_{n}\right)\right)\right)\right)\right)),
			\end{equation}
			\begin{equation}
				\label{eqn:18}
				\mathbf{R}_{n} = \text{Conv1d}\left(\text{GELU}\left(\left(\text{Conv1d}\left(\text{LN}\left(\mathbf{O}_{n}\right)\right)\right)\right)\right),
			\end{equation}
			\begin{equation}
				\label{eqn:19}
				\left\{\mathcal{R}\left(\mathbf{\tilde{S}}_{null,n}\right), \mathcal{I}\left(\mathbf{\tilde{S}}_{null,n}\right)\right\} = \text{Split}\left(\mathbf{R}_{n}\right),
			\end{equation}
			\begin{equation}
				\label{eqn:20}
				\mathbf{\tilde{\Phi}}_{null,n} = \text{Atan2}\left(\mathcal{I}\left(\mathbf{\tilde{S}}_{null,n}\right), \mathcal{R}\left(\mathbf{\tilde{S}}_{null,n}\right)\right),
			\end{equation}
			where $\text{Atan2}\left(\cdot\right)$ denotes the phase-calculation formula.
			
			Note that in the above modules, the network weights are not shared for each subband encoding. This leads to two drawbacks: First, excessive loop operation in the encoding and decoding modules can introduce additional algorithmic delay. Besides, the nonshared property results in a larger number of parameters. To this end, we propose a parameter-sharing strategy, as shown in Fig.~{\ref{fig:ende_shared}}. Specifically, in the encoding process, instead of operating within each subband, we divide the spectrogram into $I$ regions ($I\ll N$), and use a separate Conv2d layer with stride for each region to compress and encode the spectrum, followed by LN. Finally, the $I$-region features are stacked together. The process of the $i$-th region can be formulated as:
			\begin{equation}
				\label{eqn:21}
				\mathbf{F}^{in}_{i} = \text{LN}\left(\text{Conv2d}\left(\mathbf{\tilde{S}}_{range,i}\right)\right)\in\mathbb{R}^{N_{i}\times C\times T},
			\end{equation} 
			where the subscript $\left(\cdot\right)_{i}$ denotes the $i$-th region, and $N_{i}$ refers to the number of the compressed subbands in the $i$-th region. Similarly, for the decoding process, the feature map $\mathbf{O}$ is first split into $I$ regions, each of which is processed by point-wise Conv2d, LN, and GELU. Finally, the Transposed Conv2d is used for target estimation, which is formulated as:
			\begin{equation}
				\label{eqn:22}
				\mathbf{K}_{i} = \text{TrConv2d}\left(\text{GELU}\left(\text{LN}\left(\text{Conv2d1x1}\left(\mathbf{O}_{i}\right)\right)\right)\right),
			\end{equation} 
			where $\mathbf{K}_{i}\in\mathbb{R}^{\overline{F}_{i}\times T}$, and $\overline{F}_{i}$ is the resulting frequency size in the $i$-th region. To maintain the same number of subbands after compression and frequency size after reconstruction between the two schemes, the kernel/stride size in the Conv2d and TrConv2d should be specially set to satisfy:
			\begin{equation}
				\label{eqn:23}
				N = \sum_{i=1}^{I} N_{i}, F = \sum_{i=1}^{I}{\overline{F}}_{i}.
			\end{equation}

            \lad{In Sec.~{\ref{sec:results-and-analysis}}, we show that the proposed region-oriented encoding/decoding strategy significantly improves reconstruction quality. We also provide detailed comparisons between the two strategies in terms of computational cost and inference speed. Interested readers are referred to Sec.~I of the Appendix.}
            
            \begin{figure}
				\centering
				\vspace{0pt}
				\includegraphics[width=0.45\textwidth]{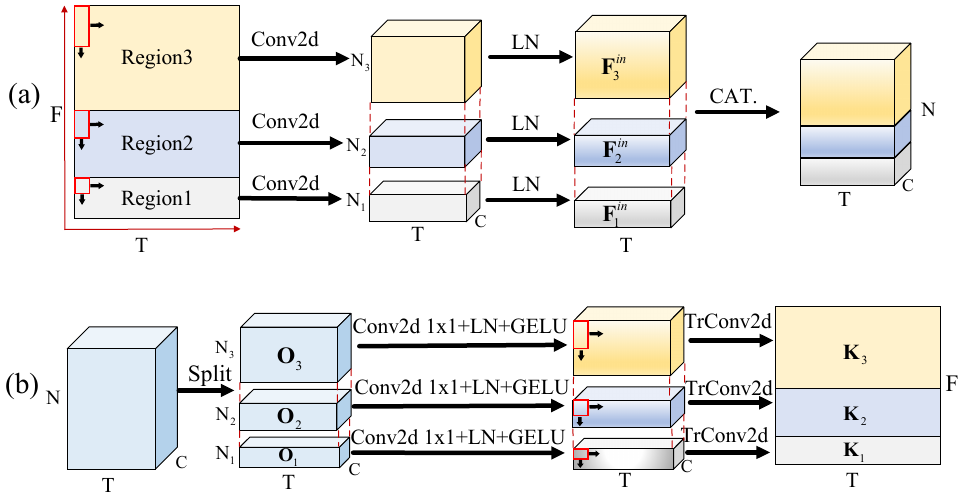}
				\vspace{-2pt}
				\caption{Illustrations of the spectral encoding and decoding module with the proposed parameter-shared strategy, and here the number of regions $I$ is set to 3 as an example. (a) Spectral encoding process. (b) Spectral decoding process, we only take one branch (magnitude or phase) as an example.}  
				\label{fig:ende_shared}
				\vspace{-0.3cm}
			\end{figure}
            \renewcommand\arraystretch{0.87}
			\begin{table*}[t]
				\centering
				\small
				\caption{Descriptions of the baseline methods used in the LJSpeech or LibriTTS benchmark.}
				\label{tbl:descriptions-baselines}
				\resizebox{0.76\textwidth}{!}{
					\begin{tabular}{ccccc}
						\toprule
						Methods &Type &Publish &Domain &Code Link\\
						\midrule
						HiFiGAN-V1~{\cite{kong2020hifi}} &GAN  &NeurIPS 2020  &T &https://github.com/jik876/hifi-gan\\
						iSTFTNet-V1~{\cite{kaneko2022istftnet}} &GAN  &ICASSP 2022 &Hybrid &https://github.com/deepvk/istftnet \\
                        \lad{PriorGrad~{\cite{leepriorgrad}}} &\lad{DDPM} &\lad{ICLR 2022} &\lad{T} &\lad{https://github.com/microsoft/NeuralSpeech/tree/master/PriorGrad-vocoder} \\
						Avocodo~{\cite{bak2023avocodo}} &GAN  &AAAI 2023 &T &https://github.com/ncsoft/avocodo\\
						HiFTNet~{\cite{li2023hiftnet}} &GAN &Arxiv 2023 &Hybrid &https://github.com/yl4579/HiFTNet\\
						BigVGAN~{\cite{leebigvgan}} &GAN &ICLR 2023 &T &https://github.com/NVIDIA/BigVGAN \\
						APNet~{\cite{ai2023apnet}} &GAN &TASLP 2023 &T-F  &https://github.com/YangAi520/APNet\\
						APNet2~{\cite{du2023apnet2}} &GAN &NCMMSC 2024 &T-F &https://github.com/redmist328/APNet2  \\
						Vocos~{\cite{siuzdakvocos}} &GAN &ICLR 2024 &T-F &https://github.com/gemelo-ai/vocos \\
                        \lad{FreGrad~{\cite{nguyen2024fregrad}}} &\lad{DDPM} &\lad{ICASSP 2024}  &\lad{T} &\lad{https://github.com/kaistmm/fregrad} \\
                        FreeV~{\cite{lv24_interspeech}} &GAN &Interspeech 2024 &T-F  &https://github.com/BakerBunker/FreeV\\
                        \lad{PeriodWave~{\cite{leeperiodwave}}} &\lad{FM} &\lad{ICLR 2025} &\lad{T} &\lad{https://github.com/sh-lee-prml/PeriodWave} \\
                        \lad{WaveFM~{\cite{luo2025wavefm}}} &\lad{FM} &\lad{NAACL 2025} &\lad{T} &\lad{https://github.com/luotianze666/WaveFM} \\
						\bottomrule
				\end{tabular}}
				\vspace{-0.35cm}
			\end{table*}
			\subsubsection{Dual-path Module}
			\label{sec:dual-path-module}
			\lad{For speech representations in the T-F domain, correlations may exist simultaneously against adjacent time frames and frequency bands. For instance, formant transitions in consonant-vowel sequences manifest across both adjacent time frames and correlated frequency bands~{\cite{story2010relation}}.} Therefore, we believe that it is beneficial to consider the joint modeling along the narrow-band and cross-band. In the DPM, we stack $B$ dual-path blocks (DPBs) to facilitate the spectral information modeling along the narrow-band and cross-band. The internal structure of each DPB is illustrated in Fig.~{\ref{fig:framework}}(d). Specifically, it consists of two sub-models, namely cross-band module and narrow-band module. 
            
			\textbf{Cross-Band Module}: Given the input feature $\mathbf{F}^{\left(b\right)}\in\mathbb{R}^{N\times C\times T}$, as this module aims to model the inter-relation among different subbands, we first reshape it to $\mathbb{R}^{T\times C\times N}$. Inspired by~{\cite{quan2024spatialnet}}, we adopt a lightweight module for subband modeling. It involves three steps. \lad{First, layer normalization (LN) is applied, followed by 
            a grouped convolution along the subband axis with kernel size of 3 and groups size of 8, and a PReLU activation~{\cite{he2015delving}}, to capture correlations between adjacent subbands, formulated as}:
			\begin{equation}
				\label{eqn:24}
				\mathbf{F}^{\left(b\right)'} = \mathbf{F}^{\left(b\right)} + \text{PReLU}\left(\text{GConv1d}\left(\text{LN}\left(\mathbf{F}^{\left(b\right)}\right)\right)\right),
			\end{equation}
			\lad{where $\mathbf{F}^{\left(b\right)'}\in\mathbb{R}^{T\times C\times N}$. Subsequently, a point-wise Conv1d layer, followed by SiLU activation~{\cite{ramachandran2017searching}}, is adopted to compress the feature channel to $C^{'}$, where $C^{'} = C / 4$. A band-mixer performs global subband shuffling, which is instantiated by a linear layer. This is followed by a point-wise Conv1d and a SiLU activation to restore the channel dimension to $C$.} This process is formulated as:
			\begin{align}
				\label{eqn:25}
				\mathbf{F}^{\left(b\right)''} = &\mathbf{F}^{\left(b\right)'} + \text{SiLU}(\text{Conv1d}(\text{Mixer}(\text{SiLU}(\nonumber\\
				&\text{Conv1d}(\text{LN}(\mathbf{F}^{(b)'})))))),
			\end{align}
			\lad{where $\mathbf{F}^{\left(b\right)''}\in\mathbb{R}^{T\times C\times N}$. Finally, to strengthen the correlation between adjacent subbands, another group convolution with kernel of 3 and group number of 8 is adopted with residual connection}:
			\begin{equation}
				\label{eqn:26}
				\mathbf{F}^{\left(b+1\right)} = \mathbf{F}^{\left(b\right)''} + \text{PReLU}\left(\text{GConv1d}\left(\text{LN}\left(\mathbf{F}^{\left(b\right)''}\right)\right)\right),
			\end{equation}
            \lad{where $\mathbf{F}^{\left(b+1\right)}\in\mathbb{R}^{T\times C\times N}$. Note that within the cross-band module, the number of subbands and frames remains unchanged, and the ``same'' padding scheme is adopted for each convolution operation.}
            
			\textbf{Narrow-Band Module}: We stack $P$ ConvNext v2 blocks~{\cite{woo2023convnext}} for time modeling, which has proven to outperform ConvNext v1 and ResNet in various computer vision and audio tasks~{\cite{liu2023survey,du2023apnet2}}. Compared to ConvNext v1~{\cite{liu2022convnet}}, it incorporates a global response normalization (GRN) layer, which increases the contrast and selectivity of channels and enriches the feature diversity. Despite ConvNext v2 blocks have been adopted in previous neural vocoder literature~{\cite{du2023apnet2}}, there are several distinctions: First, in our method, the narrow-band module processes each subband independently, and all subbands share the same network parameters. Besides, to reduce the computational cost, the number of hidden channels is set equal to the input channel.

            \lad{Specifically, given the input feature $\mathbf{F}^{\left(b\right)}\in\mathbb{R}^{N\times C\times T}$, it first passes a depth-wise Conv1d with kernel size of 7, followed by LN, point-wise Conv1d and GELU activation. Then the GRN is applied, followed by another point-wise Conv1d. The residual connection is applied, formulated as:
            }
            \lad{\begin{align}
				\label{eqn:27}
				\mathbf{F}^{\left(b+1\right)} &= \mathbf{F}^{\left(b\right)} + \text{Conv1d}(\text{GRN}(\text{GELU}((\nonumber \\ &\text{Conv1d}(\text{LN}(\text{DWConv1d}(\mathbf{F}^{\left(b\right)}))))))),
			\end{align}}
			\lad{where $\mathbf{F}^{\left(b+1\right)}\in\mathbb{R}^{N\times C\times T}$}. Note that in~{\cite{luo2023music}}, a similar dual-path structure was proposed for music source separation, in which RNN layers were used. In Sec.~{\ref{sec:ablation-studies}}, we compare the performance of RNN and our method.
            \begin{figure}
            \centering
            \vspace{0pt}
            \includegraphics[width=0.83\columnwidth]{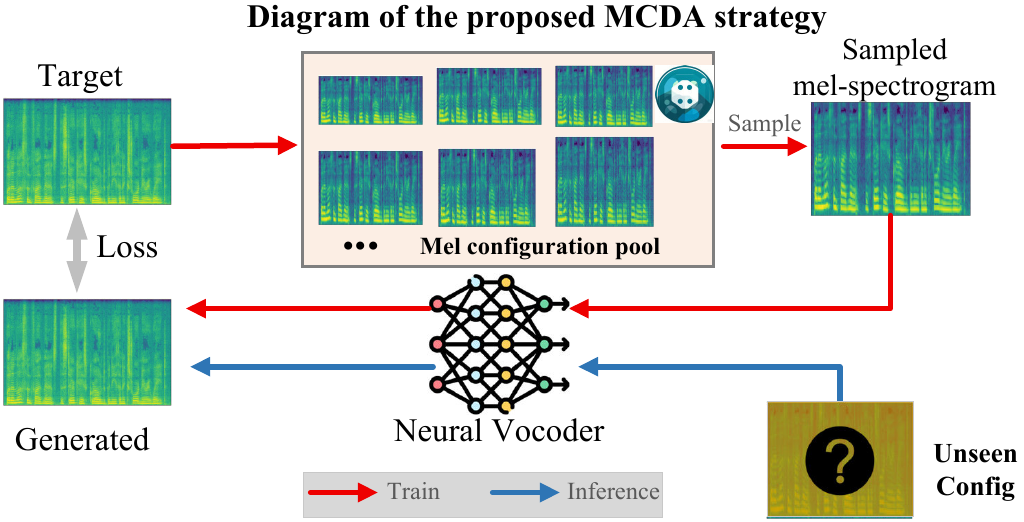}
            \vspace{-2pt}
            \caption{\lad{Illustration of the proposed MCDA strategy.}}  
            \label{fig:mcda}
            \vspace{-0.3cm}
			\end{figure}
			\subsubsection{Multi-condition-as-Data-Augmentation Strategy}
			\label{sec:multi-condition-as-data-augmentation}
			\lad{Previous neural vocoders typically require retraining to accommodate alternative mel-spectrogram configurations, which is inefficient and labor-intensive. To address this limitation, we propose a simple yet effective training strategy, termed MCDA, that enables seamless adaptation to diverse mel-spectrogram settings during inference without retraining.} 
			\lad{As shown in Fig.~{\ref{fig:mcda}}}, we first construct a mel configuration pool $\mathcal{P} = \left\{p^{jk} = \left(F_{m}^{j}, f_{\text{max}}^{k}\right)\mid j\in\left\{1,\cdots,J\right\}, k\in\left\{1,\cdots,K\right\} \right\}$, where $J$ and $K$ denote the number of candidate mel bands and \lad{the upper-bound frequency $f_{\text{max}}$ to sample}\footnote{\lad{Mel-filter bank construction primarily depends on $\left\{F_{m}, f_{\text{max}}\right\}$, which is the reason why we consider the two factors. While filter implementation (\emph{e.g.}, HTK or Slaney formula) and normalization can also affect calculation, we adopt the Slaney with region normalization, as chosen in the \texttt{librosa} package by default. Extending MCDA to include more factors is straightforward, and we leave this extension for future work. Nonetheless, we analyze its robustness to unseen implementation, and corruption by varying level of Gaussian noise.}}. A corresponding mel-filter pool $\mathcal{T}$ is then precomputed via the mapping:
			\begin{equation}
				\label{eqn:28}
				\mathcal{T} = \left\{\mathcal{A}^{jk}\right\}, \text{where }  \mathcal{A}^{jk} \text{ is calculated from } p^{jk}\noindent\in\mathcal{P}.
			\end{equation}
			
			\lad{Note that the construction of $\mathcal{T}$ can be implemented before training, with all filter weights pre-computed and stored in advance. As each condition is uniquely characterized by a linear matrix $\in\mathbb{R}^{F_{m}^{j}\times F}$, only a finite set of matrices needs to be pre-saved. During training, a mel-condition $p^{jk}$ is randomly sampled from the set $\mathcal{P}$, and the corresponding filter matrix $\mathcal{A}^{jk}$ is retrieved via direct indexing. The mel-spectrogram is then computed and fed into the range-space module to derive its representation in the linear-scale domain. In the inference stage, the pretrained model is capable of generating high-quality target waveforms from mel-spectrograms under both seen and unseen conditions. Since the proposed MCDA method is plug-and-play, the framework can be easily degraded to a single-condition counterpart by disabling the MCDA strategy, thus demonstrating superior flexibility. The pseudo-code is provided in Sec.~V of the Appendix.}
			\renewcommand\arraystretch{0.90}
			\begin{table*}
				\centering
				\Huge
				\caption{The ablation studies conducted on the LJSpeech dataset. For the inference speed, ``$a\times$'' denotes the inference ratio over the real-time. ``P'' denotes the number of ConvNext v2 blocks within each narrow-band module. \lad{The inference speed on a CPU is based on a CPU Intel(R) Core(TM) i7-14700F, and for GPU it is based on NVIDIA GeForce RTX 4060 Ti.} $\uparrow$ denotes the higher the better, and $\downarrow$ denotes the opposite situation. The best results are labeled in bold.}
				\vspace{-6pt}
				\resizebox{0.98\textwidth}{!}{
					\label{tab:ablations}
					\begin{tabular}{|cl|c|cccc|ccccc|}
						\hline
						\multicolumn{2}{|c|}{\multirow{2}*{Settings}} &\multirow{2}*{P} &\#Param. &\#MACs &\multicolumn{2}{c|}{Inference Speed} &\multirow{2}*{PESQ$\uparrow$} &\multirow{2}*{MCD$\downarrow$} &Periodicity$\downarrow$ &V/UV$\uparrow$ &Pitch$\downarrow$\\
						\cline{6-7}
						& & &(M) &(Giga/5s) &CPU &GPU & & &RMSE &F1 &RMSE\\
						\hline\hline
						(1) &Remove Narrow-Band module &- &7.88 &8.36 &0.0419(23.88$\times$) &0.0048(210$\times$) &1.697 &\lad{4.890} &\lad{0.291} &\lad{0.866} &\lad{89.166}\\
						(2) &Remove Cross-Band module &2 &8.98 &19.81  &0.0527(18.98$\times$) &0.0050(199$\times$) &3.157 &\lad{3.428} &\lad{0.130} &\lad{0.953} &\lad{32.555} \\
						(3) &Replace the proposed module with RNN  &2 &21.59 &150.81 &0.2098(4.77$\times$) &0.0211(47$\times$) &3.781 &\lad{2.402} &\lad{0.098} &\lad{0.966} &\lad{22.451} \\
						\hline\hline
						(4) &\multirow{4}*{Increase the number of ConvNext v2 blocks} &1 &8.68 &16.67 &0.0581(17.22$\times$) &0.0061(164$\times$) &3.651 &\lad{2.462} &\lad{0.104} &\lad{0.963} &\lad{24.652} \\
						(5) &  &2 &9.48 &24.98 &0.0770(12.98$\times$) &0.0063(158$\times$) &3.763 &\lad{2.394} &\lad{0.099} &\lad{0.966} &\lad{23.419} \\
						(6) & &3 &10.29  &33.29 &0.0884(11.31$\times$) &0.0069(145$\times$) &3.800 &\lad{2.349}  &\lad{0.098} &\lad{0.966} &\lad{23.500} \\
						(7) & &4 &11.09 &41.60 &0.1016(9.84$\times$) &0.0155(64$\times$) &3.851 &\lad{2.306} &\lad{0.093} &\lad{0.968} &\lad{22.174} \\
						\hline\hline
						(8) &Remove RND mode &2 &9.48 &24.98 &0.0714(14.01$\times$) &0.0063(158$\times$) &3.655 &\lad{2.498} &\lad{0.111} &\lad{0.961} &\lad{26.012} \\
						(9) &Based on (8), replace RND with null-constraint &2 &9.48 &24.98 &0.0718(13.93$\times$) &0.0063(158$\times$) &3.672 &\lad{2.496} &\lad{0.111} &\lad{0.962} &\lad{25.127} \\
                        \lad{(10)} &\lad{Use global skip-connection} &\lad{2} &\lad{9.48} &\lad{24.98} &\lad{0.0736(13.58$\times$)} &\lad{0.0065(152$\times$)} &\lad{3.695} &\lad{2.496} &\lad{0.109} &\lad{0.963} &\lad{24.678} \\ 
                        \lad{(11)} &\lad{Fix $\mathcal{A}$, replace $\mathcal{A}^{\dagger}$ with linear projection} &\lad{2} &\lad{9.48} &\lad{24.98} &\lad{0.0724(13.81$\times$)} &\lad{0.0060(167$\times$)} &\lad{3.620} &\lad{2.608} &\lad{0.115} &\lad{0.960} &\lad{26.167}\\
						(12) &Set matrices $\left\{\mathcal{A}, \mathcal{A}^{\dagger}\right\}$ as learnable &2 &9.48 &24.98 &0.0740(13.51$\times$) &0.0059(169$\times$) &3.645 &2.599 &0.116 &0.959 &25.629 \\
                        \lad{(13)} &\lad{Based on (12), add an explicit idempotent constraint} &\lad{2} &\lad{9.48} &\lad{24.98} &\lad{0.0741(13.50$\times$)} &\lad{0.0061(164$\times$)} &\lad{3.616} &\lad{2.563} &\lad{0.117} &\lad{0.956} &\lad{24.776} \\ 
						\hline\hline
						(14) &Use omnidirectional phase loss &2 &9.48 &24.98 &0.0773(12.94$\times$) &0.0063(158$\times$) &3.892 &2.214 &0.089 &0.970 &21.912 \\
						\hline\hline
						(15) &Based on (14), use shared scheme for encoding/decoding &2 &3.14 &34.10 &0.0669(14.96$\times$) &0.0043(233$\times$) &3.987 &2.047 &0.085 &0.971 &21.346\\
						\hline
				\end{tabular}}
				\vspace{-8pt}
			\end{table*}
			\vspace{-0.35cm}
			\subsection{Training Loss}
			\label{sec:training-loss}
			\subsubsection{Basic Settings}
			\label{sec:basic-setting}
			Following previous literature~{\cite{ai2023apnet,defossez2023high,du2023apnet2,ai2024apcodec}}, we incorporate both reconstruction and adversarial losses as the training criteria.
            
			\textbf{Reconstruction loss}: It consists of log-spectral amplitude loss $\mathcal{L}_{a}$, phase loss $\mathcal{L}_{p}$, real-imaginary loss $\mathcal{L}_{ri}$, mel loss $\mathcal{L}_{m}$, and consistency loss $\mathcal{L}_{c}$, which is illustrated below:
			
			\textit{(1) Log-spectral amplitude loss}: It is defined as the mean-square error (MSE) between $\mathbf{\tilde{S}}$ and $\mathbf{S}$ in the log-magnitude domain:
			\begin{equation}\label{eqn:29}
				\mathcal{L}_{a} = \frac{1}{FT}\sum_{f,t}\left\|\log{\left|\mathbf{\tilde{S}}_{f,t}\right|} - \log{\left|\mathbf{S}_{f,t}\right|}\right\|_{2}^{2}.
			\end{equation}
			
			\textit{(2) Phase loss}: It is usually non-trivial to optimize phase due to the wrapping effect. In~{\cite{ai2024apcodec}}, a specially designed anti-wrapping phase loss was proposed by incorporating instantaneous phase (IP), group delay (GD) and instantaneous frequency (IF)~{\cite{gerkmann2015phase}}, given by
			\begin{equation}
				\label{eqn:30}
				\mathcal{L}_{p} = \mathcal{L}_{IP} + \mathcal{L}_{GD} + \mathcal{L}_{IF},
			\end{equation}
			\begin{equation}
				\label{eqn:31}
				\mathcal{L}_{IP} = \frac{1}{FT}\sum_{f,t}\left\|f_{AW}\left(\mathbf{\tilde{\Phi}}_{f,t} - \mathbf{\Phi}_{f,t}\right)\right\|_{1},
			\end{equation}
			\begin{equation}
				\label{eqn:32}
				\mathcal{L}_{GD} = \frac{1}{FT}\sum_{f,t}\left\|f_{AW}\left(\Delta_{F}\mathbf{\tilde{\Phi}}_{f,t} - \Delta_{F}\mathbf{\Phi}_{f,t}\right)\right\|_{1},
			\end{equation}
			\begin{equation}
				\label{eqn:33}
				\mathcal{L}_{IF} = \frac{1}{FT}\sum_{f,t}\left\|f_{AW}\left(\Delta_{T}\mathbf{\tilde{\Phi}}_{f,t} - \Delta_{T}\mathbf{\Phi}_{f,t}\right)\right\|_{1},
			\end{equation}
			and $f_{AW}\left(x\right) = \left|x - 2\pi\text{round}\left(\frac{x}{2\pi}\right)\right|$ denotes the anti-wrapping loss, and $\left\{\Delta_{F}, \Delta_{T}\right\}$ refer to the differential along the frequency and time axes, respectively.
			
			\textit{(3) Real and imaginary loss}: It is defined as the mean absolute error (MAE) between the estimated real/imaginary components and target:
			\begin{equation}
				\small
				\label{eqn:34}
				\begin{split}
					\mathcal{L}_{ri} &=  \frac{1}{FT}\sum_{f,t}\left(\left\|\mathcal{R}(\mathbf{\tilde{S}}_{f,t}) - \mathcal{R}\left(\mathbf{S}_{f,t}\right)\right\|_{1} \right. \\
					& \left. +\left\|\mathcal{I}(\mathbf{\tilde{S}}_{f,t}) - \mathcal{I}\left(\mathbf{S}_{f,t}\right)\right\|_{1}\right)
				\end{split}
			\end{equation}
			
			\textit{(4) Mel Loss}: By calculating the mel-spectrogram from the estimated and target waveforms, the mel loss is defined as the MAE between $\mathbf{\tilde{X}}^{mel}$ and $\mathbf{X}^{mel}$:
			\begin{equation}
				\label{eqn:35}
				\mathcal{L}_{m} = \frac{1}{F_{m}T}\sum_{f,t}\left\|\mathbf{\tilde{X}}^{mel}_{f,t} - \mathbf{X}^{mel}_{f,t}\right\|_{1}.
			\end{equation}
			
			\textit{(5) Consistency loss}: Inconsistency can arise when the generated spectrogram in the T-F domain is not necessarily equal to the STFT of its time-domain version~{\cite{wisdom2019differentiable}}. To alleviate the inconsistency issue between the generation $\mathbf{\tilde{S}}$ and its consistency counterpart $\mathbf{\hat{S}} = \text{STFT}\left(\text{iSTFT}\left(\mathbf{\tilde{S}}\right)\right)$, the consistency loss is defined as:
			\begin{align}
				\label{eqn:36}
				\mathcal{L}_{c} &=  \frac{1}{FT}\sum_{f,t}\left(\left\|\text{Re}(\mathbf{\tilde{S}}_{f,t}) - \text{Re}\left(\hat{\mathbf{S}_{f,t}}\right)\right\|_{2}^{2} \right.\\ 
				& \left. +\left\|\text{Im}(\mathbf{\tilde{S}}_{f,t}) - \text{Im}(\mathbf{\hat{S}}_{f,t})\right\|_{2}^{2}\right).
			\end{align}
			
			In summary, the reconstruction loss $\mathcal{L}_{rec}$ is represented as: 
			\begin{equation}
				\label{eqn:37}
				\mathcal{L}_{rec} = \lambda_{a}\mathcal{L}_{a} + \lambda_{p}\mathcal{L}_{p} + \lambda_{ri}\mathcal{L}_{ri} + \lambda_{m}\mathcal{L}_{m} + \lambda_{c}\mathcal{L}_{c},
			\end{equation}
			where $\left\{\lambda_{a}, \lambda_{p}, \lambda_{ri}, \lambda_{m}, \lambda_{c}\right\}$ are the corresponding loss hyperparameters.
            
			\textbf{Adversarial loss}: Like prior work, we adopt the multi-period discriminator (MPD)~{\cite{kong2020hifi}} and multi-resolution spectrogram discriminator (MRSD)~{\cite{jang21_interspeech}} for adversarial training. MPD is utilized to capture varying audio periodic patterns and comprises of five sub-discriminators. For each sub-discriminator, the 1D raw audio waveform is reshaped into 2D data format with the period value of $p$, and sent to consecutive Conv2d layers and leaky ReLU to compute the score. The periods are set to $\left\{2, 3, 5, 7, 11\right\}$. For MRSD, it includes three sub-discriminators, in each of which the magnitude spectrum serves as the input and is processed by a stack of Conv2d layers to calculate the discriminative score. The $\left\{\text{window\_size}, \text{hop\_size}, \text{n\_fft}\right\}$ in three sub-discriminators are $\left(512,128,512\right)$, $\left(1024,256,1024\right)$, $\left(2048,512,2048\right)$.
			
			We adopt the hinge GAN as the adversarial loss. For discriminator, the adversarial loss can be presented as:
			\begin{equation}
				\label{eqn:38}
				\mathcal{L}_D = \frac{1}{M}\sum_{m=1}^{M}\max\left(0,1-D_{m}\left(\mathbf{s}\right)\right) + \max\left(0, 1+D_{m}\left(\mathbf{\tilde{s}}\right)\right),
			\end{equation}
			where $M$ is the number of sub-discriminators. For generator, the adversarial loss is:
			\begin{equation}
				\label{eqn:39}
				\mathcal{L}_{g} = \frac{1}{M}\sum_{m=1}^{M}\max\left(0, 1 - D_{m}\left(\mathbf{\tilde{s}}\right)\right).
			\end{equation}
			
			Besides, feature matching loss is utilized, given by:
			\begin{equation}
				\label{eqn:40}
				\mathcal{L}_{fm} = \frac{1}{LM}\sum_{l,m}\left|\mathbf{f}_{l}^{m}\left(\mathbf{\tilde{s}}\right) - \mathbf{f}_{l}^{m}\left(\mathbf{s}\right) \right|,
			\end{equation}
			where $\mathbf{f}_{l}^{m}\left(\cdot\right)$ denotes the $l$-th layer feature for $m$-th sub-discriminator.  The final loss for generator is presented as:
			\begin{equation}
				\label{eqn:41}
				\mathcal{L}_{G} = \mathcal{L}_{rec} + \lambda_{g}\mathcal{L}_{g} + \lambda_{fm}\mathcal{L}_{fm},
			\end{equation}
			where $\left\{\lambda_{g}, \lambda_{fm}\right\}$ are the corresponding hyperparameters.
            \begin{figure}
				\centering
				\vspace{0pt}
				\includegraphics[width=0.76\columnwidth]{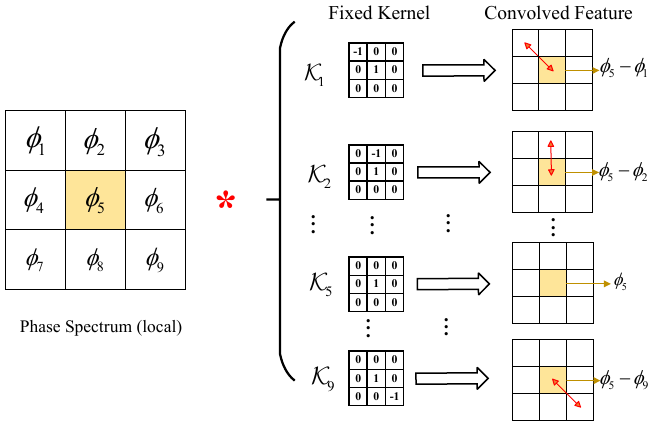}
				\vspace{-2pt}
				\caption{Illustration of the proposed omnidirectional phase loss.}  
				\label{fig:omni_loss}
				\vspace{-0.45cm}
			\end{figure}
			\begin{figure}
				\centering
				\vspace{0pt}
				\includegraphics[width=0.46\textwidth]{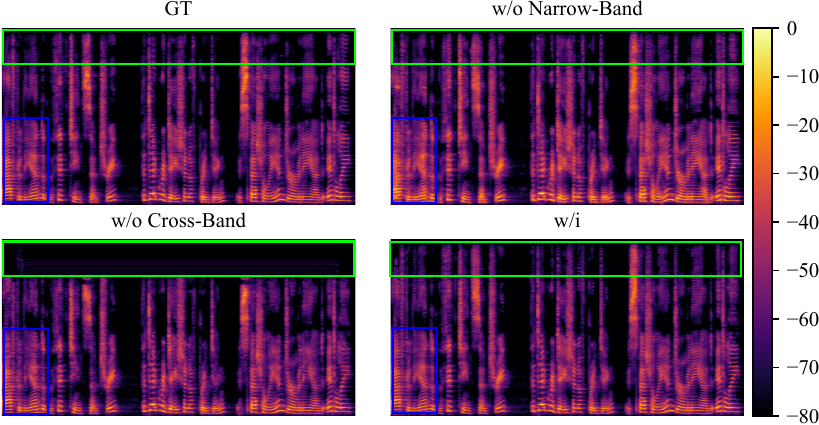}
				\vspace{-2pt}
				\caption{A typical spectral visualization where either narrow-band or cross-band module is removed. Within each figure, the x- and y-coordinates represent time and frequency, respectively.}  
				\label{fig:ablation_spec_visualization}
				\vspace{-14pt}
			\end{figure} 
            \renewcommand\arraystretch{1.00}
			\begin{table*}
				\centering
				\Huge
				\caption{Objective comparisons among different \lad{GAN-based} baselines on the LJSpeech benchmark. The best and second-best performances are respectively highlighted in bold and underlined.}
				\vspace{-6pt}
				\resizebox{0.99\textwidth}{!}{
					\label{tbl:objective-metric-ljs}
					\begin{tabular}{c|cccc|cccccccc}
						\toprule
						\multirow{2}*{Models} &\#Param. &\#MACs &\multicolumn{2}{c|}{Inference Speed} &\multirow{2}*{M-STFT} &\multirow{2}*{PESQ} &\multirow{2}*{MCD} &Periodicity &V/UV &Pitch &\multirow{2}*{\lad{UTMOS}} &\multirow{2}*{VISQOL}\\
						\cline{4-5}
						&(M) &(Giga/5s) &CPU &GPU & & & &RMSE &F1 &RMSE & & \\
						\midrule
						HiFiGAN-V1 &13.94 &152.90 &0.1669(5.99$\times$) &0.0069(145$\times$) &\lad{1.167} &3.574 &\lad{3.671} &\lad{0.134} &\lad{0.947} &\lad{33.687} &\lad{4.218} &\lad{4.771} \\
						iSTFTNet-V1 &13.26 &107.76 &0.0949(10.54$\times$) &0.0038(266$\times$) &\lad{1.188} &3.535 &\lad{3.625} &\lad{0.135} &\lad{0.947} &\lad{35.106} &\lad{4.236} &\lad{4.756} \\
						Avocodo &13.94 &152.95 &0.1611(6.21$\times$) &0.0063(158$\times$) &\lad{1.165} &3.604 &\lad{3.657} &\lad{0.139} &\lad{0.946} &\lad{32.989} &\lad{4.152} &\lad{4.771}\\
						HiFTNet &22.43 &163.05 &0.1559(6.41$\times$) &0.0085(117$\times$) &\lad{0.923} &\underline{3.992} &\lad{\underline{2.039}} &\lad{0.092} &\lad{0.966} &\lad{\underline{20.780}} &\lad{\textbf{4.343}} &\lad{\underline{4.849}} \\
						BigVGAN-base &13.94 &152.90 &0.3856(2.59$\times$) &0.0368(27$\times$) &\lad{0.978} &3.603 &\lad{2.331} &\lad{0.120} &\lad{0.956} &\lad{30.277} &\lad{4.208} &\lad{4.822} \\
						BigVGAN &112.18 &417.20 &0.6674(1.50$\times$) &0.0507(20$\times$) &\lad{\textbf{0.900}} &\lad{\textbf{4.107}} &\lad{\textbf{1.877}} &\lad{\textbf{0.084}} &\lad{\textbf{0.972}} &\lad{\textbf{20.692}} &\lad{\underline{4.314}} &\lad{\textbf{4.870}} \\
						APNet &72.19 &31.11 &0.0300(33.33$\times$) &0.0022(458$\times$) &\lad{1.266} &3.390 &\lad{3.285} &\lad{0.151} &\lad{0.945} &\lad{23.057} &\lad{3.170} &\lad{4.695} \\
						APNet2 &31.38 &13.53 &0.0187(53.48$\times$) &0.0016(643$\times$) &\lad{0.982} &3.492 &\lad{2.829} &\lad{0.113} &\lad{0.959} &\lad{25.363} &\lad{3.984} &\lad{4.752}  \\
						FreeV  &18.19 &7.84 &\underline{0.0139(71.94$\times$)} &\underline{0.0011(951$\times$)} &\lad{1.008} &3.593 &\lad{2.750} &\lad{0.112} &\lad{0.960} &\lad{25.992} &\lad{4.010} &\lad{4.743} \\
						Vocos &13.46  &5.80 &\textbf{0.0066(151.52$\times$)} &\textbf{0.0007(1400$\times$)} &\lad{1.012} &3.522 &\lad{2.670} &\lad{0.121} &\lad{0.956} &\lad{29.130} &\lad{3.968} &\lad{4.774} \\
						\hline
						\textbf{RNDVoC-nonshared(Ours)} &9.48 &24.98 &0.0770(12.98$\times$) &0.0063(158$\times$) &\lad{0.934} &3.892 &\lad{2.214} &\lad{0.089} &\lad{0.970} &\lad{21.952} &\lad{4.138} &\lad{4.812} \\
						\textbf{RNDVoC-shared(Ours)} &3.14 &34.10 &0.0669(14.96$\times$) &0.0043(233$\times$) &\lad{\underline{0.910}} &3.987 &\lad{2.047} &\lad{\underline{0.085}} &\lad{\underline{0.971}} &\lad{21.318} &\lad{4.161} &\lad{4.837} \\
						\hline
				\end{tabular}}
			\end{table*}
		      \renewcommand\arraystretch{1.00}
			\begin{table*}
				\centering
				\caption{Objective comparisons among different \lad{GAN-based} baselines on the LibriTTS benchmark. ``-'' denotes the results are not reported, and $\dagger$ denotes the results are calculated using the open-sourced model checkpoints. }
				\hspace{-0pt}
				\vspace{-4pt}
				\resizebox{0.85\textwidth}{!}{
					\label{tbl:objective-metric-libritts}
					\begin{tabular}{c|cc|cccccccc}
						\toprule
						\multirow{2}*{Models} &\#Param. &\#MACs  &\multirow{2}*{M-STFT} &\multirow{2}*{PESQ} &\multirow{2}*{MCD} &Periodicity &V/UV &Pitch &\multirow{2}*{\lad{UTMOS}}  &\multirow{2}*{VISQOL} \\
						&(M) &(Giga/5s) & & & &RMSE &F1 &RMSE & & \\
						\midrule
						HiFiGAN-V1 &14.01 &166.41  &\lad{1.104} &3.056 &\lad{4.284} &\lad{0.167} &\lad{0.921} &\lad{52.529} &\lad{3.511} &\lad{4.721}  \\
						iSTFTNet-V1 &13.33 &117.30 &\lad{1.157} &2.880 &\lad{4.480} &\lad{0.167} &\lad{0.918} &\lad{53.072} &\lad{3.464} &\lad{4.655}  \\
						Avocodo &14.01 &166.47  &\lad{1.113} &3.217 &\lad{4.314} &\lad{0.161} &\lad{0.913} &\lad{51.600} &\lad{3.358} &\lad{4.762} \\
						BigVGAN-base (1M steps)$\dagger$ &14.01 &166.41  &\lad{0.879} &3.519 &- &\lad{0.129} &\lad{0.946} &- &- &-  \\
						BigVGAN (1M steps)$\dagger$ &112.39 &454.08  &\lad{0.800} &4.027 &- &\lad{0.102} &\lad{0.960} &-
						&- &- \\
						BigVGAN-base (5M steps)$\dagger$ &14.01 &166.41  &\lad{0.829} &3.841 &\lad{2.431} &\lad{0.107} &\lad{0.954} &\lad{32.541} &\lad{3.617} &\lad{4.907} \\
						BigVGAN (5M steps)$\dagger$ &112.39 &454.08  &\textbf{0.736} &\textbf{4.269} &\lad{\textbf{1.871}} &\lad{\underline{0.079}} &\lad{\underline{0.967}} &\lad{\underline{24.281}} &\lad{\textbf{3.744}} 
						&\lad{\textbf{4.963}} \\
						APNet &73.33 &33.92  &\lad{1.273} &2.897 &\lad{4.150} &\lad{0.159} &\lad{0.927} &\lad{39.663} &\lad{2.416} &\lad{4.666}\\
						APNet2 &31.52 &\underline{14.79} &\lad{1.143} &2.834 &\lad{4.153} &\lad{0.153} &\lad{0.923} &\lad{46.373} &\lad{2.956} &\lad{4.582}  \\
						Vocos$\dagger$ &13.53  &\textbf{6.35} &\lad{0.854} &3.615 &\lad{3.105} &\lad{0.115} &\lad{0.948} &\lad{35.584} &\lad{3.566} &\lad{4.879} \\
						\hline
						\textbf{RNDVoC-nonshared(Ours)} &\underline{9.48} &28.29  &\lad{0.798} &4.085 &\lad{2.251} &\lad{0.090} &\lad{0.961} &\lad{27.041} &\lad{3.607} &\lad{4.873}  \\
						\textbf{RNDVoC-shared(Ours)} &\textbf{3.14} &37.20  &\lad{\underline{0.746}} &\underline{4.226} &\lad{\underline{1.914}} &\lad{\textbf{0.074}} &\lad{\textbf{0.970}} &\lad{\textbf{23.966}} &\lad{\underline{3.657}} &\lad{\underline{4.915}} \\
						\hline
				\end{tabular}}
                \vspace{-0.3cm}
			\end{table*}
			\subsubsection{Omnidirectional Phase Loss}\label{sec:omnidirectional}
			In the previous phase loss~{\cite{ai2024apcodec}}, the differential operations $\Delta_{F}$ and $\Delta_{T}$ are realized via first constructing sparse differential matrices and then implementing matrix multiplication{\footnote{https://github.com/YangAi520/APCodec/blob/main/models.py}}. This incurs significant computational overhead as most of the elements in the sparse matrix are zero and have no impact on the result. Besides, the previous phase loss only considers the differential relations on merely two directions, neglecting rich relations with more adjacent T-F bins. To this end, we propose a novel omnidirectional phase loss, which regards differential operation as fixed $3\times 3$ convolutional kernels in Conv2d, 
			
			As indicated in Fig.~{\ref{fig:omni_loss}}, we specially design nine $3\times3$ kernels with fixed parameters $\mathcal{K} = \text{Cat}\left(\mathcal{K}_{1},\cdots,\mathcal{K}_{9}\right)\in\mathbb{R}^{9\times3\times3}$ to traverse the differential relations with adjacent eight T-F bins, and the fifth kernel is to return the instantaneous phase. Therefore, with a simple convolution operation, the phase differential can be efficiently implemented as:
			\begin{equation}
				\label{eqn:42}
				\Delta\mathbf{\Phi} = \mathbf{\Phi}*\mathcal{K}, \Delta\mathbf{\tilde{\Phi}} = \mathbf{\tilde{\Phi}}*\mathcal{K},
			\end{equation}
			where $\left\{\Delta\mathbf{\Phi}, \Delta\mathbf{\tilde{\Phi}}\right\}\in\mathbb{R}^{9\times F\times T}$, and $*$ denotes the convolution operation. The omnidirectional phase loss can be defined as:
			\begin{equation}
				\label{eqn:43}
				\mathcal{L}_{p} = \frac{1}{FT}\sum_{f,t}\left\|f_{AW}\left(\Delta\mathbf{\Phi}_{f,t} - \Delta\mathbf{\tilde{\Phi}}_{f,t}\right)\right\|_{1}.
			\end{equation}
			
			Compared with previous phase loss, the differential operation is instantiated with a Conv2D layer, which is highly optimized by PyTorch platform and thus enables faster implementation speed. Besides, our method enables the phase modeling with neighboring eight T-F bins, leading to better phase reconstruction quality.
			\vspace{-4pt}
			\section{Experimental Setups}\label{sec:experimental-setup}
			\subsection{Dataset Preparations}
			\label{sec:dataset-preparation}
			In this study, two benchmarks, namely LJSpeech~{\cite{ljspeech17}} and LibriTTS~{\cite{zen19_interspeech}}, are employed for training. The LJSpeech dataset is a relatively small benchmark with 13,100 clean speech clips by a single female, and the sampling rate is 22.05 kHz. Following the division in the open-sourced VITS repository{\footnote{https://github.com/jaywalnut310/vits/tree/main/filelists}}, $\left\{12500, 100, 500\right\}$ clips are used for training, validation, and testing, respectively. The LibriTTS dataset covers diverse recording scenarios with a sampling rate of 24 kHz. Following~{\cite{leebigvgan}}, all training data (including \textit{train-clean-100}, \textit{train-clean-300}, and \textit{train-other-500}) are used for model training, and \textit{val-full} for validation. The subsets \textit{dev-clean}+\textit{dev-other} are for model comparisons in objective metrics, while the subsets \textit{test-clean}+\textit{test-other} are for subjective quality ratings. We follow the same division as in the open-sourced BigVGAN repository{\footnote{https://github.com/NVIDIA/BigVGAN/tree/main/filelists}}. To assess the generalization capability of neural vocoders, three out-of-distributions datasets are adopted, namely the EARS~{\cite{richter2024ears}}, VCTK~{\cite{vctk2012}}, and MUSDB18~{\cite{rafii2017musdb18}}. The EARS and VCTK datasets mainly consider the speech scenario and 200 clips are randomly selected. The MUSDB18 is a music dataset including bass, drums, and other instructments, as well as their mixture, and we only use the test set for evaluations. 
			\vspace{-10pt}
			\subsection{Mel Configurations}\label{sec:mel-configuration}
			For the LJSpeech benchmark, the number of mel-spectrogram $F_{m}$ is set to 80, and the upper-bound frequency $f_{\text{max}}$ is set to 8 kHz, \emph{i.e.}, the model is required to conduct super-resolution task to generate the spectral components over 8 kHz. \lad{For the LibriTTS benchmark, a 100-band mel-spectrogram is extracted with a fixed $f_{\text{max}}$ of 12 kHz{\footnote{\lad{the function \texttt{librosa.filters.mel} is adopted for mel-spectrogram extraction by default.}}.}} For both benchmarks, the STFT parameters are set as in previous work~{\cite{kong2020hifi}}, with 1024 FFT size, 1024 Hann window, and 256 hop size. \lad{Otherwise stated, we do not adopt the MCDA strategy for training.}
			\begin{figure}[t]
				\centering
				\vspace{0pt}
				\includegraphics[width=0.98\columnwidth]{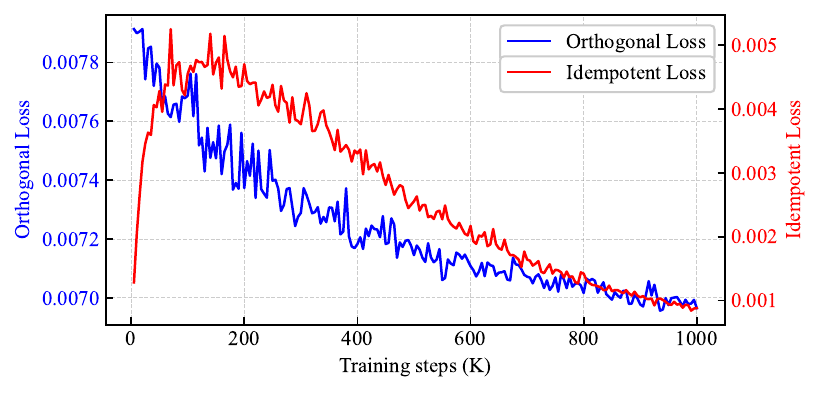}
				\vspace{-2pt}
				\caption{\lad{Loss curves of the null-constraint loss and idempotent loss.}}  
				\label{fig:orthogonal-loss}
				\vspace{-12pt}
			\end{figure}
			\vspace{-0.3cm}
			\subsection{Network Configurations}\label{sec:net-configuration}
			For the proposed method, in BAEM/BAMM/BAPM, if the nonshared scheme is adopted, totally $N = 24$ subbands are obtained. In the shared case, $I = 3$ regions are set, and the kernel and stride size in the Conv2d and TrConv2d are set to satisfy Eq.~{\eqref{eqn:23}}. In DPM, $B = 6$ DPBs are utilized, and the channel size $C$ in the narrow-band and cross-bands are set to 256 if not otherwise specified. Within each narrow-band module, $P = 2$ ConvNext v2 blocks are stacked. Consequently, the number of trainable parameters for RNDVoC-nonshared and RNDVoC-shared are 9.48 M and 3.14 M, respectively. \lad{To facilitate reproducibility, hyperparameter configurations are provided in Sec.~VII of the Appendix.}
			\vspace{-0.3cm}
			\subsection{Training Configurations}\label{sec:training-configuration}
			A batch size of 16, a segment size of 16384, and an initial learning rate of 2e-4 were used for training. The AdamW optimizer~{\cite{loshchilov2017decoupled}} is employed, with $\beta_{1} = 0.8$, $\beta_{2} = 0.99$. The generator and discriminator were updated for 1 million steps, respectively.
			\vspace{-0.3cm}
			\subsection{Model Baselines}\label{sec:model-baselines}
			\lad{In this work, various GAN- and diffusion-based baselines are utilized for comparison with the proposed RNDVoC, including HiFiGAN-V1~{\cite{kong2020hifi}}, iSTFTNet-V1~{\cite{kaneko2022istftnet}}, PriorGrad~{\cite{leepriorgrad}}, Avocodo~{\cite{bak2023avocodo}}, HiFTNet~{\cite{li2023hiftnet}}, BigVGAN~{\cite{leebigvgan}}, APNet~{\cite{ai2023apnet}}, APNet2~{\cite{du2023apnet2}}, Vocos~{\cite{siuzdakvocos}}, FreGrad~{\cite{nguyen2024fregrad}}, FreeV~{\cite{lv24_interspeech}}, PeriodWave~{\cite{leeperiodwave}}, and WaveFM~{\cite{luo2025wavefm}}.} Detailed baseline information is provided in Table~{\ref{tbl:descriptions-baselines}}. \lad{Notably, to ensure fair comparison, we retrained baselines on the LJSpeech benchmark to mitigate bias induced by inconsistent dataset splits in the original literature. For the LibriTTS benchmark, we used official pretrained checkpoints for BigVGAN{\footnote{\lad{Only the 5M-step version checkpoint is officially provided.}}}, Vocos, PeriodWave, and WaveFM, and retrained the remaining baselines.}
		\begin{figure}
			\centering
			\vspace{0pt}
			\includegraphics[width=0.9\columnwidth]{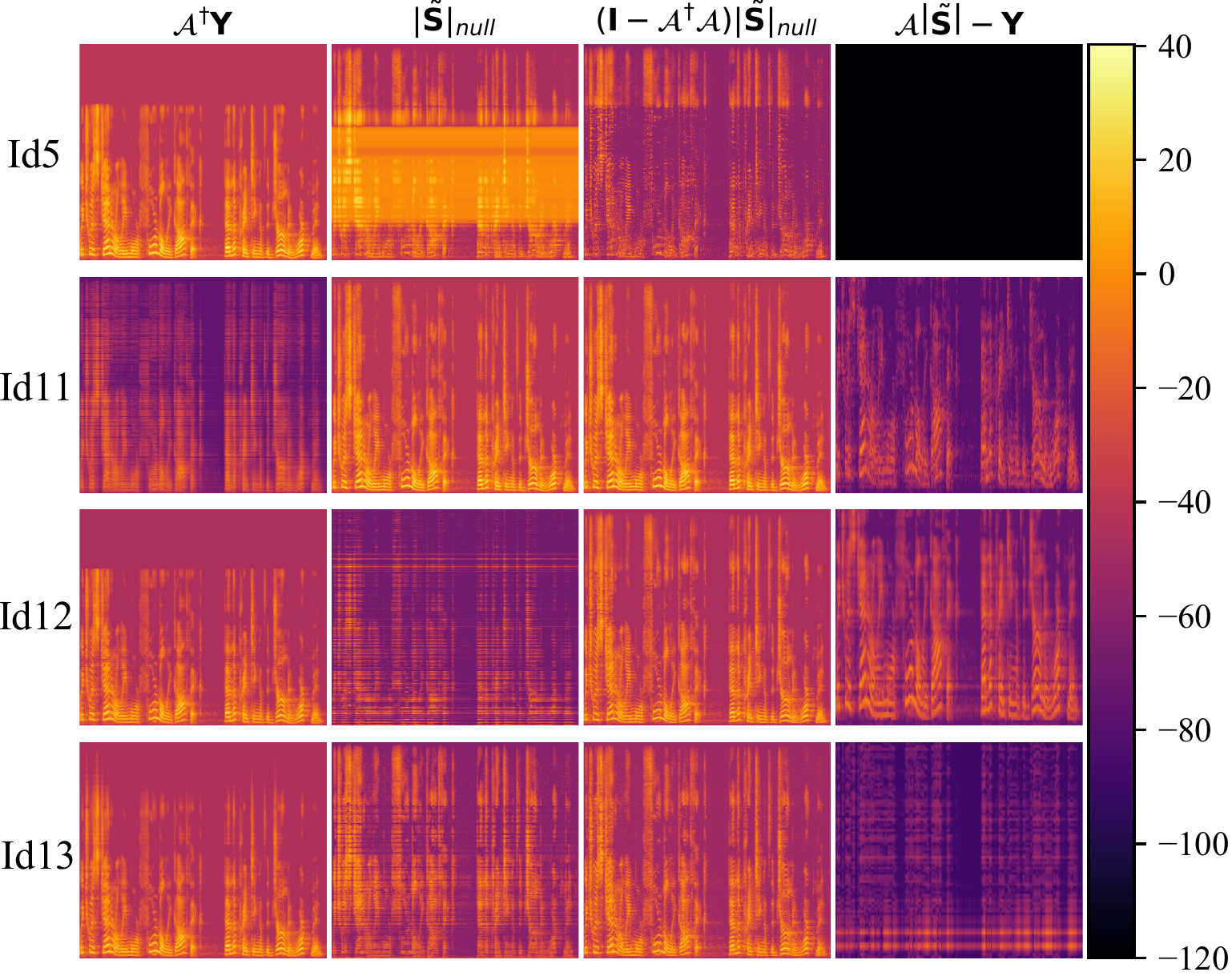}
			\vspace{-8pt}
			\caption{\lad{Spectral visualization of the range-space, null-space and the reconstruction error between the estimated and target mel-spectrograms for Id5, Id11-Id13.}}
			\label{fig:rnd-visualization}
			\vspace{-0.3cm}
		\end{figure}
        \begin{figure}
				\centering
				\includegraphics[width=0.85\columnwidth]{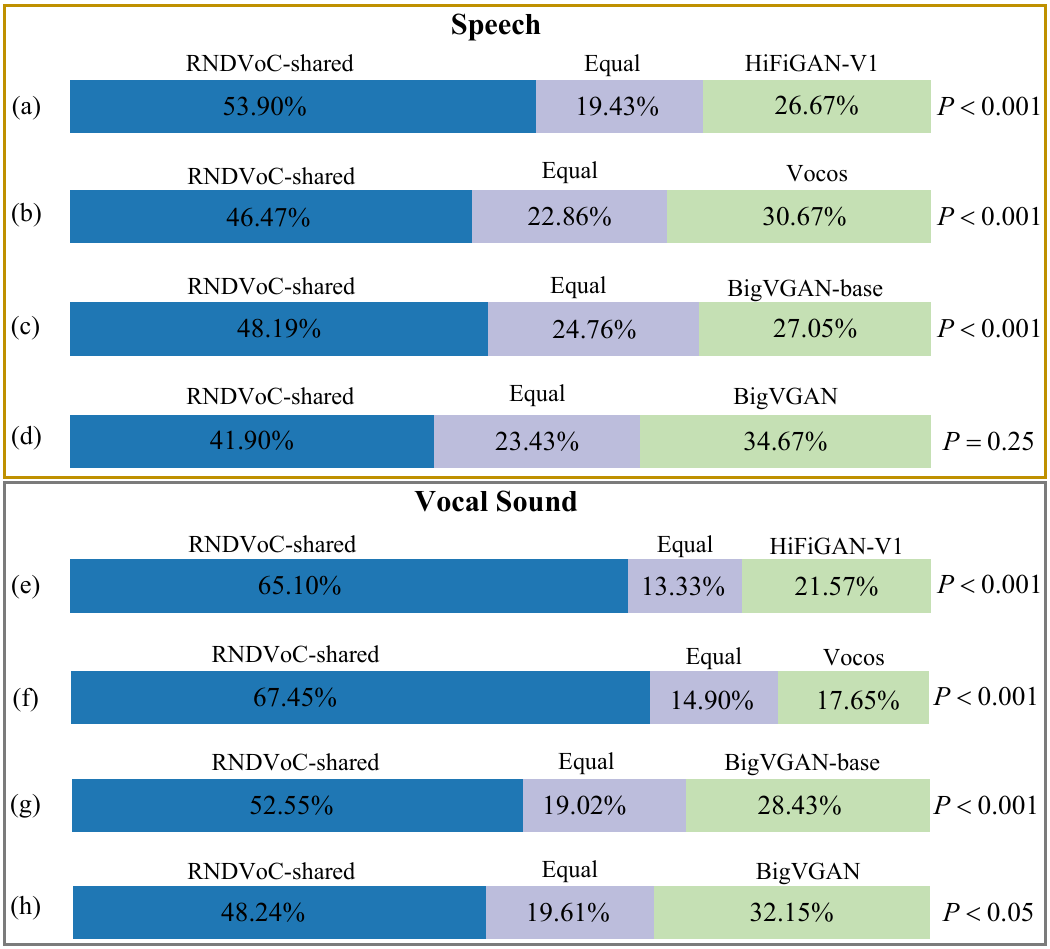}
				\vspace{-8pt}
				\caption{\lad{Average preference scores (in \%) of AB tests between RNDVoC-shared and three other GAN-based baselines. (a)-(d) Mel-spectrograms are obtained from speech clips in the LibriTTS test set. (e)-(h) Mel-spectrograms are obtained from vocal clips in the MUSDB18 test set.}}
				\label{fig:abtest}
				\vspace{-0.5cm}
			\end{figure}
		\vspace{-8pt}
		\section{Results and Analysis}\label{sec:results-and-analysis}
		\subsection{Evaluation Metrics}
		For objective evaluations, we employ \lad{eight} metrics: (1) Multi-resolution STFT (M-STFT){\footnote{https://github.com/csteinmetz1/auraloss}}~{\cite{yamamoto2020parallel}} is used to evaluate the spectral distance across multiple resolutions. (2) Wide-band version of Perceptual evaluation of speech quality (PESQ){\footnote{https://github.com/ludlows/PESQ}}~{\cite{rec2005p}} serves to assess the objective speech quality. (3) Mel-cepstral distortion (MCD){\footnote{https://github.com/chenqi008/pymcd}}~{\cite{kubichek1993mel}} measures the difference between mel-spectrograms through dynamic time wrapping (DTW). (4) Periodicity RMSE, V/UV F1 score, and pitch RMSE~{\cite{morrisonchunked}}{\footnote{https://github.com/gemelo-ai/vocos/blob/main}}, which are regarded as major artifacts for non-autoregressive neural vocoders. \lad{(5) UTMOS~{\cite{saeki2022utmos}}, a non-intrusive automatic MOS predictor that evaluates quality without requiring a reference signal.} \lad{(6)} Virtual Speech Quality Objective Listener (VISQOL){\footnote{https://github.com/google/visqol}}~{\cite{hines2015visqol}} predicts the Mean Opinion Score-Listening Quality Objective (MOS-LQO) score by evaluating the spectro-temporal similarity.
		
		\lad{In addition to objective metrics, we conduct subjective evaluations, namely MUSHRA and A/B preference tests, which are implemented on the BeaqleJS platform~{\cite{kraft2014beaqlejs}}. A total of 43 participants with backgrounds in audio signal processing are recruited for testing. We apply the data filtering by excluded participants who completed partial evaluations, as well as those who give anomalously low scores to ground-truth samples. After filtering, 35 valid responses are collected, comprising 21 male and 14 female participants with a mean age of 29.6 years. Each participant will receive reasonable remuneration for their participation. For the MUSHRA test, each participant is instructed to rate the speech processed by different models on a scale ranging from 0 to 100 in terms of its similarity to the provided reference. And each test group consists of 15 listening pairs. For the A/B preference test, each participant should select the clip with superior overall quality from two candidates, or choose an ``equal'' option if no preference can be given. Each test group includes 15 listening pairs. All test sequences and labels are randomly shuffled to eliminate selection bias. To assess the significance in subjective quality, we conduct one-tailed two-paired Kolmogorov-Smirnov (KS) tests{\footnote{\lad{We implement it with \texttt{scipy.stats.ks\_2samp}.}}}, and the one-tailed KS tests reject the null hypothesis if the $p$-value is lower than 0.05. All subjective evaluation audio samples and platform are publicly available at \url{https://huggingface.co/AndongLi/RNDVoC/tree/main}, and a screenshot of the listening interface is provided in Sec.~XI of the Appendix. }
			\renewcommand\arraystretch{1.10}
			\begin{table*}[t]
				\caption{MUSHRA scores among different methods on the LibriTTS benchmark. The confidence level is 95\%, and we performed a t-test comparing RNDVoc-shared with BigVGAN.}
				\centering
				\Huge
				\resizebox{0.94\textwidth}{!}{
					\begin{tabular}{cccccccccc}
						\toprule
						Models &GT &HiFiGAN-V1 &Avocodo &BigVGAN-base &BigVGAN &APNet2 &Vocos &RNDVoC-nonshared &RNDVoC-shared\\
						MUSHRA &89.45$\pm$0.45 &69.99$\pm$1.15 &68.16$\pm$1.26 &74.27$\pm$1.12 &79.33$\pm$0.92 &54.95$\pm$1.11 &73.18$\pm$1.15 &78.76$\pm$0.94 &\textbf{**80.74$\pm$0.99}\\
						\bottomrule\\[-0.8em] \multicolumn{9}{l}{Note: $^{**}p<0.05$, $^{*}p<0.1$}\\
				\end{tabular}}
				\label{tbl:libritts-mushra}
				\vspace{-0.45cm}
			\end{table*}
            \renewcommand\arraystretch{1.00}
			\begin{table*}
				\centering
				\huge
				\caption{\lad{Metric comparisons on two out-of-domain datasets among different GAN-based methods. All the models are pretrained on the LibriTTS dataset, and BigVGAN-base and BigVGAN are pretrained for 5M steps. For the MUSHRA test, the confidence level is 95\%, and we conducted one-tailed two-paired KS-test comparing RNDVoC-shared with BigVGAN.} }
				\vspace{-6pt}
				\resizebox{0.98\textwidth}{!}{
					\label{tbl:objective-metric-out-domain}
					\begin{tabular}{cccccccc|ccccccc}
						\toprule
						\multirow{2}*{Models} &\multicolumn{7}{c|}{EARS  \lad{(English Corpus with Diverse Emotions)}} &\multicolumn{6}{c}{VCTK \lad{(English Corpus with Unseen Speakers)}}\\
						&PESQ &Periodicity &V/UV F1 &Pitch &\lad{UTMOS} &VISQOL &MUSHRA &PESQ &Periodicity &V/UV F1 &Pitch &\lad{UTMOS} &VISQOL &MUSHRA \\
						\cline{1-1}\cline{2-8}\cline{9-15}
						GT &- &- &- &- &\lad{3.300} &- &89.06$\pm$0.48 &- &- &- &- &\lad{4.117} &- &89.52$\pm$0.45\\
						HiFiGAN-V1 &2.907 &\lad{0.149} &\lad{0.876} &\lad{48.417} &\lad{2.835} &\lad{4.570} &70.82$\pm$1.12 &3.090 &\lad{0.115} &\lad{0.943} &\lad{33.289} &\lad{3.952} &\lad{4.723} &78.49$\pm$1.10\\
						Avocodo &3.065 &\lad{0.149} &\lad{0.866} &\lad{48.494} &\lad{2.809} &\lad{4.641} &69.15$\pm$1.29 &3.337 &\lad{0.114} &\lad{0.943} &\lad{33.041} &\lad{3.905} &\lad{4.758} &73.14$\pm$1.13\\
						BigVGAN-base &3.797 &\lad{0.087} &\lad{0.946} &\lad{26.050} &\lad{3.040} &\lad{\underline{4.871}} &78.19$\pm$0.95 &3.859 &\lad{0.079} &\lad{0.965} &\lad{28.847} &\lad{3.970} &\lad{4.893} &79.73$\pm$0.86\\
						BigVGAN &\textbf{4.236} &\lad{\textbf{0.064}} &\lad{\textbf{0.963}} &\lad{\textbf{17.655}} &\textbf{\lad{3.190}} &\lad{\textbf{4.946}} &\underline{80.80$\pm$0.84} &\lad{\textbf{4.282}} &\lad{\textbf{0.062}} &\lad{\textbf{0.972}} &\lad{\underline{20.316}} &\lad{\textbf{4.034}} &\lad{\textbf{4.959}} &\underline{80.59$\pm$0.84}\\
						APNet2 &2.450 &\lad{0.149} &\lad{0.862} &\lad{39.627}  &\lad{2.259} &\lad{4.371} &55.08$\pm$1.16 &2.650 &\lad{0.110} &\lad{0.947} &\lad{30.294} &\lad{3.280} &\lad{4.532} &57.71$\pm$1.14\\
						Vocos &3.522 &\lad{0.094} &\lad{0.937} &\lad{25.904} &\lad{3.007} &\lad{4.803} &74.10$\pm$1.16 &\lad{3.684} &\lad{0.079} &\lad{0.965} &\lad{23.462} &\lad{3.916} &\lad{4.866} &75.06$\pm$1.10 \\
						RNDVoC-nonshared &3.856 &\lad{0.083} &\lad{0.949} &\lad{18.946} &\lad{3.078} &\lad{4.821} &79.88$\pm$0.84 &\lad{4.022} &\lad{0.069} &\lad{0.970} &\lad{21.467} &\lad{3.937} &\lad{4.863} &78.48$\pm$0.86\\
						RNDVoC-shared &\underline{3.989} &\lad{\underline{0.073}} &\lad{\underline{0.958}} &\lad{\underline{17.725}}  &\underline{\lad{3.110}} &\lad{4.869} &*\textbf{81.51$\pm$0.90} &\lad{\underline{4.136}} &\lad{\underline{0.065}} &\lad{\underline{0.972}} &\lad{\textbf{19.828}} &\lad{\underline{3.975}}  &\lad{\underline{4.905}} &**\textbf{81.73$\pm$0.89}\\
						\bottomrule\\[-0.8em] \multicolumn{13}{l}{Note: $^{**}p<0.05$, $^{*}p<0.1$}\\
				\end{tabular}}
                \vspace{-0.35cm}
			\end{table*}
			\vspace{-0.44cm}
			\subsection{Ablation Studies}\label{sec:ablation-studies}
			\subsubsection{Model Structure}
            \vspace{-0.1cm}
			In this part, we investigate the efficacy of different modules within the DPB. \lad{Given that the dual-path structure models spectral features along both subband and time axes, we ablate the narrow-band and cross-band modules separately, denoted as settings Id1-Id2 in Table~{\ref{tab:ablations}}}. Evidently, in either case, significant performance degradation occurs. For example, from Id5 to Id1, a degradation of up to 2.066 in PESQ is observed, while that of Id2 is 0.606. This verifies the significance of both time and subband modeling. Moreover, when the narrow-band module is removed, a substantial lower performance is attained, suggesting that the narrow-band module can contribute more than the cross-band module in the overall modeling process. This conclusion is consistent with findings in~{\cite{quan2024spatialnet}}. Fig.~{\ref{fig:ablation_spec_visualization}} presents the spectrograms processed by the methods in Id1-Id2 and Id5. \lad{Removing the narrow-band module introduces noticeable blurring of harmonic structures, whereas omitting the cross-band module leads to failed high-frequency region reconstruction, as marked by the blue and green boxes, respectively.}
			
			Subsequently, we replace the proposed modules in the narrow- and cross-band parts with residual RNNs~{\cite{yu23b_interspeech}}. \lad{While comparable performance is attained between Id3 and Id5, the latter incurs substantially more trainable parameters, higher computational cost, and slower inference speed, which is attributed to the autoregressive nature of RNNs.}
			
			Afterward, we analyze the impact of varying the numbers of ConvNext v2 blocks $P$ in the narrow-band module, with $P$ increasing from 1 to 4, as shown in Id4-Id7. It can be seen that the increase in ConvNext v2 blocks can enlarge the time receptive filed, leading to improved metric score, but at the cost of higher computational complexity and reduced inference efficiency. Therefore, we select $P = 2$ by default to strike a better balance between efficiency and performance.
			\subsubsection{RND Settings}
			In this part, we investigate the settings related to range-null decomposition. First, we eliminate the RND mode in target reconstruction, \emph{i.e.,} the explicit superimposition operation between $|\tilde{\mathbf{S}}|_{range}$ and $\left(\mathbf{I} - \mathcal{A}^{\dagger}\mathcal{A}\right)|\tilde{\mathbf{S}}|_{null}$ in Eq.~({\ref{eqn:11}}) is changed to an implicit target mapping. From Id8 to Id5, except for PESQ and MCD, significant performance improvements can be observed in Periodicity RMSE, V/UV F1, and Pitch RMSE, which are recognized as typical acoustic artifacts in non-autoregressive neural vocoders. The underlying reasons might be attributed to two-fold. First, the superimposition operation between the range-space and null-space can effectively preserve the prior acoustic information embedded in the range-space. Besides, the orthogonality characteristic between the range-space and null-space minimizes the degree of acoustic distortions. 
			
			In Id9, based on Id8, we devise a null-constraint loss. Specifically, given the magnitude estimation $|\mathbf{\tilde{S}}|$ from the neural network, we subtract the range-space component $|\mathbf{\tilde{S}}|_{range}$ and compel the remaining component to be orthogonal to the range-space, \emph{i.e.},
			\begin{align}
				\label{eqn:44}
				\mathcal{L}_{ort} = \frac{1}{FT}\sum_{f,t}\left\|\mathcal{A}\left(|\mathbf{\tilde{S}}| - \mathcal{A}^{\dagger}\overline{\mathbf{X}}^{mel}\right)\right\|_{1}.
			\end{align}
			
			In contrast to the proposed method in Sec.~{\ref{sec:connection-between-rnd-and-vocoder}}, here the network is required to output the overall estimation and also learn to separate them via the null-constraint loss.
			
			\lad{In Fig.~{\ref{fig:orthogonal-loss}}, we plot the curve of the null-constraint loss (see blue curve). The loss value decreases gradually with the increase in training steps, suggesting that this loss term effectively drives the orthogonal separation of the range and null subspaces. However, the magnitude of the decrease is marginal (from 0.008 to 0.007), indicating that implicit modeling of these two orthogonal subspaces remains highly challenging. This observation partly explains why the performance of Id9 is only modestly improved over Id8, and why both configurations still lag behind Id5.} 

            \lad{To further validate the efficacy of the RND operation, we evaluate four additional variants. In Id10, we modify the target spectral magnitude reconstruction formula  (in Eq.~{(\ref{eqn:10})}) by introducing a global skip connection, defined as $|\tilde{\mathbf{S}}| = \mathcal{A}^{\dagger}\mathcal{A}\left|\mathbf{S}\right| + |\tilde{\mathbf{S}}|_{null}$. In Id11, we replace the fixed pseudo-inverse $\mathcal{A}^{\dagger}$ with a linear layer from scratch. In Id12, both matrices $\left\{\mathcal{A}^{\dagger}, \mathcal{A}\right\}$ are set to be learnable. Based on Id12, Id13 incorporates an explicit idempotent constraint to enforce the idempotent property of the subspace projection matrix, formulated as $\mathcal{L}_{ide} = \frac{1}{\mathcal{C}}\left\|\mathcal{A}^{\dagger}\mathcal{A}\mathcal{A}^{\dagger}\mathcal{A} - \mathcal{A}^{\dagger}\mathcal{A}\right\|_{1}$, where $\mathcal{C}$ denotes the number of matrix elements.}

            \lad{Overall, Id11 achieves performance closest to that of Id5. The key distinction lies in the null-space estimation target, that is, Id11 requires the null-space module to predict the complete term $\left(\mathbf{I} - \mathcal{A}^{\dagger}\mathcal{A}\right)\left|\mathbf{S}\right|$, whereas Id5 can exploit the component prior for null-space filtering. This difference means Id5 can better leverage the orthogonal signal prior of the null space. Performance degradation is observed when either $\mathcal{A}^{\dagger}$ or both $\left\{\mathcal{A}, \mathcal{A}^{\dagger}\right\}$ are made learnable, which underscores the importance of maintaining strict orthogonality between the range and null subspaces. Notably, introducing idempotent constraint yields no significant performance improvement. In Fig.~{\ref{fig:orthogonal-loss}}, we plot the idempotent loss curve (see red curve). The loss increases initially during training and then gradually decreases after approximately 200 k steps, indicating that the network prioritizes target spectral reconstruction over enforcing subspace orthogonality in the early training phase. These findings confirm that fixed orthogonal projection weights outperform their learnable counterparts.}

            \lad{Fig.~{\ref{fig:rnd-visualization}}
            visualizes the estimated range-space, null-space components, and the mel-spectrogram error between the reconstructed spectrum and that of the target for Id5 and Id11-Id13. Several significant observations can be made. First, when $\mathcal{A}$ and $\mathcal{A}^{\dagger}$ are fixed, the null-space component $\left(\mathbf{I} - \mathcal{A}^{\dagger}\mathcal{A}\right)|\tilde{\mathbf{S}}|_{null}$ can effectively capture sparse, fine-grained harmonic details. In contrast, when either $\mathcal{A}^{\dagger}$ or $\left\{\mathcal{A}^{\dagger}, \mathcal{A}\right\}$ are learnable, the estimation from the null-space is no longer sparse. This confirms that the null-space filter $\mathbf{I} - \mathcal{A}^{\dagger}\mathcal{A}$ essentially acts as a physical prior filter, extracting detailed spectral information from $|\tilde{\mathbf{S}}|_{null}$ to complement final spectral reconstruction. Second, for the fixed type, the mel-spectrogram error between the generated and target is near zero. This demonstrates that all the acoustic information from the input can be transmitted losslessly, which is attributed to the strong orthogonality property of the RND theory. Even with explicit idempotent constraint (\emph{i.e.}, Id13), the error cannot be fully eliminated.}
			\subsubsection{Phase Loss}
			In Id14, we replace the phase loss defined in Eq.~({\ref{eqn:30}}) with the proposed omnidirectional phase loss in Sec.~{\ref{sec:omnidirectional}, and notable improvements are achieved in multiple objective metrics. For example, from Id5 to Id14, around 0.129 improvement in PESQ and 0.180 decrease in MCD are achieved, validating the effectiveness of the proposed phase loss. Note that when the number of kernels $\mathcal{K}$ decreases to 3, in which only two adjacent T-F bins and itself are considered, the proposed loss can degrade to that of Eq.~({\ref{eqn:30}}), indicating that our loss is a more general format by aggregating more adjacent relations.
            \subsubsection{Weights Sharing in the Encoder and Decoder Modules}
            Finally, we explore the impact of the weight-sharing scheme in the spectral encoder/decoder modules, as stated in Sec.~{\ref{sec:band-aware-encoding-module}}. By comparing Id15 and Id14, we observe further performance improvements with much fewer trainable parameters and faster inference speed. For instance, while PESQ score is improved by 0.095, the parameters are decreased by 66.87\%, and the inference speed is increased by 15.25\% on CPU and that of 47.47\% on GPU. As the shared scheme can incur more computational complexity, we regard it as a trade-off for optional use.
				\begin{table}
					\centering
					\huge
					\caption{VISQOL scores of multiple methods on the test set of MUSDB18 test set. Note that BigVGAN-base and BigVGAN are trained for 5M steps while other models are trained for 1M steps.}
					\vspace{-10pt}
					\resizebox{0.9\columnwidth}{!}{
						\label{tbl:objective-metric-mushdb18}
						\begin{tabular}{c|ccccc|c}
							\toprule
							Models &Bass &Drums &Mixture &Other &Vocals &Mean \\
							\midrule
							HiFiGAN-V1 &4.068 &4.452 &4.349 &4.346 &4.631 &4.3691 \\
							Avocodo &4.491 &4.568 &4.479 &4.480 &4.685 &4.5406\\
							BigVGAN-base &4.674 &4.792 &4.742 &4.727 &4.851 &4.757\\
							BigVGAN &\underline{4.825} &\textbf{4.885} &\textbf{4.874} &\textbf{4.853} &\textbf{4.924} &\textbf{4.872}\\
							APNet2 &4.274 &4.230 &4.122 &4.246 &4.539 &4.282\\
							Vocos &4.7035 &4.785 &4.684 &4.677 &4.835 &4.737 \\
							RNDVoC-nonshared &4.798 &4.773 &4.774 &4.768 &4.855 &4.793\\
							RNDVoC-shared &\textbf{4.842} &\underline{4.832} &\underline{4.829} &\underline{4.810} &\underline{4.888} &\underline{4.840}\\
							\bottomrule
					\end{tabular}}
				\end{table}
                \begin{figure}
				\centering
				\vspace{0pt}
				\includegraphics[width=0.45\textwidth]{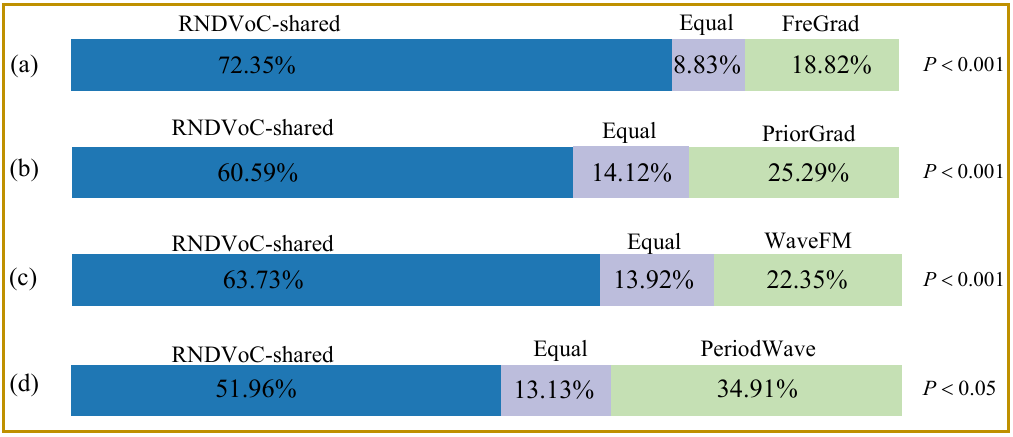}
				\vspace{-4pt}
				\caption{\lad{Average preference scores (in \%) of AB tests between RNDVoC-shared and three other diffusion-based baselines. Clips are from the vocal type of the MUSDB18 test set.}}
				\label{fig:abtest-diffusion}
				\vspace{-0.35cm}
			\end{figure}
                \begin{figure*}
				\centering
				\includegraphics[width=0.85\textwidth]{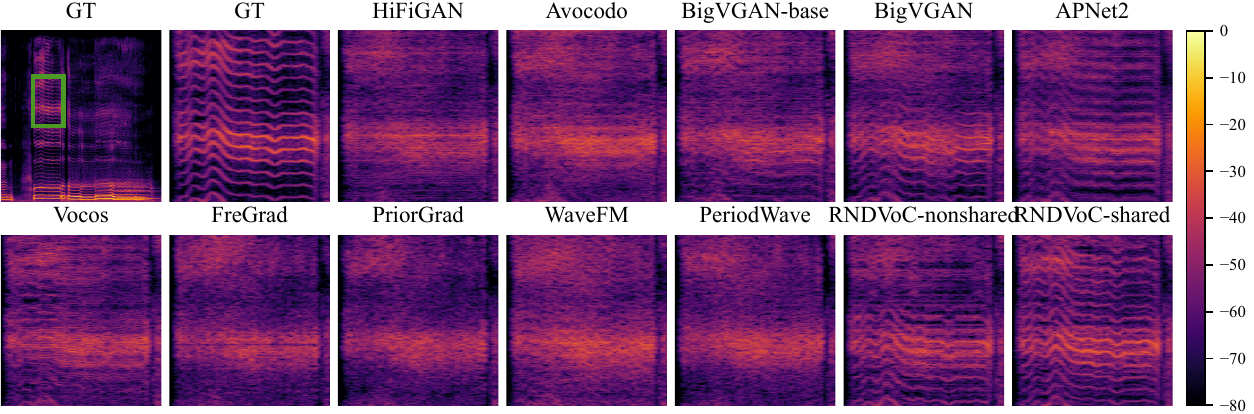}
				\vspace{-2pt}
				\caption{\lad{Spectral visualization of different vocoder methods. The audio clip is a singing voice from the MUSDB18 test set.}}
				\label{fig:sota-comparison-vocals-visualization}
				\vspace{-10pt}
			\end{figure*}
			\begin{figure*}
				\centering	\includegraphics[width=0.85\textwidth]{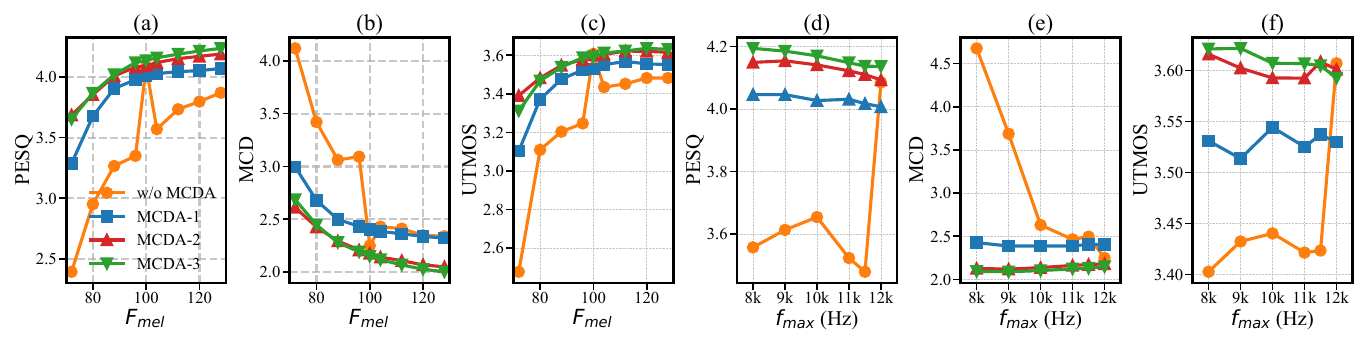}
				\vspace{-2pt}
				\caption{\lad{Metric comparisons in terms of whether the proposed MCDA strategy is adopted. (a)-(c): Metric performance under varying number of mel filters $F_{m}$, and $f_{\text{max}}$ is fixed to 12 kHz. (d)-(f): Metric performance under varying $f_{\text{max}}$, and $F_{m}$ is fixed to 100.}}
				\label{fig:mcda-nmel}
				\vspace{-10pt}
			\end{figure*}
            \renewcommand\arraystretch{0.90}
			\begin{table*}
				\centering
				\Huge
				\caption{\lad{Metric comparisons on multiple datasets between our method and advanced diffusion-based methods. For EARS and VCTK evaluations, models are pretrained on the LibriTTS benchmark. ``NFE'' denotes the number of function evaluations, and we choose the default NFE setup in the original literature.} }
				\vspace{-6pt}
				\resizebox{0.92\textwidth}{!}{
					\label{tbl:objective-metric-diffusion}
					\begin{tabular}{cccc|cccccccccccc}
						\toprule
						\multirow{2}*{\lad{Models}} &\multirow{2}*{\lad{NFE}} &\lad{\#Param.} &\lad{\#MACs} &\multicolumn{3}{c}{\lad{LJSpeech}} &\multicolumn{3}{c}{\lad{LibriTTS}} &\multicolumn{3}{c}{\lad{EARS}} &\multicolumn{3}{c}{\lad{VCTK }} \\
                         & &\lad{(M)} &\lad{(Giga/5s)} &\lad{PESQ} &\lad{UTMOS} &\lad{VISQOL} &\lad{PESQ} &\lad{UTMOS} &\lad{VISQOL} &\lad{PESQ} &\lad{UTMOS} &\lad{VISQOL} &\lad{PESQ} &\lad{UTMOS} &\lad{VISQOL} \\
                         \midrule
                          \lad{FreGrad} &\lad{50} &\lad{1.82} &\lad{1589} &\lad{3.775} &\lad{3.934} &\lad{4.574} &\lad{3.793} &\lad{3.217} &\lad{4.699} &\lad{3.495} &\lad{2.753} &\lad{4.526} &\lad{3.803} &\lad{3.670} &\lad{4.579} \\
                          \lad{PriorGrad} &\lad{50} &\lad{2.70} &\lad{8364} &\lad{\underline{3.955}}  &\lad{4.006} &\lad{4.659} &\lad{4.017} &\lad{3.329} &\lad{4.737} &\lad{3.847} &\lad{2.863} &\lad{4.636} &\lad{4.030} &\lad{3.740} &\lad{4.685} \\
                          \lad{WaveFM} &\lad{6} &\lad{19.50} &\lad{1188} &\lad{-} &\lad{-} &\lad{-} &\lad{3.954} &\lad{3.191} &\lad{\textbf{4.943}} &\lad{3.856} &\lad{2.709} &\lad{\textbf{4.926}} &\lad{3.960} &\lad{3.631} &\lad{\textbf{4.928}} \\
                          \lad{PeriodWave} &\lad{16} &\lad{29.81} &\lad{4986.56} &\lad{\textbf{4.235}} &\lad{\textbf{4.374}} &\lad{4.722} &\lad{\textbf{4.240}} &\lad{\underline{3.652}} &\lad{4.749} &\lad{\textbf{4.238}} &\lad{\textbf{3.218}} &\lad{4.686} &\lad{\textbf{4.248}} &\lad{\textbf{4.082}}  &\lad{4.671} \\
                          \midrule
                          \lad{RNDVoC-nonshared} &\lad{1} &\lad{9.48} &\lad{\textbf{28.29}} &\lad{3.892} &\lad{4.138} &\lad{\underline{4.812}} &\lad{4.085} &\lad{3.607} &\lad{4.873} &\lad{3.856} &\lad{3.078} &\lad{4.821} &\lad{4.022} &\lad{3.937} &\lad{4.863} \\
                          \lad{RNDVoC-shared} &\lad{1} &\lad{\textbf{3.14}} &\lad{37.20} &\lad{3.987} &\lad{\underline{4.161}} &\lad{\textbf{4.837}} &\lad{\underline{4.226}} &\lad{\textbf{3.657}} &\lad{\underline{4.915}} &\lad{\underline{3.989}} &\lad{\underline{3.110}} &\lad{\underline{4.869}} &\lad{\underline{4.136}} &\lad{\underline{3.975}} &\lad{\underline{4.905}}\\
                          \bottomrule
				\end{tabular}}
                \vspace{-0.2cm}
			\end{table*}
\renewcommand\arraystretch{0.80}
\begin{table*}[t]
	\centering
    \Huge
	\caption{\lad{Objective performance under different number of subbands. $N =1$ denotes the full-band model is adopted, and all subband-related operations are removed. Results are based on the LibriTTS benchmark.}}
	\resizebox{0.82\textwidth}{!}{
		\begin{tabular}{ccccccccccccc}
			\toprule
			\lad{Subband} &\multirow{2}*{\lad{w/i Even}} &\lad{\#Pram.} &\multicolumn{3}{c}{\lad{\#MACs (Giga/5s)}} &\multirow{2}*{\lad{M-STFT}} &\multirow{2}*{\lad{PESQ}} &\multirow{2}*{\lad{MCD}} &\lad{V/UV} &\lad{Periodicity}  &\multirow{2}*{\lad{UTMOS}} &\multirow{2}*{\lad{VISQOL}} \\
			\cmidrule{4-6}
			\lad{Number $N$} & &\lad{(M)}  &\lad{BAEM} &\lad{LKCAM} &\lad{BAMM+BAPM} & & & &\lad{F1} &\lad{RMSE}  & & \\
			\midrule
            \lad{1} &\lad{-} &\lad{200.89} &\lad{4.30} &\lad{73.74} &\lad{15.97} &\lad{0.912} &\lad{3.490} &\lad{3.220} &\lad{0.947} &\lad{0.117} &\lad{3.423} &\lad{4.810} \\
			\lad{6} &\lad{$\times$} &\lad{3.88} &\lad{0.37} &\lad{5.91} &\lad{41.23} &\lad{0.891} &\lad{3.680} &\lad{3.025} &\lad{0.949} &\lad{0.116} &\lad{3.398} &\lad{4.860} \\
			\lad{12} &\lad{$\times$} &\lad{3.38}  &\lad{0.37} &\lad{11.82} &\lad{21.73} &\lad{0.836} &\lad{3.939} &\lad{2.500} &\lad{0.958} &\lad{0.097} &\lad{3.552} &\lad{4.904} \\
			\lad{24} &\lad{$\times$} &\lad{3.14} &\lad{0.37} &\lad{23.61} &\lad{13.09} &\lad{0.746} &\lad{4.226} &\lad{1.913} &\lad{0.970} &\lad{0.074} &\lad{3.657} &\lad{4.915} \\
			\lad{24} &\lad{\checkmark} &\lad{3.09} &\lad{0.37} &\lad{23.64} &\lad{10.58} &\lad{0.830} &\lad{4.018} &\lad{2.319} &\lad{0.959} &\lad{0.090} &\lad{3.623} &\lad{4.904} \\
			\lad{48} &\lad{$\times$} &\lad{3.02} &\lad{0.37} &\lad{47.34} &\lad{11.00} &\lad{\underline{0.738}} &\lad{\underline{4.291}} &\lad{\underline{1.808}} &\lad{\underline{0.972}} &\lad{\underline{0.070}} &\lad{\underline{3.680}} &\lad{\textbf{4.952}} \\
            \lad{96} &\lad{$\times$} &\lad{2.95} &\lad{0.38} &\lad{94.62} &\lad{14.40} &\lad{\textbf{0.701}} &\lad{\textbf{4.409}} &\lad{\textbf{1.538}} &\lad{\textbf{0.978}} &\lad{\textbf{0.057}} &\lad{\textbf{3.754}} &\lad{\textbf{4.960}}\\
			\bottomrule
	\end{tabular}}
	\label{tbl:objective-different-subband}
	\vspace{-6pt}
\end{table*}
				\vspace{-0.45cm}
				\subsection{\lad{Comparisons with State-of-the-Art Methods}}\label{sec:comparisons-with-sota}
                \label{-0.2cm}
                \subsubsection{Comparisons with GAN-based Methods}
				\lad{Table~{\ref{tbl:objective-metric-ljs}} presents an objective performance comparison of RNDVoC against state-of-the-art GAN-based vocoders on the LJSpeech benchmark. As observed, T-F domain GAN methods like APNet and Vocos, achieves substantially fast inference speeds than time-domain counterparts, which is due to the removal of consecutive upsampling layers along the time axis. However, the former typically lag behind time-domain models in reconstruction quality. For example, scaling BigVGAN to 112 M parameters yields significantly higher PESQ and VISQOL scores than Vocos. Notably, our proposed method addresses this trade-off by fully exploiting spectral priors and modeling correlations along the time and subband axes. To be specific, both RNDVoC-nonshared and RNDVoC-shared outperforms Vocos, a leading T-F domain baseline, by a large margin across all objective metrics. When compared to BigVGAN, RNDVoC-nonshared exceeds the base BigVGAN variant in both reconstruction quality and inference efficiency. For the shared scheme, the performance is also comparable to BigVGAN while requiring only 8.17\% of BigVGAN's computational complexity, thereby underscoring its potential for resource-constrained scenarios.}
				
				Table~{\ref{tbl:objective-metric-libritts}} showcases the comparison results on the LibriTTS benchmark. Compared with the LJSpeech dataset, the LibriTTS benchmark exhibits more diverse acoustic environments. \lad{Among baselines,  BigVGAN and Vocos deliver the most competitive performance. Even so, RNDVoC maintains its superiority. For instance, under the same training step, RNDVoC-shared outperforms both BigVGAN-base and BigVGAN across multiple objective metrics, and achieves performance comparable to BigVGAN trained for 5M steps. These results demonstrate the efficacy of our approach in balancing high-fidelity audio generation and efficient inference.}
				
				\lad{Table~{\ref{tbl:libritts-mushra}} reports MUSHRA scores for nine representative GAN-based vocoders on the LibriTTS test set. RNDVoC-nonshared achieves higher scores than most baselines, while the shared version further enhances subjective quality and outperforms BigVGAN with statistical significance ($p<0.05$). Notably, APNet2’s subjective score is substantially lower than other methods. According to the feedback from the listeners, APNet2 can generate audible husky buzzing artifacts, which leads to biased low score.}  

                \lad{Fig.~{\ref{fig:abtest}}(a)-(d) presents A/B preference results comparing RNDVoC-shared against four advanced GAN-based baselines. For speech type, our method achieves a statistically significant preference margin over HiFiGAN, Vocos and BigVGAN-base ($p < 0.001$), while exhibiting no significant difference relative to BigVGAN ($p > 0.05$). To this end, we also consider the music scenario. Table~{\ref{tbl:objective-metric-mushdb18}} reports VISQOL scores for multiple methods on the MUSDB18 test set. In Fig.~{\ref{fig:abtest}}(e)-(h), we show A/B preference results on the vocal sound of the MUSDB18 test set. All the models are pretrained on the LibriTTS benchmark. Notably, although RNDVoC-shared is slightly inferior to BigVGAN in VISQOL, it achieves a statistically significant subjective preference advantage ($p < 0.05$). This discrepancy likely stems from richer harmonic content of music signals compared to speech, thus highlighting the effectiveness of RNDVoC's explicit subband division and modeling strategy for capturing complex harmonic structures. 
                } 
                
				Table~{\ref{tbl:objective-metric-out-domain}} presents the objective and subjective comparisons on the out-of-distribution samples from the EARS and VCTK datasets among various methods, where all the models are pretrained on the LibriTTS benchmark. Since the official checkpoints of BigVGAN-base and BigVGAN with 1M steps are unavailable, we adopt the 5M-step version instead. Evidently, while RNDVoC-nonshared performs slightly worse than BigVGAN, RNDVoC-shared achieves additional performance gains and outperforms BigVGAN in the MUSHRA.
                	\begin{table}
					\centering
					\huge
					\caption{\lad{Detailed setups of three MCDA strategies with varying configuration pools. $\left\{\Delta F_{m}, \Delta f_{\text{max}}\right\}$ denote the sampling interval of $\left\{F_{m}, f_{\text{max}}\right\}$, respectively.}}
					\vspace{-10pt}
					\resizebox{0.45\textwidth}{!}{
						\label{tbl:mcda-setups}
						\begin{tabular}{ccccc}
							\toprule
							\lad{Sets} &\lad{$F_{m}$ range} &\lad{$\Delta F_{m}$} &\lad{$f_{\text{max}}$ range (Hz)} &\lad{$\Delta f_{\text{max}}$ (Hz)} \\
							\midrule
							\lad{MCDA-1} &\lad{$\{88, 96, 100\}$} &\lad{-} &\lad{$[9\text{k}, 10\text{k}]$} &\lad{100}\\
                            \lad{MCDA-2} &\lad{$[64, 128]$} &\lad{16} &\lad{$[8\text{k}, 12\text{k}]$} &\lad{100}\\
                            \lad{MCDA-3} &\lad{$[64, 128]$} &\lad{1} &\lad{$[8\text{k}, 12\text{k}]$} &\lad{50}\\
							\bottomrule
					\end{tabular}}
                    \label{tbl:mcda-setups}
                    \vspace{-0.25cm}
				\end{table}
                \subsubsection{\lad{Comparisons with diffusion-based Methods}}
                \label{sec:comparison-diffusion}
                \lad{Table~{\ref{tbl:objective-metric-diffusion}} compares the objective performance of RNDVoC against four advanced diffusion-based vocoders. Owing to their iterative sampling strategy, diffusion models generally yield better reconstruction quality. For example, PeriodWave, a SoTA flow-matching vocoder, achieves the best overall metric scores. Thanks to the superiority of subband modeling, RNDVoC-shared reduces computational complexity by over 99\% while delivering performance competitive with PeriodWave. This result fully demonstrates the remarkable efficiency-quality trade-off of our approach.} 

                \lad{Figure~{\ref{fig:abtest-diffusion}} shows the subjective preference between RNDVoC-shared and four diffusion approaches, and all clips are from the vocal type of MUSDB18 test set. Again, our method yields notable preference with statistical difference ($<0.05$), further revealing the superiority of our method in subjective ratings.}
                \vspace{-2pt}
                \lad{\subsubsection{Qualitative Comparisons} Fig.~{\ref{fig:sota-comparison-vocals-visualization}} provides qualitative comparisons of different vocoders, and the clip is a vocal sound from the MUSDB18 test set. Notably, baseline methods inevitably introduce varying degrees of distortion in harmonic-rich regions, whereas RNDVoC can accurately restore fine-grained spectral details, as clearly illustrated in the zoomed-in regions of the figure. Additional qualitative visualizations are provided in Sec.~X of the Appendix.
                }
                \renewcommand\arraystretch{1.40}
				\begin{table}[t]
					\centering
					\Huge
					\caption{Objective comparisons among lightweight neural vocoders.}
					\vspace{-2pt}
					\resizebox{0.485\textwidth}{!}{
						\label{tbl:light-weight}
						\begin{tabular}{cc|c|cc|cc}
							\toprule
							\multirow{2}*{Models} &\#Parm. &\multirow{2}*{Sets} &\multicolumn{2}{c|}{Inference Speed}
							&\multirow{2}*{PESQ}  &\multirow{2}*{VISQOL} \\
							\cline{4-5}
							&(M) & &CPU &GPU & & \\
							\midrule
							\multirow{2}*{HiFiGAN-V2} &\multirow{2}*{0.92} &LJSpeech &0.0257(38.71$\times$)  &0.0014(725$\times$) &2.848 &\lad{4.542}\\
							\cline{3-3}
							& &LibriTTS &0.0288(34.72$\times$) &0.0014(704$\times$) &2.195 &\lad{4.370} \\
							\midrule
							\multirow{2}*{RNDVoC-Lite} &\multirow{2}*{0.71} &LJSpeech &0.0267(37.43$\times$) &0.0019(531$\times$) &3.769 &\lad{4.782} \\
							\cline{3-3}
							& &LibriTTS &0.0261(38.33$\times$) &0.0020(502$\times$) &3.834 &\lad{4.874}\\
							\midrule
							\multirow{2}*{RNDVoC-UltraLite} &\multirow{2}*{0.08} &LJSpeech &0.0251(39.04$\times$) &0.0018(557$\times$) &3.264 &\lad{4.701} \\
							\cline{3-3}
							& &LibriTTS &0.0253(39.46$\times$) &0.0018(543$\times$) &3.499 &\lad{4.804}\\
							\bottomrule
					\end{tabular}}
                \vspace{-0.2cm}
				\end{table}
                \vspace{-0.3cm}
                \subsection{\lad{The Effect of Subband Scaling}}\label{sec:subband-scaling}
                \lad{Recall that in Sec.~{\ref{sec:architecture-design-of-rndvoc}}, subband-based network is employed. To validate the efficacy of this subband design, we evaluate performance across diverse subband configurations. Specifically, we vary the number of subbands $N$ from 6 to 96. To verify the superiority of the proposed uneven band splitting strategy, we also consider the even-splitting case. Besides, we include a full-band baseline, \emph{i.e.}, $N = 1$. Concretely, we remove the cross-band module and retain only the narrow-band module. The number of channels $C$ is set to 2056, resulting in 200 M parameters, which is 64$\times$ larger than RNDVoC-shared's default configuration. Detailed configurations are provided in Sec.~VIII of the Appendix.}
                
                \lad{As shown in Table~{\ref{tbl:objective-different-subband}}, several observations can be made. First, increasing the number of subbands $N$ yields significant improvements in generation quality, as indicated by consistent gains across various objective metrics. And a new SoTA performance is achieved for $N=96$. This confirms that finer-grained spectral modeling is essential for high-fidelity audio generation. Notably, unlike previous model-scaling paradigms, increasing $N$ does not increase the network parameters, and we therefore formally define this strategy as subband-scaling, a scaling approach that has been largely underexplored in the neural-vocoder literature. Second, despite having more than 200 M parameters, the full-band model performs substantially worse than all subband versions, which further underscoring the efficacy of subband modeling. Third, under the same $N$, our proposed uneven band-splitting strategy outperforms the even-splitting baseline by a notable margin, validating the advantage of tailoring subband partitions to the inherent structure of audio spectra.}  
            \begin{figure}[t]
                \includegraphics[width=0.44\textwidth]{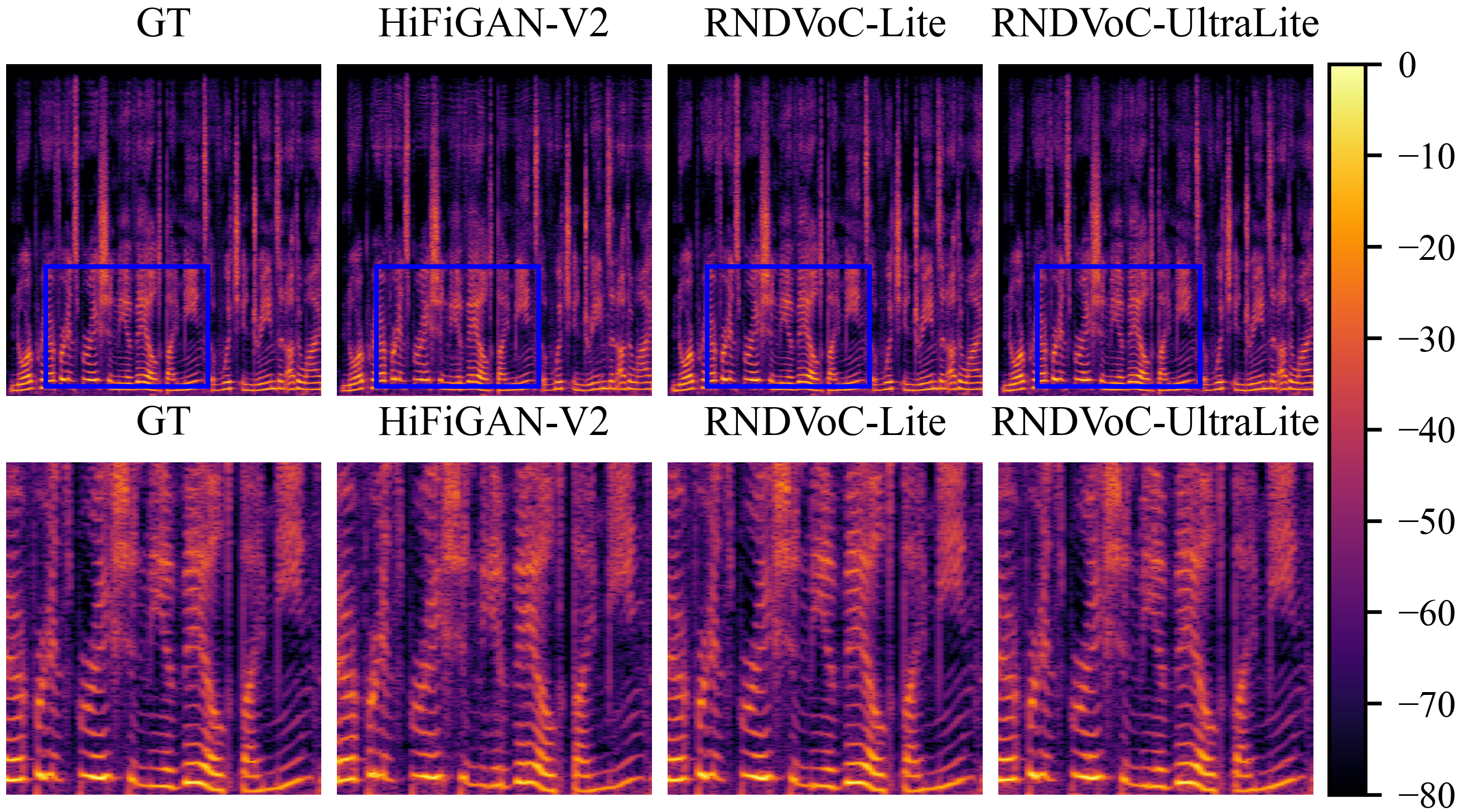}
                \centering
				\vspace{-4pt}
				\caption{\lad{Spectral visualizations between lightweight neural vocodes. Audio clips are from the LibriTTS test set.}}
				\label{fig:light-weight-visualization}
				\vspace{-0.35cm}
			\end{figure}
            \begin{figure}[t]
				\centering
				\vspace{0pt}
				\includegraphics[width=0.83\columnwidth]{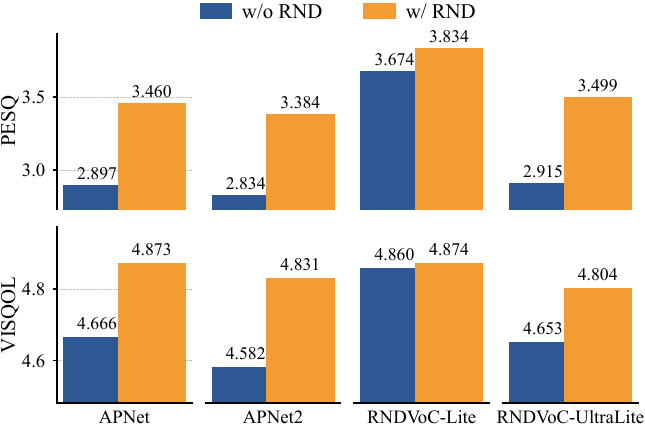}
				\vspace{-2pt}
				\caption{\lad{Effect of the proposed RND strategy for different neural vocoders.}}  
				\label{fig:rnd-strategy-performance}
				\vspace{-16pt}
			\end{figure}
                \vspace{-0.3cm}
				\subsection{The Effect of MCDA Strategy}\label{sec:the-effect-of-mcda-strategy}
                \lad{To validate the effectiveness of the proposed MCDA strategy, we consider four variants, and RNDVoC-nonshared with 1 M training steps is adopted by default: 
                For the first case, no MCDA is adopted, \emph{i.e.}, fixed mel-spectrogram setup with $F_{m}=100$ and $f_{\text{max}} = 12$  \text{kHz}. For another three cases, \emph{i.e.}, $\left\{\text{MCDA-1}, \text{MCDA-2}, \text{MCDA-3}\right\}$, we employ MCDA with enlarged sampling pool $\mathcal{T}$, as detailed in Table~{\ref{tbl:mcda-setups}}. Fig.~{\ref{fig:mcda-nmel}} presents the performance under varying $F_{m}$ and $f_{\text{max}}$, and several important findings can be made. First, the configuration without MCDA suffers from significant performance mismatch when confronted with unseen mel configurations, achieving optimal performance only for the pretrained setup, as indicated in orange curve. In sharp contrast, models with MCDA exhibit substantially enhanced robustness to both seen and unseen configurations, with overall smoother performance curves. For instance, on the unseen $\left\{F_{m}=72, f_{\text{max}}=12 \text{kHz}\right\}$ setup, the ``w/o MCDA'' case yields a UTMOS score below 2.6, while MCDA-1 improves the score by over 0.5.  Similar trends are observed across other objective metrics. Further expanding the sampling pool, \emph{e.g.}, from MCDA-1 to MCDA-2 and MCDA-3, brings consistent performance gains across nearly all configurations, including both seen and unseen setups. Notably, even for the fixed mel configuration used in the ``w/o MCDA'' training, \emph{e.g.}, $F_{m}=100$, $f_{\text{max}} = $ 12 \text{kHz}, MCDA-3 still achieves slightly better performance than the non-MCDA counterpart. Collectively, these results demonstrate that the proposed MCDA strategy  can effectively enhance the adaptability of neural vocoders to diverse mel-spectrogram configurations, enabling unified inference mode after a single training stage. We provide detailed quantitative results of different mel-spectrogram configurations in Sec.~IX of the Appendix.}
                \vspace{-0.3cm}
                \subsection{Toward Lightweight Vocoder Design}\label{sec:toward-more-light-weight}
				Although the proposed RNDVoC only exhibits 3.4 M parameters for the shared version, which is significantly smaller than mainstream schemes, \emph{e.g.}, HiFiGAN and BigVGAN, it is still challenging when applied to edge-devices. To this end, we devise a lightweight version called RNDVoC-Lite, where the number of DPBs $B$ is reduced to 4 and the channel size $C$ is squeezed to 128. With the shared scheme, the number of network parameters is reduced to 0.7 M. By further reducing $C$ to 32, we yield an ultralite counterpart whose number of trainable parameters is only 0.08 M. To our best knowledge, this is the smallest end-to-end neural vocoder up to now. In Table~{\ref{tbl:light-weight}}, we compare the performance of RNDVoC-Lite, RNDVoC-UltraLite with HiFiGAN-V2, whose parameters are around 0.92 M. Despite being slightly slower over HiFiGAN-V2, our method outperforms it by a large margin in PESQ and VISQOL, revealing the promising potential of our method in the lightweight scenarios. \lad{Fig.~{\ref{fig:light-weight-visualization}} presents qualitative comparison results among the three lightweight vocoders. Both RNDVoC-Lite and its UltraLite counterpart can better reconstruct harmonic components of relatively low energy.}
				\vspace{-0.5cm}
				\subsection{\lad{Effect of the Proposed RND Strategy}}\label{sec:apply-rnd-to-other-baselines}
				\lad{To further validate the inherent superiority of the proposed RND framework itself, we integrate it into four existing T-F domain neural vocoders: APNet, APNet2, RNDVoC-Lite and RNDVoC-UltraLite. For APNet and APNet2, two key modifications are implemented: $\Circled{\footnotesize{1}}$ the mel-spectrogram is transformed into the linear-scale domain via $\mathcal{F}_{range}\left(\cdot\right)$ prior to network input; $\Circled{\footnotesize{2}}$ the network output is used for target reconstruction following Eq.({\ref{eqn:11}}). These modifications introduce only negligible additional parameters and computational cost. For RNDVoC-Lite and its UltraLite variant, we replace Eq.~({\ref{eqn:11}}) with direct prediction of target magnitude and phase components. As shown in Fig.~{\ref{fig:rnd-strategy-performance}}, integrating the RND strategy yields significant performance improvements for both APNet and APNet2, indicating that RND can serve as a plug-and-play tool to empower existing T-F domain neural vocoders. Besides, when intrinsic generation capability of the vocoder is weakened, as shown from RNDVoC-Lite to RNDVoC-UltraLite, the performance gap becomes more pronounced. This demonstrates that the advantage of RND strategy in enabling lossless acoustic information transmission becomes more impactful when the generation capability of the vocoder is limited, which further underscores its value as a lightweight yet effective performance booster.}
				\vspace{-6pt}
				\section{Concluding Remarks}
				\label{sec:conclusion}
				In this paper, we propose a scalable T-F domain based neural vocoder motivated by range-null space decomposition theory. Specifically, the reconstruction process of the target spectrogram can be decomposed into the explicit superimposition between two orthogonal sub-spaces, namely range-space and null-space, where the former can be realized via the pseudo-inverse operation to project the original acoustic representation in the mel-domain into target linear-scale domain, and the latter is responsible for spectral detail generation. Based on that, we propose a novel generation framework, where the spectrogram is hierarchically encoded and decoded, and the correlations among frames and subbands can be effectively modeled by an elaborately devised dual-path structure. Besides, the  multiple-condition-as-data-augmentation (MCDA) strategy, a plug-and-play tactic, is proposed to support for scalable inference under various mel configurations. \lad{We conduct extensive experiments on various benchmarks and out-of-distribution evaluations.} Results show that while enjoying significantly smaller parameters, computational complexity and competitive inference efficiency, the proposed framework outshines existing state-of-the-art vocoders in both objective and subjective evaluations.

                \lad{While this paper introduces the range-null decomposition theory to enhance interpretability in audio generation, it mainly adopts mel-spectrogram as acoustic features. This requires further investigation for more general acoustic features, where the linear degradation formulation may no longer hold. Besides, the proposed MCDA strategy only considers two factors, \emph{i.e.}, $F_{m}$ and $f_{\text{max}}$, and its generalization to more general mel-spectrogram extraction pipelines remains to be explored. In future work, we plan to extend the RND framework to more audio tasks, such as  speech restoration and neural audio codec.}

\clearpage
\appendix
\section{Appendix Summary}\label{sec:summary}
\lad{In this appendix, additional details and supplementary experimental results are provided to facilitate a deeper understanding of the core contributions of this paper, which are summarized as follows:}
\begin{itemize}
    \item \lad{We conduct a theoretical analysis of the computational complexity for the nonshared and shared subband encoding/decoding schemes, and further compare the practical inference efficiency under different number of feature channels.}
    \item \lad{We design and compare a cascaded architecture, where the magnitude spectrum is estimated in the first submodule and the phase spectrum is predicted subsequently in the second submodule.}
    \item \lad{We extend the proposed range-null decomposition (RND) framework to the complex-valued domain, and compare the objective performance.}
    \item \lad{We present the pseudo-code for the proposed multiple-condition-as-data-augmentation (MCDA) strategy.}
    \item \lad{We analyze the sensitivity of the proposed MCDA strategy to deviations in the mel-filter matrix $\mathcal{A}$ in practical implementations.}
    \item \lad{We provide detailed illustrations of the network parameters and input and output feature size.}
    \item \lad{We report objective performance comparisons of different approaches on additional evaluation benchmarks.} 
    \item \lad{We elaborate on the parameter configurations under different number of subbands.}
    \item \lad{We present detailed quantitative results for the baseline ``w/o MCDA'' and another three MCDA variants.}
    \item \lad{We apply the neural vocoder to the speech enhancement (SE) task and compare the reconstruction performance for both discriminative and generative SE methods.}
    \item \lad{We provide additional spectral visualizations by different vocoder approaches.}
    \item \lad{We present a screenshot of the subjective evaluation platform used in our experiments.}
    \end{itemize}     

    \begin{figure*}[t]
		\centering	\includegraphics[width=0.98\textwidth]{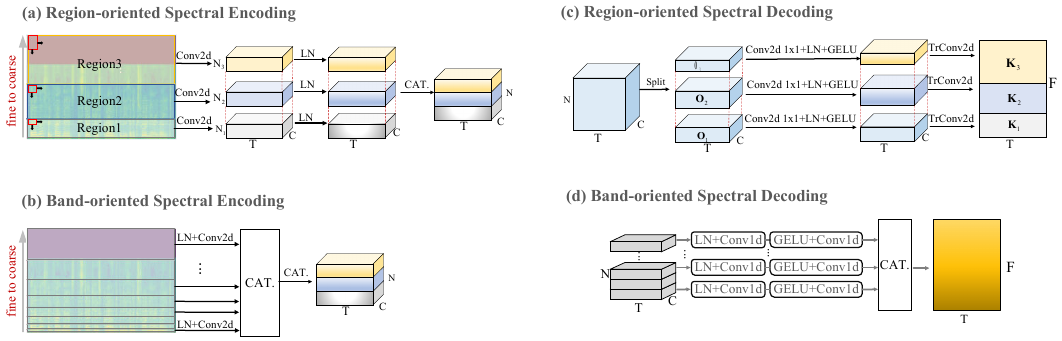}
		\vspace{-2pt}
		\caption{\lad{Implementation process of the region-oriented and subband-oriented spectral encoding and decoding, respectively.}}
        \label{fig:appendix_nonshared_shared_sheme}
		\vspace{-1pt}
	\end{figure*}
    
	\subsection{Complexity Analysis of the non-shared and shared schemes}\label{sec:complexity-analysis}
    \lad{Recall from the main text that, we introduce region-oriented band-split and merging strategies for spectral encoding and decoding. For the splitting strategy, the spectrum is partitioned into $I$ regions, with a dedicated encoding module employed for feature encoding in each region. For the merging strategy, the resulting 3-D feature map is fist split into $I$ regions, in each of which the feature is gradually recovered to the spectral target. The detailed implementation is illustrated in Fig.~{\ref{fig:appendix_nonshared_shared_sheme}}(a) and (c). In contrast, for the subband-oriented scheme, each subband is separately encoded and decoded, as shown in Fig.~{\ref{fig:appendix_nonshared_shared_sheme}}(b) and (d).} Notably, encoding/decoding parameters are shared across subbands within a single region for the region-oriented strategy, but not shared across subbands for the subband-oriented strategy. Therefore, we term these two strategies the \textbf{shared} and \textbf{nonshared} schemes, respectively. A theoretical analysis of their computational complexity is provided below:

\begin{figure}[t]
		\centering
		\includegraphics[width=0.99\columnwidth]{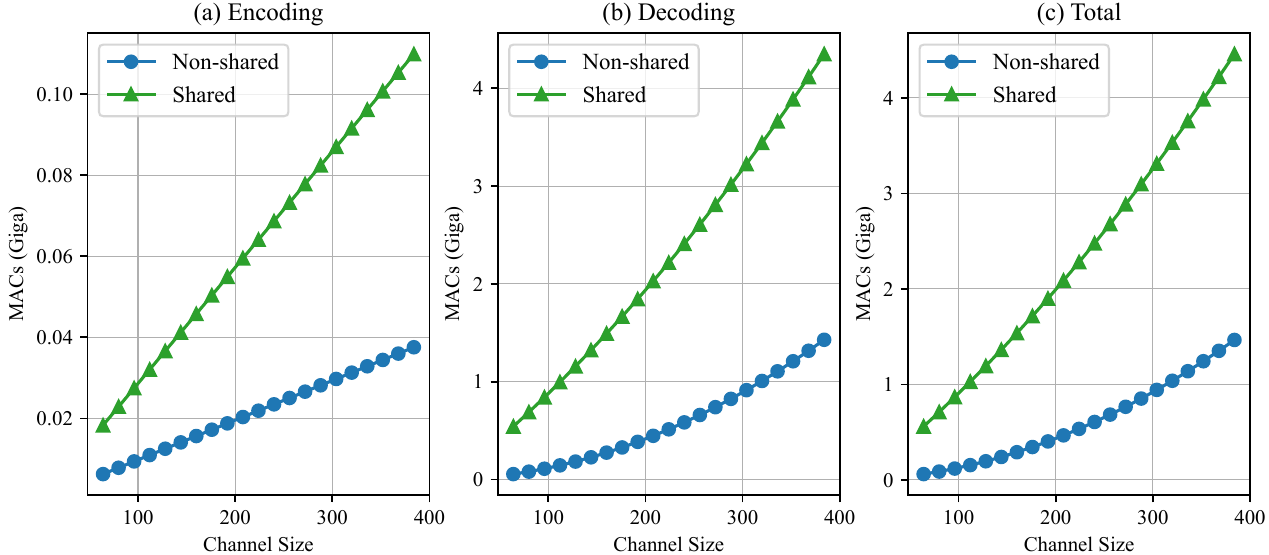}
		\vspace{-2pt}
		\caption{Complexity comparisons between the non-shared and shared schemes for encoding/decoding modules.}  
		\label{fig:appendix_nonshared_shared}
		\vspace{-0.85cm}
	\end{figure}
    
\begin{itemize}
	\item \textbf{Non-shared scheme}: For the $n$-th sub-band, the encoding module consist of a LN and a Conv1d layer with a kernel size of 1. It is computational complexity is calculated as:
	\begin{equation}
		\label{eqn:44}
		g_{e,n} = (2F_{n} + 1)CT,
	\end{equation}
	where $F_{n}$ denotes the frequency dimension of the $n$-th subband, $T$ is the time frame size, and $C$ is the number of feature channels after encoding. The term $\left(2F_{n}+1\right)$ is due to use of gain representation~{\cite{luo2024gull}}. Here normalization layers and bias terms are neglected in complexity calculation, as their computational cost is negligible compared to convolution layers. Besides, all complexity metrics are reported in multiply-accumulate operations (MACs) instead of floating-point operations (FLOPs). The total encoding complexity is thus:
	\begin{align}
		\label{eqn:45}
		G_{e} = \sum_{n=1}^{N}g_{e,n} = (2F + N)CT.
	\end{align}
	
	For the decoding process, taking the magnitude branch as an example. Reconstruction of the $n$-th subband involves a LN, a point-wise Conv1d that doubles the channel dimension, a GELU activation, and an additional Conv1d. The complexity $g_{dm}$ can thus be expressed as:
	\begin{equation}
		\label{eqn:46}
		g_{dm,n} = 2C^{2}T + 2CF_{n}T,
	\end{equation}
	and the total complexity of the magnitude decoding module is:
	\begin{align}
		\label{eqn:47}
		G_{dm} = \sum_{n=1}^{N}g_{dm,n}=2NC^{2}T + 2FCT.
	\end{align}
	
	Similarly, the complexity of the phase decoding module can be calculated as:
	\begin{align}
		\label{eqn:48}
		G_{dp} = \sum_{n=1}^{N}g_{dp,n} = 2NC^{2}T + 4FCT.
	\end{align}
	The total complexity of spectral encoding/decoding for the non-shared scheme is thus $\mathcal{O}\left(4NC^{2}T+(8F+N)CT\right) = \mathcal{O}\left(4NC^{2}T+8FCT\right)$.
	\item \textbf{Shared scheme}: For the $i$-th region, the encoding module consists of a Conv2d layer and an LN. Let the kernel size of the Conv2d be $\left(K_{t,i}^{e}, K_{f,i}^{e}\right)$, where $K_{t,i}^{e}$ and $K_{f,i}^{e}$ denote the kernel size along the time and frequency axes, respectively, and the superscript $\left(\cdot\right)^{e}$ denotes the encoding process. The complexity $h_{e,i}$ for the $i$-th region can be calculated as:
	\begin{equation}
		\label{eqn:49}
		h_{e,i} = 2CK_{t,i}^{e}K_{f,i}^{e}N_{i}T.
	\end{equation}
    
	The total complexity of the encoding module is thus:
	\begin{align}
		\label{eqn:50}
		H_{e} = \sum_{i=1}^{I}h_{e,i} = 2CT\sum_{i=1}^{I}K_{t,i}K_{f,i}N_{i}.
	\end{align}
    
	For the decoding process, taking the magnitude branch as an example, we assume the kernel size of the TrConv2d to be $\left(K_{t,i}^{d}, K_{f,i}^{d}\right)$. The complexity for the $i$-th region can be calculated as:
	\begin{align}
		\label{eqn:51}
		h_{dm,i} = 2C^{2}TN_{i} + 2CTK_{t,i}^{d}K_{f,i}^{d}\overline{F}_{i},
	\end{align} 
	and the complexity of the magnitude decoding module is:
	\begin{align}
		\label{eqn:52}
		H_{dm} = \sum_{i=1}^{I}h_{dm,i} = 2NC^{2}T + 2CT\sum_{i=1}^{I}K_{t,i}^{d}K_{f,i}^{d}\overline{F}_{i}.
	\end{align}
    
	Similarly, the complexity of the phase decoding module is:
	\begin{align}
		\label{eqn:53}
		H_{dp} = \sum_{i=1}^{I}h_{dp,i} = 2NC^{2}T + 4CT\sum_{i=1}^{I}K_{t,i}^{d}K_{f,i}^{d}\overline{F}_{i}.
	\end{align}
	The total complexity of the spectral encoding/decoding for the shared scheme is thus $\mathcal{O}\left(4NC^{2}T + 2CT\sum_{i=1}^{I}K_{t,i}K_{f,i}N_{i} + 6CT\sum_{i=1}^{I}K_{t,i}^{d}K_{f,i}^{d}\overline{F}_{i}\right)$. 
\end{itemize}

Fig.~{\ref{fig:appendix_nonshared_shared}} compares the computational complexity of the spectral encoding and decoding modules for the non-shared and shared schemes, with feature channel sizes ranging from 64 to 384. Two key observations can be made. First, the decoder accounts for a higher computational complexity than the encoder for both schemes. Second, the shared scheme exhibits a notably higher complexity than the non-shared scheme. By comparing Eqs.~(\ref{eqn:47})-(\ref{eqn:48}) and Eqs.~(\ref{eqn:52})-(\ref{eqn:53}), it is evident that the additional computational cost of the shared schemes stem essentially from the introduction of TrConv2d layers in the decoding module.

\lad{\textbf{Default parameter configurations}: For the nonshared scheme, the spectrum is split into 24 subbands with non-uniform bandwidths: twelve subbands of 250 Hz, eight of 500 Hz, three of 1000 Hz, and one subband covering the remaining frequency range. For the shared scheme, we empirically partition the spectrum into $I=3$ regions with frequency dimensions $\left\{144, 192, 176\right\}$, respectively. For the encoding module, the Conv2d kernel sizes for the three regions are $\left(K_{t,1}^{e}, K_{f,1}^{e}\right)=\left(3, 12\right)$, $\left(K_{t,2}^{e}, K_{f,2}^{e}\right)=\left(3, 24\right)$, and $\left(K_{t,3}^{e}, K_{f,3}^{e}\right)=\left(3, 44\right)$, with corresponding stride sizes along the time and frequency axes of $\left\{\left(1, 12\right), \left(1, 24\right), \left(1, 44\right)\right\}$. For the decoding module, the TrConv2d kernel sizes for the three regions are $\left(K_{t,1}^{d}, K_{f,1}^{d}\right)=\left(1, 12\right)$, $\left(K_{t,2}^{d}, K_{f,2}^{d}\right)=\left(1, 24\right)$, $\left(K_{t,3}^{d}, K_{f,3}^{d}\right)=\left(1, 44\right)$, with stride size of $\left(K_{t,1}^{d}, K_{f,1}^{d}\right)=\left(1, 12\right)$, $\left(K_{t,2}^{d}, K_{f,2}^{d}\right)=\left(1, 24\right)$, and $\left(K_{t,3}^{d}, K_{f,3}^{d}\right)=\left(1, 44\right)$. Zero-padding is applied in all convolution and transposed convolution operations to preserve the time dimension, ensuring frame-level spectral generation. This configuration results in a total of $12+8+4=24$ subbands, matching the nonshared scheme for fair comparison. Note that this subband division and merging configuration is by no means optimal for performance. Nonetheless, we empirically found that it works well and we leave the joint optimization of division strategy and target reconstruction in high-fidelity as future work.}

\renewcommand\arraystretch{1.00}
    \begin{table}[t]
        \caption{A case comparison in terms of trainable parameters (in million), computational complexity (in Giga), and processing time (in second) under different channel settings.}
        \centering
        \large
        \resizebox{0.99\columnwidth}{!}{
            \begin{tabular}{ccccccc}
                \toprule
                Scheme &\multicolumn{3}{c}{Non-shared Scheme}  &\multicolumn{3}{c}{Shared Scheme}\\
                \cmidrule(lr){1-1} \cmidrule(lr){2-4}\cmidrule(lr){5-7}
                Channels $C$ &\#Param.(M) &\#MACs(G) &Time(s) &\#Param.(M) &\#MACs(G) &Time(s)\\
                \hline
                64 &0.67  &0.06 &24.61 &0.11 &0.52 &12.23 \\
                128 &2.12 &0.19 &39.30 &0.32 &1.11 &13.30 \\
                256 &7.38 &0.64 &59.06 &0.93 &2.50 &18.04 \\
                384 &15.79 &1.38 &62.53 &2.15 &4.16 &21.53\\
                \bottomrule
        \end{tabular}}
        \label{tbl:pre_exp_param_macs_time}
        \vspace{-8pt}
    \end{table}

\begin{figure}[h]
	\centering
	\vspace{0pt}
	\includegraphics[width=0.988\columnwidth]{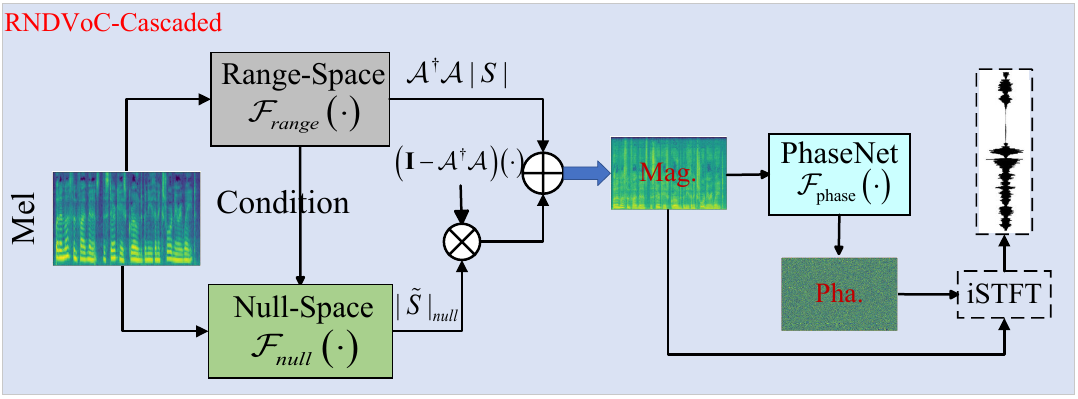}
	\vspace{-2pt}
	\caption{\lad{Framework diagrams of the RNDVoC with a cascaded generation style.}}  
	\label{fig:appendix_rnd_cascaded}
	\vspace{-6pt}
\end{figure}

\lad{Table~{\ref{tbl:pre_exp_param_macs_time}} presents a case comparison in terms of the number of trainable parameters, computational complexity, and processing time for the encoding and decoding modules. The MACs are calculated by processing a mel-spectrogram whose target waveform is approximately 1 second, and the processing time is accumulated by repeating the step 1000 times on a CPU Intel(R) Core(TM) i7-14700F. It is evident that by reducing the loop time from $\mathcal{O}\left(N\right)$ to $\mathcal{O}\left(I\right)$, the shared scheme significantly reduces the processing time. Besides, with the parameter sharing scheme, the number of trainable parameters is also notably decreased.}

\begin{algorithm}[h]
		\caption{Training and inference procedure of MCDA.}
		\label{alg:mel_adapt}
		\begin{algorithmic}[1]
			\small
			\Statex \textbf{Input}: Speech dataset $\mathcal{D}_s$, Model parameters $\Theta$, Learning rate $r$
			\Statex \textbf{Define}: Mel configuration pool $\mathcal{P}=\left\{\left(F_{mel}^{j}, f_{max}^{k}\right)\mid j\in\left\{1,\cdots,J\right\}, k\in\left\{1,\cdots,K\right\}\right\}$
			\Statex \textbf{Construct}: Mel-filter pool $\mathcal{T}=\left\{\mathcal{A}^{jk}\right\}$, where $\mathcal{A}^{jk}$ is calculated from $p^{jk}\in\mathcal{P}$ (constructed prior to training)
			\Statex \textbf{Training Procedure}:
			\While{not converged}
			\State $p^{jk} \gets$ RandomSample($\mathcal{P}$) \Comment{Randomly sample a mel-condition from $\mathcal{P}$}
			\State $\mathcal{A}^{jk} \gets$ ObtainByIndexing($\mathcal{T}$, $p^{jk}$) \Comment{Obtain the corresponding mel filter from $\mathcal{T}$ via indexing}
			\State $B_{s} \gets$ SampleBatch($\mathcal{D}_s$, BatchSize) \Comment{Sample a batch from $\mathcal{D}_s$}
			\State $mel \gets$ ApplyMelFilter($B_{s}$, $\mathcal{A}^{jk}$) \Comment{Apply the filter to obtain the output mel-spectrogram}
			\State $input \gets$ Range-Space-Module($mel$) \Comment{Project the mel-spectrogram to linear-scale domain}
			\State $loss \gets$ ForwardPass($input$) \Comment{Calculate the loss through ForwardPass}
			\State $\Theta \gets$ UpdateParameters($\Theta$, $loss$, $r$) \Comment{Update the model parameters}
			\EndWhile
			\Statex \textbf{Inference Procedure}:
			\State Given a mel-spectrogram $mel$ with a preset condition
			\State $input \gets$ Range-Space-Module($mel$) \Comment{Transform the mel-spectrogram to linear-scale domain}
			\State $output \gets$ GenerateTargetWaveform($input$) \Comment{Generate the target waveform}
			\Statex \textbf{Output}: In training, updated model parameters $\Theta$. In inference, generated output
		\end{algorithmic}
	\end{algorithm}

\begin{figure}[h]
	\centering
	\vspace{0pt}
	\includegraphics[width=0.988\columnwidth]{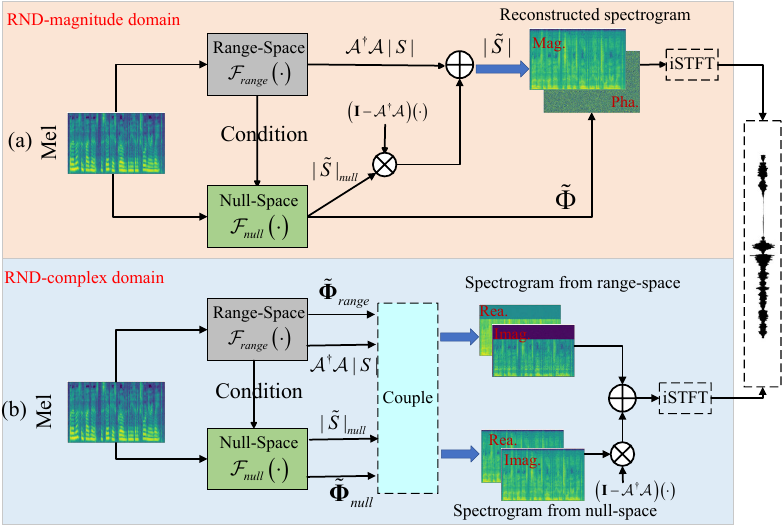}
	\vspace{-2pt}
	\caption{Framework diagrams of the RNDVoC defined in the magnitude and complex-valued domains. (a) The proposed RNDVoC defined in the magnitude domain. (b) The proposed RNDVoC defined in the complex domain.}  
	\label{fig:appendix_rnd_complex}
	\vspace{-6pt}
\end{figure}

\renewcommand\arraystretch{1.20}
\begin{table}[t]
	\centering
	\caption{\lad{Performance comparisons between RNDVoC-Cascaded and the adopted format in the main text. Results are based on the LibriTTS benchmark.}}
	\vspace{-6pt}
	\resizebox{0.95\columnwidth}{!}{
		\label{tbl:ablations-multi-task}
		\begin{tabular}{ccccc}
			\toprule
			\lad{Model Type} &\lad{M-STFT$\downarrow$} &\lad{PESQ$\uparrow$} &\lad{MCD$\downarrow$} &\lad{UTMOS$\uparrow$} \\
			\midrule 
			  \lad{RNDVoC-Cascaded} &\lad{0.827} &\lad{3.949} &\lad{2.623} &\lad{3.437} \\
			  \lad{Ref. (Main Text)} &\lad{\textbf{0.746}} &\lad{\textbf{4.226}} &\lad{\textbf{1.913}} &\lad{\textbf{3.657}} \\
			\bottomrule
	\end{tabular}}
	\vspace{-0.3cm}
\end{table}

\renewcommand\arraystretch{1.30}
\begin{table*}
	\centering
	\caption{The comparisons between RNDVoC-Mag and RNDVoC-Com on the LJSpeech dataset.}
	\vspace{-6pt}
	\resizebox{0.98\textwidth}{!}{
		\label{tab:ablationss}
		\begin{tabular}{|cl|c|cccc|ccccc|}
			\hline
			\multicolumn{2}{|c|}{\multirow{2}*{Setting}} &\multirow{2}*{P} &\#Param. &\#MACs &\multicolumn{2}{c|}{Inference Speed} &\multirow{2}*{PESQ$\uparrow$} &\multirow{2}*{MCD$\downarrow$} &Periodicity$\downarrow$ &V/UV$\uparrow$ &Pitch$\downarrow$\\
			\cline{6-7}
			& & &(M) &(Giga/5s) &CPU &GPU & & &RMSE &F1 &RMSE\\
			\hline\hline
			(1) &RNDVoC-Mag &2 &9.48 &24.98 &0.0770(12.98$\times$) &0.0063(158$\times$) &\textbf{3.763} &\textbf{2.394} &\textbf{0.099} &\textbf{0.966} &\textbf{23.419}\\
			(2) &RNDVoC-Com &2 &\lad{11.07} &\lad{26.39} &\lad{0.0988(10.12$\times$)} &\lad{0.0076(131$\times$)}  &\lad{3.335} &\lad{2.814} &\lad{0.120} &\lad{0.958} &\lad{27.867} \\
			\hline
	\end{tabular}}
	\vspace{-8pt}
\end{table*}

\renewcommand\arraystretch{1.00}
\begin{table*}[t]
    \centering
    \caption{\lad{Objective performance between ``w/o MCDA'' and three MCDA variants in terms of different number of mel-filters. $f_{max}$ is fixed to 12 kHz.}}
    \resizebox{0.65\textwidth}{!}{
        \label{tbl:objective-mcda-Fm}
        \begin{tabular}{cccccccc}
            \toprule
            \lad{Type} &\lad{$F_{m}$} &\lad{$f_{min}$ (Hz)}  &\lad{$f_{max}$ (kHz)} &\lad{M-STFT} &\lad{PESQ$\uparrow$}  &\lad{MCD$\downarrow$} &\lad{UTMOS$\uparrow$} \\
            \midrule
            \multirow{9}*{\lad{$\text{w/o MCDA}$}} &\lad{72} &\lad{0} &\lad{12} &\lad{1.505} &\lad{2.393} &\lad{4.118} &\lad{2.477} \\
            &\lad{80} &\lad{0} &\lad{12} &\lad{1.315} &\lad{2.951} &\lad{3.420} &\lad{3.109}\\
            &\lad{88} &\lad{0} &\lad{12} &\lad{1.226} &\lad{3.266} &\lad{3.062} &\lad{3.204}\\
            &\lad{96} &\lad{0} &\lad{12} &\lad{1.087} &\lad{3.350} &\lad{3.093} &\lad{3.246} \\
            &\lad{100} &\lad{0} &\lad{12} &\lad{0.780} &\lad{4.085} &\lad{2.082} &\lad{3.552}\\
            &\lad{104} &\lad{0} &\lad{12} &\lad{1.023} &\lad{3.570} &\lad{2.430} &\lad{3.434} \\
            &\lad{112} &\lad{0} &\lad{12} &\lad{1.073} &\lad{3.734} &\lad{2.411} &\lad{3.452}\\
            &\lad{120} &\lad{0} &\lad{12} &\lad{1.048} &\lad{3.798} &\lad{2.349} &\lad{3.482} \\
            &\lad{128} &\lad{0} &\lad{12} &\lad{1.056} &\lad{3.871} &\lad{2.340} &\lad{3.482} \\
            \midrule
            \multirow{9}*{\lad{$\text{MCDA-1}$}} &\lad{72} &\lad{0} &\lad{12} &\lad{0.947} &\lad{3.287} &\lad{2.996} &\lad{3.103} \\
            &\lad{80} &\lad{0} &\lad{12} &\lad{0.885} &\lad{3.677} &\lad{2.680} &\lad{3.371} \\
            &\lad{88} &\lad{0} &\lad{12} &\lad{0.849} &\lad{3.905} &\lad{2.500} &\lad{3.476} \\
            &\lad{96} &\lad{0} &\lad{12} &\lad{0.832} &\lad{3.979} &\lad{2.432} &\lad{3.524} \\
            &\lad{100} &\lad{0} &\lad{12} &\lad{0.824} &\lad{4.007} &\lad{2.405} &\lad{3.530} \\
            &\lad{104} &\lad{0} &\lad{12} &\lad{0.818} &\lad{4.028} &\lad{2.386} &\lad{3.551} \\
            &\lad{112} &\lad{0} &\lad{12} &\lad{0.810} &\lad{4.045} &\lad{2.363} &\lad{3.568} \\
            &\lad{120} &\lad{0} &\lad{12} &\lad{0.804} &\lad{4.053} &\lad{2.338} &\lad{3.555} \\
            &\lad{128} &\lad{0} &\lad{12} &\lad{0.799} &\lad{4.071} &\lad{2.329} &\lad{3.552} \\
            \midrule
            \multirow{9}*{\lad{$\text{MCDA-2}$}} &\lad{72} &\lad{0} &\lad{12} &\lad{0.872} &\lad{3.686} &\lad{2.612} &\lad{3.391} \\
            &\lad{80} &\lad{0} &\lad{12} &\lad{0.832} &\lad{3.854} &\lad{2.427} &\lad{3.482} \\
            &\lad{88} &\lad{0} &\lad{12} &\lad{0.805} &\lad{4.009} &\lad{2.295} &\lad{3.547}\\
            &\lad{96} &\lad{0} &\lad{12} &\lad{0.785} &\lad{4.082} &\lad{2.210} &\lad{3.578} \\
            &\lad{100} &\lad{0} &\lad{12} &\lad{0.776} &\lad{4.093} &\lad{2.178} &\lad{3.601} \\
            &\lad{104} &\lad{0} &\lad{12} &\lad{0.769} &\lad{4.120} &\lad{2.142} &\lad{3.604} \\
            &\lad{112} &\lad{0} &\lad{12} &\lad{0.756} &\lad{4.152} &\lad{2.109} &\lad{3.621} \\
            &\lad{120} &\lad{0} &\lad{12} &\lad{0.756} &\lad{4.171} &\lad{2.069} &\lad{3.621} \\
            &\lad{128} &\lad{0} &\lad{12} &\lad{0.737} &\lad{4.190} &\lad{2.051} &\lad{3.615} \\
            \midrule
            \multirow{9}*{\lad{$\text{MCDA-3}$}} &\lad{72} &\lad{0} &\lad{12} &\lad{0.894} &\lad{3.652} &\lad{2.687} &\lad{3.311} \\
            &\lad{80} &\lad{0} &\lad{12} &\lad{0.844} &\lad{3.867} &\lad{2.448} &\lad{3.467} \\
            &\lad{88} &\lad{0} &\lad{12} &\lad{0.815} &\lad{4.020} &\lad{2.281} &\lad{3.538} \\
            &\lad{96} &\lad{0} &\lad{12} &\lad{0.795} &\lad{4.120} &\lad{2.192} &\lad{3.588} \\
            &\lad{100} &\lad{0} &\lad{12} &\lad{0.786} &\lad{4.136} &\lad{2.149} &\lad{3.593}\\
            &\lad{104} &\lad{0} &\lad{12} &\lad{0.777} &\lad{4.159} &\lad{2.118} &\lad{3.612}\\
            &\lad{112} &\lad{0} &\lad{12} &\lad{0.766} &\lad{4.189} &\lad{2.068} &\lad{3.622}\\
            &\lad{120} &\lad{0} &\lad{12} &\lad{0.756} &\lad{4.215} &\lad{2.027} &\lad{3.636} \\
            &\lad{128} &\lad{0} &\lad{12} &\lad{0.746} &\lad{4.237} &\lad{2.004} &\lad{3.631} \\
            \bottomrule
    \end{tabular}}
    \vspace{-0.3cm}
\end{table*}

\renewcommand\arraystretch{1.00}
\begin{table*}[t]
    \centering
    \caption{\lad{Objective performance between ``w/o MCDA'' and three MCDA variants in terms of different upper-bound frequency values. $F_{m}$ is fixed to 100.}}
    \resizebox{0.65\textwidth}{!}{
        \label{tbl:objective-mcda-fmax}
        \begin{tabular}{cccccccc}
            \toprule
            \lad{Type} &\lad{$F_{m}$} &\lad{$f_{min}$ (Hz)}  &\lad{$f_{max}$ (kHz)} &\lad{M-STFT} &\lad{PESQ$\uparrow$}  &\lad{MCD$\downarrow$} &\lad{UTMOS$\uparrow$} \\
            \midrule
            \multirow{6}*{\lad{$\text{w/o MCDA}$}} &\lad{100} &\lad{0} &\lad{8} &\lad{2.549} &\lad{3.557} &\lad{4.676} &\lad{3.403} \\
            &\lad{100} &\lad{0} &\lad{9} &\lad{2.220} &\lad{3.613} &\lad{3.687} &\lad{3.433} \\
            &\lad{100} &\lad{0} &\lad{10} &\lad{1.897} &\lad{3.654} &\lad{2.631} &\lad{3.440} \\
            &\lad{100} &\lad{0} &\lad{11} &\lad{1.502} &\lad{3.523} &\lad{2.464} &\lad{3.422} \\
            &\lad{100} &\lad{0} &\lad{11.5} &\lad{1.343} &\lad{3.479} &\lad{2.496} &\lad{3.424} \\
            &\lad{100} &\lad{0} &\lad{12} &\lad{0.798} &\lad{4.085} &\lad{2.250} &\lad{3.552} \\
            \midrule
            \multirow{6}*{\lad{$\text{MCDA-1}$}} &\lad{100} &\lad{0} &\lad{8} &\lad{1.595} &\lad{4.046} &\lad{2.428} &\lad{3.531} \\
            &\lad{100} &\lad{0} &\lad{9} &\lad{0.877} &\lad{4.046} &\lad{2.387} &\lad{3.514} \\
            &\lad{100} &\lad{0} &\lad{10} &\lad{0.848} &\lad{4.027} &\lad{2.389} &\lad{3.544} \\
            &\lad{100} &\lad{0} &\lad{11} &\lad{0.837} &\lad{4.031} &\lad{2.389} &\lad{3.536} \\
            &\lad{100} &\lad{0} &\lad{11.5} &\lad{0.831} &\lad{4.017} &\lad{2.407} &\lad{3.537} \\
            &\lad{100} &\lad{0} &\lad{12} &\lad{0.824} &\lad{4.007} &\lad{2.405} &\lad{3.530} \\
            \midrule
            \multirow{6}*{\lad{$\text{MCDA-2}$}} &\lad{100} &\lad{0} &\lad{8} &\lad{0.871} &\lad{4.149} &\lad{2.129} &\lad{3.616} \\
            &\lad{100} &\lad{0} &\lad{9} &\lad{0.840} &\lad{4.154} &\lad{2.120} &\lad{3.602} \\
            &\lad{100} &\lad{0} &\lad{10} &\lad{0.809} &\lad{4.141} &\lad{2.138} &\lad{3.593} \\
            &\lad{100} &\lad{0} &\lad{11} &\lad{0.788} &\lad{4.122} &\lad{2.163} &\lad{3.593} \\
            &\lad{100} &\lad{0} &\lad{11.5} &\lad{0.782} &\lad{4.110} &\lad{2.174} &\lad{3.609} \\
            &\lad{100} &\lad{0} &\lad{12} &\lad{0.780} &\lad{4.093} &\lad{2.176} &\lad{3.601} \\
            \midrule
            \multirow{6}*{\lad{$\text{MCDA-3}$}} &\lad{100} &\lad{0} &\lad{8} &\lad{0.882} &\lad{4.194} &\lad{2.094} &\lad{3.621} \\
            &\lad{100} &\lad{0} &\lad{9} &\lad{0.839} &\lad{4.185} &\lad{2.090} &\lad{3.622} \\
            &\lad{100} &\lad{0} &\lad{10} &\lad{0.813} &\lad{4.170} &\lad{2.103} &\lad{3.607} \\
            &\lad{100} &\lad{0} &\lad{11} &\lad{0.796} &\lad{4.148} &\lad{2.123} &\lad{3.607} \\
            &\lad{100} &\lad{0} &\lad{11.5} &\lad{0.789} &\lad{4.137} &\lad{2.131} &\lad{3.605} \\
            &\lad{100} &\lad{0} &\lad{12} &\lad{0.786} &\lad{4.136} &\lad{2.149} &\lad{3.593} \\
            \bottomrule
    \end{tabular}}
    \vspace{-0.3cm}
\end{table*}

\begin{figure*}[t]
\centering
\vspace{0pt}
\includegraphics[width=0.98\textwidth]{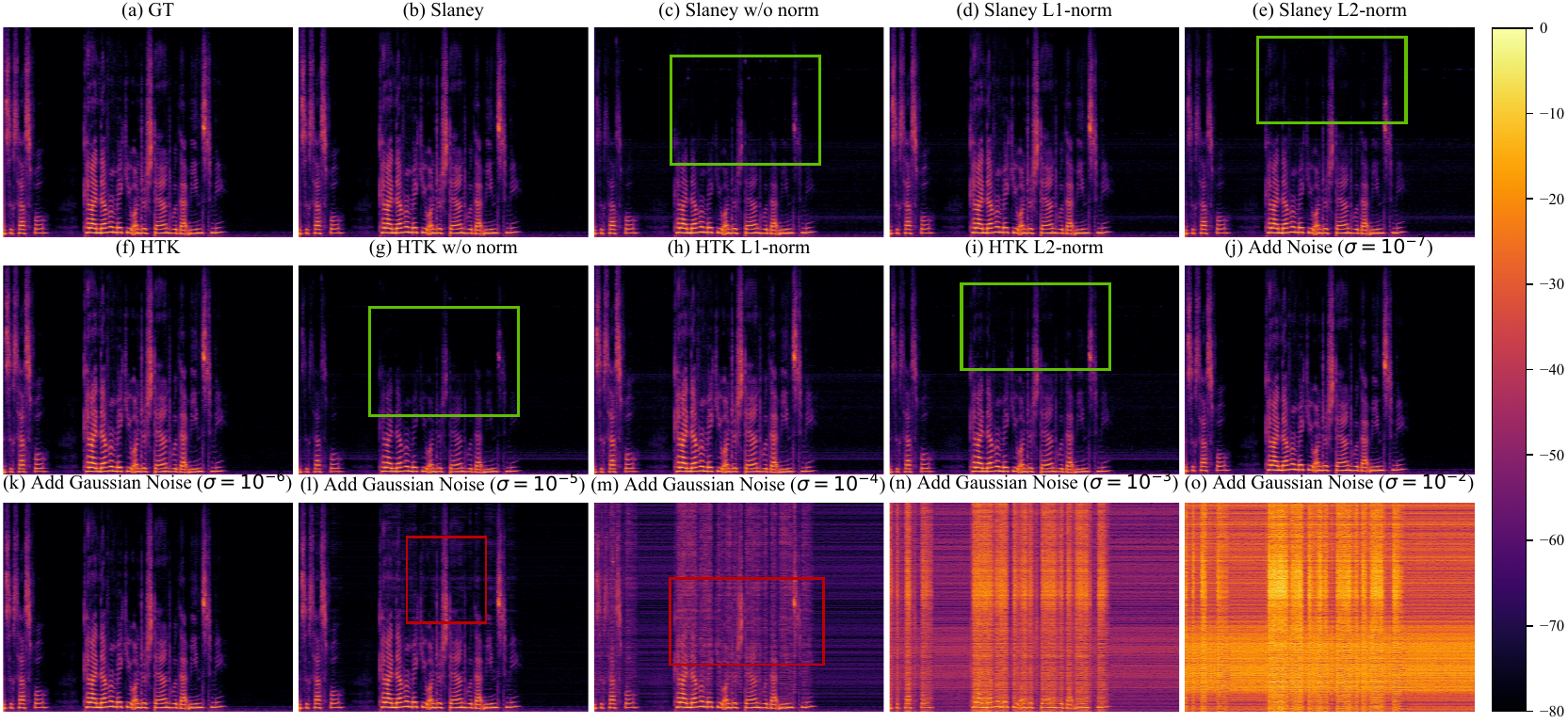}
\vspace{-2pt}
\caption{\lad{Spectral visualizations given different deviations of mel-filter. (a): Target. (b): Default configuration (Slaney strategy with region normalizaiton). (c): Slaney strategy without normalization. (d): Slaney strategy with $L_{1}$ normalization. (e): Slaney strategy with $L_{2}$ normalization. (f): HTK strategy with region normalization. (g): HTK without normalization. (h): HTK with $L_{1}$ normalization. (i): HTK with $L_{2}$ normalization. (j)-(o): Based on (b), add Gaussian noise with varying strength $\sigma$, ranging from $10^{-7}$ to $10^{-2}$. }}  
\label{fig:appendix_sensitivity}
\vspace{-6pt}
\end{figure*}

\begin{figure}[h]
	\centering
	\vspace{0pt}
	\includegraphics[width=0.99\columnwidth]{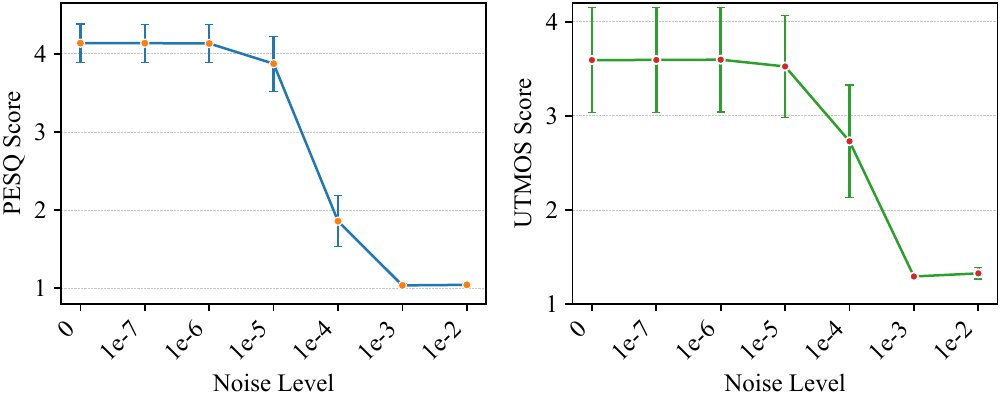}
	\vspace{-2pt}
	\caption{\lad{Sensitivity analysis of mel-filter in terms of PESQ and UTMOS. Results are calculated based on the LibriTTS benchmark.}}  
	\label{fig:appendix_sensitivity_performance}
	\vspace{-6pt}
\end{figure}

\renewcommand\arraystretch{1.0}
\setlength{\tabcolsep}{0.5pt}
\begin{table}[t]
    \large
    \caption{\lad{Detailed parameter illustrations of the RNDVoC-Shared. For BAEM, hyper-parameters denote kernel sizes for three spectral regions along the time and frequency axes, respectively. For conv1d layers, hyper-parameters denote (output\_channel, kernel\_size, stride\_size, group\_num). For band\_mixer layer, hyper-parameters denotes (output\_size, group\_num).}}
    \begin{center}
        \resizebox{0.95\columnwidth}{!}{
            \begin{tabular}{cc|c|c|c}
                \hline
                &\lad{layer name} & \lad{input size} & \lad{hyper-parameters} & \lad{output size}\\
                \hline
                &\lad{RSM} & \lad{$100\times T$} &\lad{-} &\lad{$513\times T$} \\
                \hline\hline
                &\lad{BAEM} &\lad{$513\times T$}&\lad{$(3,12),(3,24),(3,44)$} &\lad{$24\times 256\times T$}\\
                \hline\hline
                \multirow{5}*{\rotatebox{90}{\lad{\textbf{Cross}}}}
                &\lad{reshape\_1} &\lad{$24\times 256\times T$} &\lad{-} &\lad{$T\times 256\times 24$}\\
                \cline{2-5}
                &\lad{gconv1d\_1} &\lad{$T\times 256\times 24$} &\lad{$(256, 3, 1, 8)$} &\lad{$T\times 256\times 24$} \\
                \cline{2-5}
                &\lad{pconv1d\_1} &\lad{$T\times 256\times 24$} &\lad{$(64, 1, 1, 1)$ }&\lad{$T\times 64\times 24$} \\
                \cline{2-5}
                &\lad{band\_mixer} &\lad{$T\times 64\times 24$} &\lad{$(24, 64)$} &\lad{$T\times 64\times 24$} \\
                \cline{2-5}
                &\lad{pconv1d\_2} &\lad{$T\times 64\times 24$} &\lad{$(256, 1, 1, 1)$} &\lad{$T\times 256\times 24$} \\
                \hline\hline
                \multirow{5}*{\rotatebox{90}{\lad{\textbf{Narrow}}}}
                &\lad{reshape\_2} &\lad{$24\times 256\times T$} &\lad{-} &\lad{$24\times 256\times T$} \\
                \cline{2-5}
                &\lad{dwconv\_1} &\lad{$24\times 256\times T$} &\lad{$(256, 7, 1, 256)$} &\lad{$24\times 256\times T$} \\
                \cline{2-5}
                &\lad{pconv1d\_3} &\lad{$T\times 256\times 24$} &\lad{$(256, 1, 1, 1)$} &\lad{$24\times 256\times T$} \\
                \cline{2-5}
                &\lad{grn} &\lad{$T\times 256\times 24$} &\lad{-} &\lad{$24\times 256\times T$} \\
                \cline{2-5}
                &\lad{pconv1d\_4} &\lad{$T\times 256\times 24$} &\lad{$(256, 1, 1, 1)$} &\lad{$24\times 256\times T$} \\
                \hline\hline
                &\lad{BAMM} &\lad{$24\times 256\times T$} &\lad{$(1, 12), (1, 24), (1, 44)$} &\lad{$512\times T$}\\
                \hline
                &\lad{zero\_pad\_1} &\lad{$512\times T$} &\lad{-} &\lad{$513\times T$} \\
                \hline
                &\lad{BAPM} &\lad{$24\times 256\times T$} &\lad{$(1, 12), (1, 24), (1, 44)$} &\lad{$512\times T$}\\
                \hline
                &\lad{zero\_pad\_2} &\lad{$512\times T$} &- &\lad{$513\times T$} \\
                \hline\hline
        \end{tabular}}
        \label{tbl:hyper-parameter}
    \end{center}
    \vspace{-0.2cm}
\end{table}

\renewcommand\arraystretch{0.85}
\begin{table*}[t]
	\centering
	\caption{\lad{Objective comparisons among different baselines on the out-of-distribution sound effect evaluation set from~{[8]} and AISHELL3 datasets. All models are pretrained on the LibriTTS dataset.}}
	\resizebox{0.85\textwidth}{!}{
		\begin{tabular}{ccccccccccc}
			\toprule
			\lad{Test sets} &\multicolumn{3}{c}{\lad{Sound Effect from~{\cite{liu2024rfwave}}}} &\multicolumn{7}{c}{\lad{AISHELL3 (Mandarin)}}\\
			\midrule
			\multirow{2}*{\lad{Models}} &\multirow{2}*{\lad{M-STFT$\downarrow$}} &\multirow{2}*{\lad{MCD$\downarrow$}} &\multirow{2}*{\lad{VISQOL$\uparrow$}} &\multirow{2}*{\lad{M-STFT$\downarrow$}} &\multirow{2}*{\lad{PESQ$\uparrow$}} &\multirow{2}*{\lad{MCD$\downarrow$}} &\lad{V/UV$\uparrow$} &\lad{Period.$\downarrow$} &\multirow{2}*{\lad{UTMOS$\uparrow$}}  &\multirow{2}*{\lad{VISQOL$\uparrow$}}\\
			& & & & & & &\lad{F1} &\lad{RMSE} &  &\\
			\midrule
			\lad{GT} &- &- &- &- &- &- &- &-  &\lad{2.700} &-\\
			\lad{HiFiGAN-V1} &\lad{1.462} &\lad{6.971} &\lad{4.291} &\lad{1.059} &\lad{2.823} &\lad{2.827} &\lad{0.934} &\lad{0.163} &\lad{2.432} &\lad{4.651} \\
			\lad{iSTFTNet-V1} &\lad{1.540} &\lad{7.352} &\lad{4.150} &\lad{1.132} &\lad{2.595} &\lad{3.061} &\lad{0.932} &\lad{0.164} &\lad{2.351} &\lad{4.486}\\
			\lad{Avocodo} &\lad{1.488} &\lad{6.795} &\lad{4.426} &\lad{1.040} &\lad{3.003} &\lad{2.864} &\lad{0.937} &\lad{0.154} &\lad{2.126} &\lad{4.731}\\
			\lad{BigVGAN-base$^{\ddagger}$ (5M steps)} &\lad{1.156} &\lad{5.577} &\lad{4.716} &\lad{0.811} &\lad{3.642} &\lad{1.688} &\lad{0.960} &\lad{0.111} &\lad{2.428} &\lad{4.885} \\
			\lad{BigVGAN$^{\ddagger}$ (5M steps)} &\lad{0.998} &\lad{\underline{4.837}} &\lad{\underline{4.851}}  &\lad{\underline{0.720}} &\lad{\underline{4.110}} &\lad{\underline{1.313}} &\lad{\textbf{0.970}} &\lad{\textbf{0.090}} &\lad{\textbf{2.573}} &\lad{\textbf{4.941}} \\
			\lad{APNet2} &\lad{2.138} &\lad{8.318} &\lad{4.303} &\lad{0.975} &\lad{2.846} &\lad{2.743} &\lad{0.950} &\lad{0.136} &\lad{2.129} &\lad{4.715}\\
			\lad{Vocos$^{\ddagger}$} &\lad{1.110} &\lad{5.715} &\lad{4.745} &\lad{0.845} &\lad{3.342} &\lad{2.046} &\lad{0.952} &\lad{0.129} &\lad{2.409} &\lad{4.836}\\
			\lad{FreGrad} &\lad{1.669} &\lad{8.343} &\lad{4.193} &\lad{1.061} &\lad{3.667} &\lad{1.930} &\lad{0.937} &\lad{0.148} &\lad{2.145} &\lad{4.621}\\
			\lad{PriorGrad} &\lad{1.610} &\lad{8.106} &\lad{4.289} &\lad{1.054} &\lad{3.995} &\lad{1.717} &\lad{0.917} &\lad{0.160} &\lad{2.197} &\lad{4.675}\\
			\lad{WaveFM$^{\ddagger}$} &\lad{1.001} &\lad{4.983} &\lad{\underline{4.851}} &\lad{0.832} &\lad{3.810} &\lad{1.599} &\lad{0.964} &\lad{0.108} &\lad{2.037} &\lad{\underline{4.929}}\\
			\lad{PeriodWave$^{\ddagger}$} &\lad{1.232} &\lad{5.197} &\lad{4.595} &\lad{0.958} &\lad{\textbf{4.195}} &\lad{1.598} &\lad{0.953} &\lad{0.119} &\lad{\underline{2.565}} &\lad{4.616}\\
			\midrule
			\lad{RNDVoC-nonshared} &\lad{\underline{0.968}} &\lad{4.843} &\lad{4.821} &\lad{0.752} &\lad{3.938} &\lad{1.445} &\lad{\underline{0.968}} &\lad{0.093} &\lad{2.402} &\lad{4.886}\\
			\lad{RNDVoC-shared} &\lad{\textbf{0.913}} &\lad{\textbf{4.542}} &\lad{\textbf{4.870}} &\lad{\textbf{0.702}} &\lad{4.049} &\lad{\textbf{1.267}} &\lad{\underline{0.968}} &\lad{\underline{0.091}} &\lad{2.486} &\lad{4.912}\\
			\bottomrule
	\end{tabular}}
	\vspace{-0.3cm}
	\label{tbl:objective-metric-sound-effect-aishell3}
\end{table*}

\begin{table*}[t]
	\centering
	\caption{\lad{Different subband division schemes, including total number of sub-bands, frequency range, bandwidths, kernel size, and stride size.}}
	\resizebox{0.72\textwidth}{!}{
		\begin{tabular}{ccccccc}
			\hline
			\lad{Split-Type} &\lad{Total Subbands} &\lad{Range (Hz)} &\lad{Width (Hz)} &\lad{Sub-band number} &\lad{Kernel Size} &\lad{Stride Size}\\
			\hline
			\multirow{15}*{\lad{Uneven}} &\multirow{3}*{\lad{6}} &\lad{0-3375} &\lad{1125} &\lad{3} &\lad{(48, 3)} &\lad{(48, 1)}\\
			& &\lad{3375-7875} &\lad{2250} &\lad{2} &\lad{(96, 3)} &\lad{(96, 1)}\\
			& &\lad{7875-12000} &\lad{4125} &\lad{1} &\lad{(176, 3)} &\lad{(176, 1)} \\
			\cline{2-7}
			&\multirow{3}*{\lad{12}} &\lad{0-3375} &\lad{562.5} &\lad{6} &\lad{(24, 3)} &\lad{(24, 3)}\\
			& &\lad{3375-7825} &\lad{1125} &\lad{4} &\lad{(48, 3)} &\lad{(48, 1)}\\
			& &\lad{7825-12000} &\lad{2062.5} &\lad{2} &\lad{(88, 3)} &\lad{(88, 1)}\\
			\cline{2-7}
			&\multirow{3}*{\lad{24}} &\lad{0-3375} &\lad{281.25} &\lad{12} &\lad{(12, 3)} &\lad{(12, 1)}\\
			& &\lad{3375-7825} &\lad{562.5} &\lad{8} &\lad{(24, 3)} &\lad{(24, 1)} \\
			& &\lad{7825-12000} &\lad{1031.25} &\lad{4} &\lad{(44, 3)} &\lad{(44, 1)}\\
			\cline{2-7}
			&\multirow{3}*{\lad{48}} &\lad{0-3375} &\lad{140.625} &\lad{24} &\lad{(6, 3)} &\lad{(6, 1)}\\
			& &\lad{3375-7825} &\lad{281.25} &\lad{16} &\lad{(12, 3)} &\lad{(12, 1)} \\
			& &\lad{7825-12000} &\lad{515.625} &\lad{8} &\lad{(22, 3)} &\lad{(22, 1)}\\
			\cline{2-7}
			&\multirow{3}*{\lad{96}} &\lad{0-3375} &\lad{70.3125} &\lad{48} &\lad{(3, 3)} &\lad{(3, 1)}\\
			& &\lad{3375-7825} &\lad{140.625} &\lad{32} &\lad{(6, 3)} &\lad{(6, 1)} \\
			& &\lad{7825-12000} &\lad{257.8125} &\lad{16} &\lad{(11, 3)} &\lad{(11, 1)}\\
			\cline{1-7}
			\multirow{3}*{\lad{Even}} &\multirow{3}*{\lad{24}} &\lad{0-4000} &\lad{500} &\lad{8} &\lad{(21, 3)} &\lad{(21, 1)}\\
			& &\lad{4000-8000} &\lad{500} &\lad{8} &\lad{(21, 3)} &\lad{(21, 1)}\\
			& &\lad{8000-12000} &\lad{500} &\lad{8} &\lad{(21, 3)} &\lad{(21, 1)}\\
			\hline
	\end{tabular}}
	\label{tbl:subband-division-other}
\end{table*}

\renewcommand\arraystretch{0.9}
\begin{table*}
	\centering
	\huge
	\caption{Comparisons among different neural vocoders on the SE task. Note that BigVGAN-base and BigVGAN are pretrained for 5M steps while other models are trained for 1M steps.}
	\vspace{-2pt}
	\resizebox{0.92\textwidth}{!}{
        \label{tbl:denoising-performance}
        \begin{tabular}{c|cccc|cccc}
                \toprule
                SE Models &\multicolumn{4}{c|}{Discriminative} &\multicolumn{4}{c}{Generative}\\
                \cline{1-1}\cmidrule(lr){2-5}\cmidrule(lr){6-9}
                Vocoders &PESQ ($\uparrow$) &eSTOI (in \%)$\uparrow$ &P808-DNSMOS $\uparrow$ &VISQOL $\uparrow$ &PESQ $\uparrow$ &eSTOI (in \%)$\uparrow$ &P808-DNSMOS$\uparrow$ &VISQOL $\uparrow$ \\
                \midrule
                Mixture &1.163 &57.48 &2.580 &2.512 &1.163 &27.48 &2.580 &2.512\\
                HiFiGAN-V1 &1.903 &75.89 &3.268 &3.601 &1.739 &71.78 &\underline{3.440} &3.318 \\
                iSTFTNet-V1 &1.818 &74.76 &3.275 &3.551 &1.677 &70.87 &3.416 &3.268\\
                Avocodo &1.892 &75.47 &3.212 &3.613 &1.720 &71.48 &3.389 &3.311\\
                BigVGAN-base &2.051 &76.62 &3.252 &\underline{3.698} &1.839 &72.77 &3.422 &3.375\\
                BigVGAN &\underline{2.149} &77.41 &3.281 &\textbf{3.729} &\textbf{1.915} &73.37 &3.409 &\textbf{3.402} \\
                APNet &1.800 &77.31 &3.176 &3.595 &1.630 &73.19 &3.335 &3.290 \\
                APNet2 &1.976  &76.60 &\textbf{3.348} &3.521 &1.762 &72.56 &\textbf{3.460} &3.301 \\
                Vocos &2.060 &80.48 &3.290 &3.685 &1.829 &75.97 &3.467 &3.373 \\
                RNDVoC-nonshared &2.127 &\underline{81.62} &3.266 &3.678 &1.887 &\underline{77.11} &3.404 &3.361 \\
                RNDVoC-shared &\textbf{2.155} &\textbf{81.87} &\underline{3.292} &3.695 &\underline{1.910} &\textbf{77.41} &3.429 &\underline{3.383}\\
                \bottomrule
        \end{tabular}}
\end{table*}

\begin{figure*}[t]
	\centering
	\vspace{0pt}
	\includegraphics[width=0.82\textwidth]{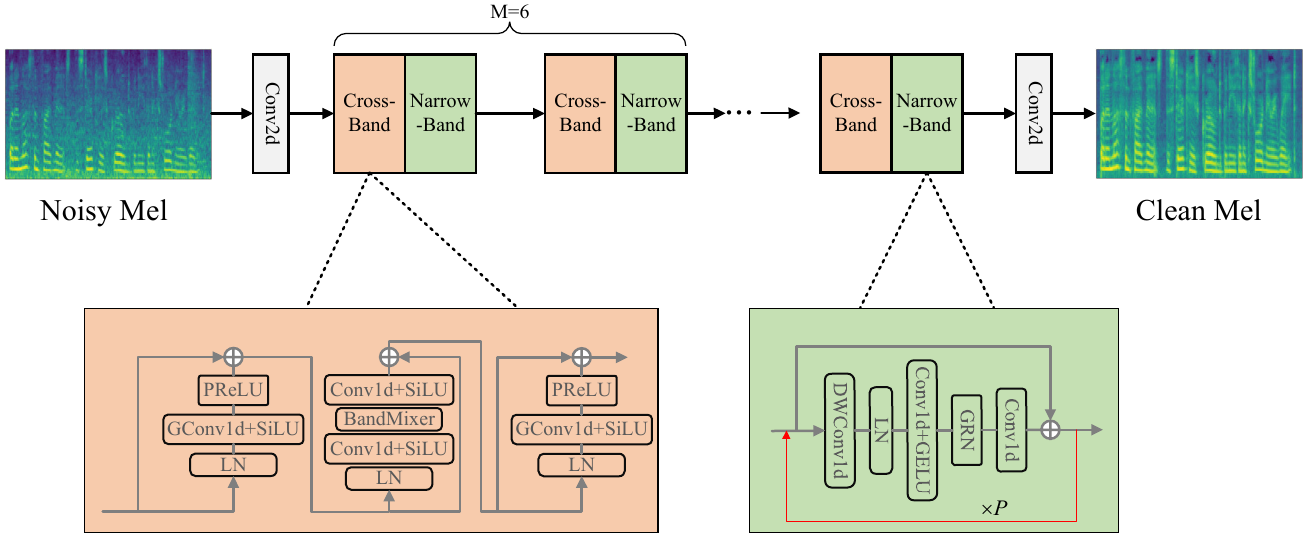}
	\vspace{-2pt}
	\caption{Detailed structure of the adopted discriminative SE networks.}  
	\label{fig:appendix_discriminative}
	\vspace{-1pt}
\end{figure*}	
\begin{figure*}[h]
	\centering
	\vspace{0pt}
	\includegraphics[width=0.82\textwidth]{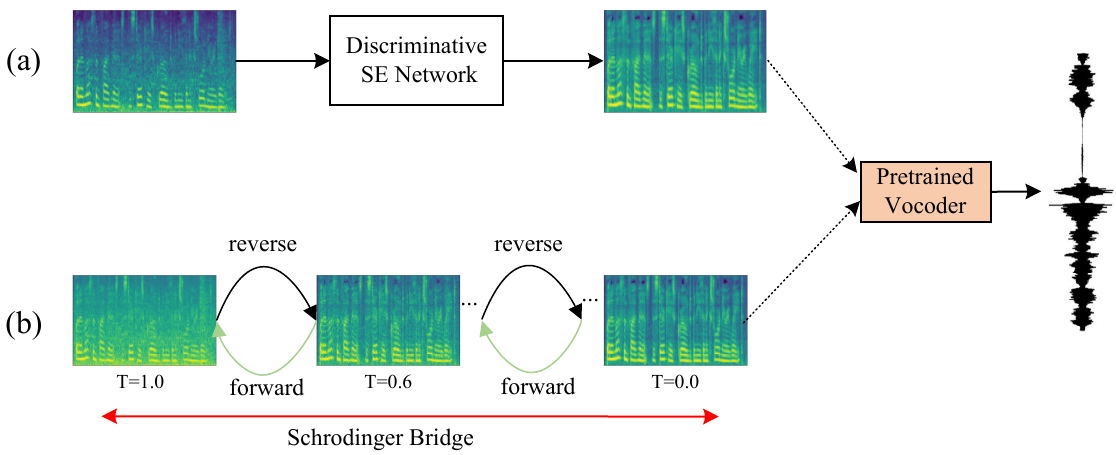}
	\vspace{-2pt}
	\caption{Framework diagram of the adopted discriminative and generative networks for SE task. For both methods, a pre-trained neural vocoder is followed for target waveform generation. (a) Discriminative method, where a discriminative SE network is utilized for target mel-spectrogram mapping. (b) Generative method, where a Schr\" odinger bridge is established between noisy and target mel-spectrograms.}  
	\label{fig:appendix_discriminative_generative}
	\vspace{-1pt}
\end{figure*}

\begin{figure*}[t]
	\centering{\includegraphics[width=\textwidth]{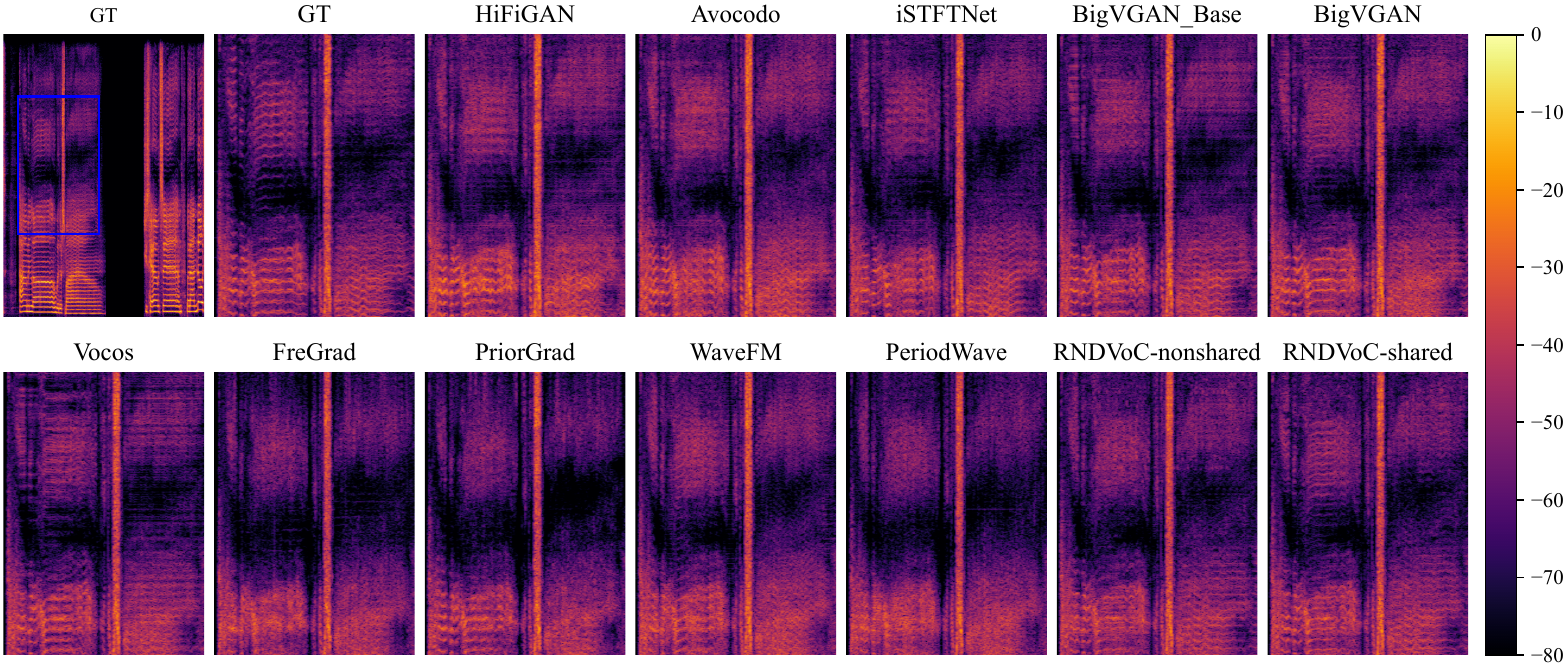}}
	\caption{\lad{Spectral Visualizations generated by different neural vocoders. The file is a vocal sound from the song ``Al James - Schoolboy Facination'' in the MUSDB18 corpus.}}
	\label{fig:spectral-visualization-1}
\end{figure*}
\begin{figure*}[t]
	\centering{\includegraphics[width=\textwidth]{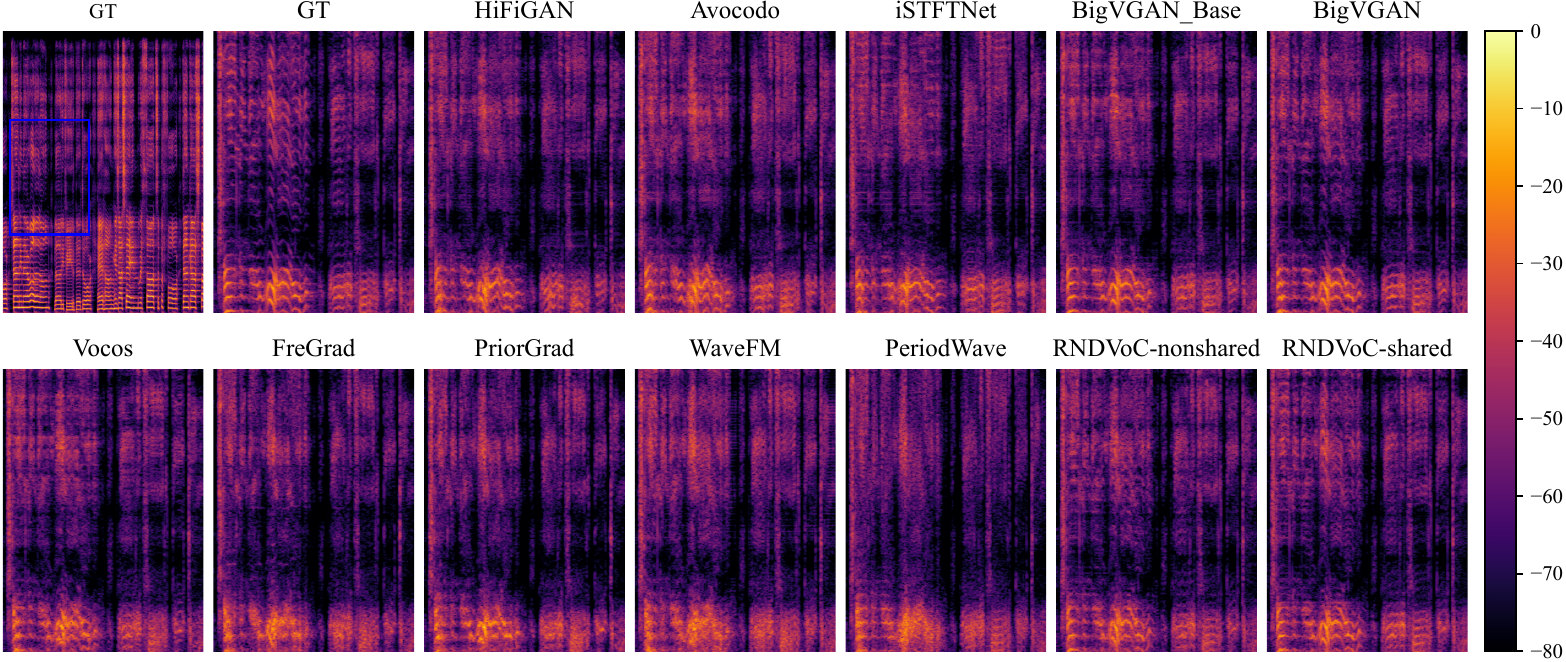}}
	\caption{\lad{Spectral Visualizations generated by different neural vocoders. The file is a vocal sound from the song ``Ben Carrigan - We'll Talk About It All Tonight'' in the MUSDB18 corpus.}}
	\label{fig:spectral-visualization-2}
\end{figure*}
\begin{figure*}[t]
	\centering{\includegraphics[width=\textwidth]{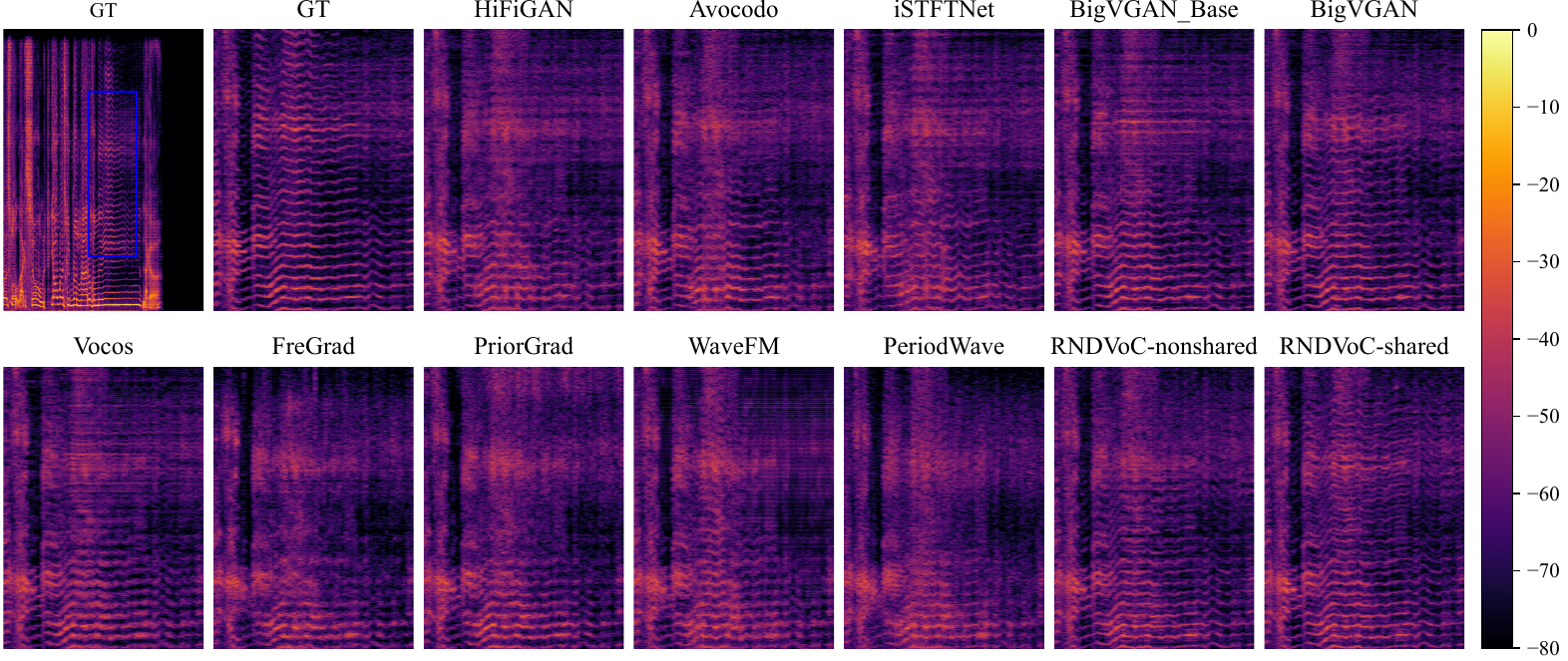}}
	\caption{\lad{Spectral Visualizations generated by different neural vocoders. The file is a vocal sound from the song ``BKS - Bulldozer - We'll Talk About It All Tonight'' in the MUSDB18 corpus.}}
	\label{fig:spectral-visualization-3}
\end{figure*}
\begin{figure*}[htbp]
	\centering{\includegraphics[width=\textwidth]{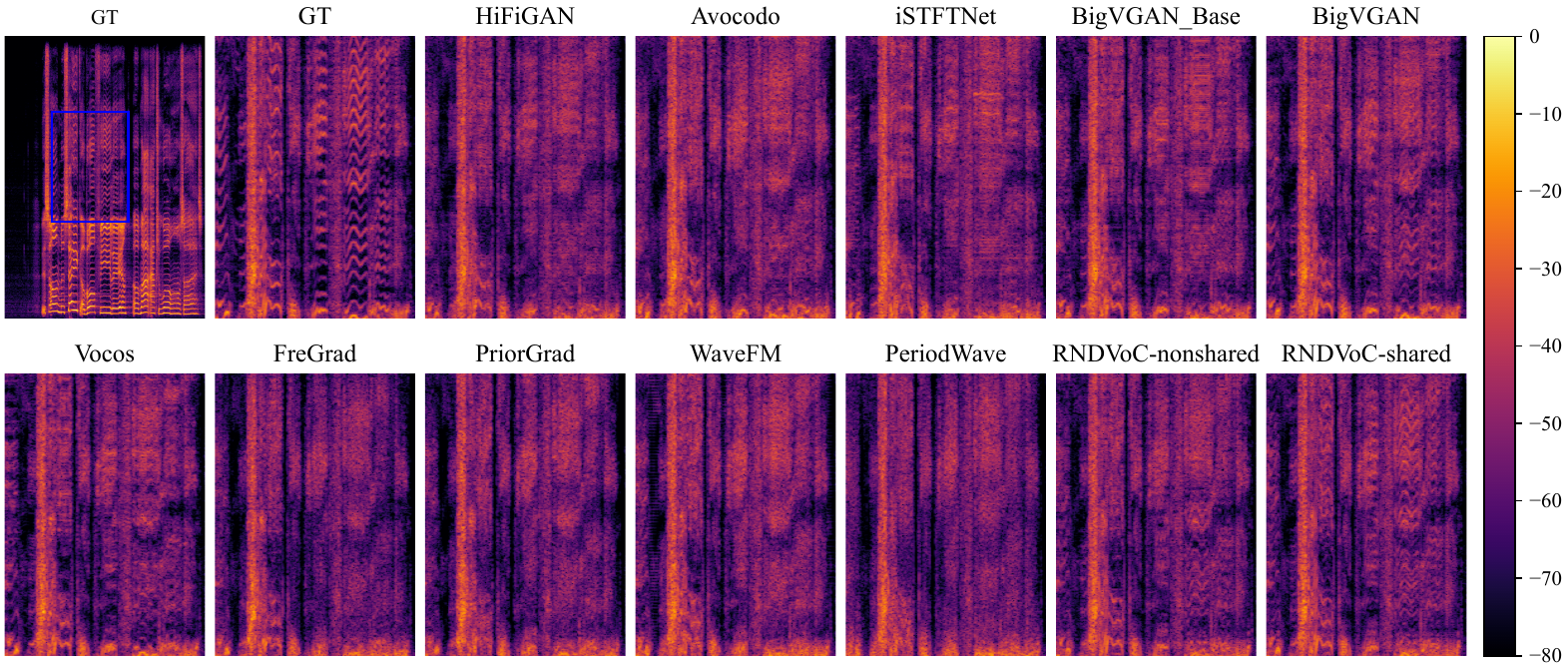}}
	\caption{\lad{Spectral Visualizations generated by different neural vocoders. The file is a vocal sound from the song ``Mu - Too Bright'' in the MUSDB18 corpus.}}
	\label{fig:spectral-visualization-4}
\end{figure*}

\subsection{RND in the cascaded format}\label{sec:rnd-cascade}
\lad{Recall from Sec.~III-C of the main text that superimposing the range-space and null-space components only yields the magnitude spectrum. To predict the phase component, a straightforward engineering approach is to enable multi-task learning in the null-space module, that is, adopting two output heads for magnitude and phase estimation, respectively. With the rapid development of phase retrieval methods~{\cite{peer2023diffphase}}, an alternative approach is to decouple magnitude and phase estimation, and we thus propose a cascaded variant of RNDVoC, denoted as \textbf{RNDVoC-Cascaded}. As illustrated in Fig.~{\ref{fig:appendix_rnd_cascaded}}, this architecture consists of two core parts. The first one is a RND module, which estimates the range-space and null-space components. After the superimposition operation, we can recover the magnitude information. Then, the estimated magnitude is sent to the phase-net to predict the phase information. The two components are combined into a complex-valued spectrum, followed by an iSTFT operation to generate the reconstructed waveform. The whole process can be formulated as:}
\begin{equation}
    \label{eqn:54}
    \lad{\left|\tilde{\mathbf{S}}\right|_{range} = \mathcal{F}_{range}\left(\mathbf{Y}\right) = \mathcal{A}^{\dagger}\mathbf{Y},}
\end{equation}
\begin{equation}
    \label{eqn:55}
    \lad{\left|\tilde{\mathbf{S}}\right|_{null} = \mathcal{F}_{null}\left(\mathbf{Y}\right),}
\end{equation}
\begin{equation}
\label{eqn:56}
\lad{\left|\tilde{\mathbf{S}}\right| = \left|\tilde{\mathbf{S}}\right|_{range} + \left(\mathbf{I} - \mathcal{A}^{\dagger}\mathcal{A}\right)\left|\tilde{\mathbf{S}}\right|_{null},}
\end{equation}
\begin{equation}
\label{eqn:157}
\lad{\tilde{\mathbf{\Phi}} = \mathcal{F}_{phase}\left(\left|\tilde{\mathbf{S}}\right|\right),}
\end{equation}
\begin{equation}
    \label{eqn:58}
    \lad{\tilde{\mathbf{s}} = \text{iSTFT}\left(\left|\tilde{\mathbf{S}}\right|e^{j\tilde{\mathbf{\Phi}}}\right).}
\end{equation}

\lad{Table~{\ref{tbl:ablations-multi-task}} reports the objective performance of RNDVoC-Cascaded and the default setting in the main text. For fair comparison, in RNDVoC-Cascaded, we adopt the shared-scheme and reduce the number of channels $C$ in both the null-space module and phase-net to 192, resulting in 3.08M trainable parameters and a computational cost of 37.17 GMACs/5s, which closely match those of RNDVoC-shared. Notably, \textbf{the joint modeling but with separate prediction heads} of magnitude and phase yield substantial performance improvements. This reveals the intrinsic coupling relations between the magnitude and phase.}
\subsection{RND in the complex-valued Scenario}\label{sec:rnd-in-complex-valued-scenario}
As demonstrated in Eq.~(5) of the main text, linear degradation is only applied to the magnitude, leading to real-valued reconstruction via the RND framework. In this section, we empirically investigate an extended RND formulation defined in the complex-valued domain. As neural networks are usually defined for the real-valued operations, to adapt to it, we transform the spectrogram in the complex domain into the real-valued format by concatenating the real and imaginary (RI) parts along the channel axis~{\cite{tan2019learning,li2021two}}. Similar to Eqs.~(9)-(11) in the main text, we design a complex-domain inverse mapping scheme for RND, formulated as:
\begin{align}
	\label{eqn:59}
	\left\{|\tilde{\mathbf{S}}|_{range}, \mathbf{\tilde{\Phi}}_{range}\right\} = \mathcal{F}_{range}\left(\mathbf{\overline{X}}^{mel}\right),
\end{align}
\begin{align}
	\label{eqn:60}
	\left\{|\tilde{\mathbf{S}}|_{null}, \mathbf{\tilde{\Phi}}_{null}\right\} = \mathcal{F}_{null}\left(|\tilde{\mathbf{S}}|_{range}\right),
\end{align}
\begin{align}
	\label{eqn:61}
	\tilde{\mathbf{S}} &= |\tilde{\mathbf{S}}|_{range}e^{j\mathbf{\tilde{\Phi}}_{range}} + \left(\mathbf{I} - \mathcal{A}^{\dagger}\mathcal{A}\right)|\tilde{\mathbf{S}}|_{null}e^{j\mathbf{\tilde{\Phi}}_{null}}\nonumber\\
	&= \mathcal{A}^{\dagger}\mathcal{A}\left|\mathbf{S}\right|e^{j\mathbf{\tilde{\Phi}}_{range}} + \left(\mathbf{I} - \mathcal{A}^{\dagger}\mathcal{A}\right)|\tilde{\mathbf{S}}|_{null}e^{j\mathbf{\tilde{\Phi}}_{null}}.
\end{align}

By comparing the two schemes in Fig.~{\ref{fig:appendix_rnd_complex}}, several differences can be observed. First, unlike the magnitude-only case, the complex-domain variant requires explicit estimation of the range-space phase term $\tilde{\boldsymbol{\Phi}}_{range}$. Second, after $\left\{|\tilde{\mathbf{S}}|_{range}, \mathbf{\tilde{\Phi}}_{range}\right\}$ and $|\tilde{\mathbf{S}}|_{null}, \mathbf{\tilde{\Phi}}_{null}$, both component pairs are obtained and converted to their RI representations before performing the spectral superimposition operation. For clear distinction, the above two processing schemes are dubbed \textbf{RNDVoC-Mag} and \textbf{RNDVoC-Com}, respectively.

In network design, for the range-space module, to estimate the phase $\tilde{\boldsymbol{\Phi}}_{range}$, we adopt a similar framework to the null-space module, except only two DPBs are adopted and the BAMM is removed. For null-space module, the network is nearly the same except the input phase is substituted from a all-zero tensor to $\tilde{\boldsymbol{\Phi}}_{range}$. For network training, compared to the magnitude case, we add an extra phase loss to restrict the phase distribution in the range-space, whose definitions are the same to Eqs.~(41)-(42) in the main text.

We compare the two schemes on the LJSpeech benchmark, with results reported in Table~{\ref{tab:ablationss}}. Notably, RNDVoC-Com achieves significantly inferior performance to RNDVoC-Mag across all metrics. We can explain the phenomenon from several aspects. First, in the problem formulation, the compression matrix $\mathcal{A}$ is applied into the magnitude spectrum, and the phase term is not considered in the physical signal model. Therefore, it seems not suitable to consider the phase in the range- and null-spaces. Second, Eq.~({\ref{eqn:18}}) can be converted into the magnitude case if $\tilde{\boldsymbol{\Phi}}_{range}$ is equal to $\tilde{\boldsymbol{\Phi}}_{null}$, \emph{i.e.}, the range- and null-spaces share the same phase term. In our settings, as only the phase term needs to be estimated by network, and both magnitude and phase are estimated in the null-space module, the network size of the range-space module is set to be relatively smaller than the null-space module. Therefore, the estimation accuracy of $\tilde{\boldsymbol{\Phi}}_{range}$ can be inferior to $\tilde{\boldsymbol{\Phi}}_{null}$. Although scaling up the number of parameters of the range-space module can improve the phase estimation performance, it can incur higher computational cost and slower inference speed. Therefore, we think the magnitude-based RND scheme seems more reasonable, which is adopted by default.

\subsection{Pseudo Procedure of the proposed MCDA strategy}\label{sec:pseudo-procedure}
In Algorithm~{\ref{alg:mel_adapt}}, we provide the pseudo procedure of the proposed MCDA strategy.
\subsection{Sensitivity of the MCDA strategy toward unseen mel implementations}\label{sec:sensitivity-mcda}
\lad{Mel-filter bank extraction typically involves three key steps: First, given the minimum and maximum frequency bounds $\left\{f_{\text{min}}, f_{\text{max}}\right\}$, the full spectrum is partitioned into $F_{m}$ regions following a perceptual scale. Two common partitioning strategies can be chosen: Slaney~{\cite{slaney1998auditory}} and HTK~{\cite{young2002htk}}, which differ in their implementation details. Second, raw filter weights are calculated, and normalization is applied to derive the final mel-filter weights. Recall that in Sec.~IV-B of the main text, \texttt{librosa.filters.mel} is adopted for mel-filter calculation by default, and this function is configured with two parameters set to fixed default values: the Slaney method for division and region-based normalization for filter weight normalization. For our MCDA strategy, we focus on adapting the vocoder to varying configurations of $\left\{F_{m}, f_{\text{max}}\right\}$, with $f_{\text{min}}$ fixed to 0 Hz by default.}

\lad{In this section, we analyze the sensitivity of our method when mel-filter $\mathcal{A}$ deviates the distribution of the sampling pool $\mathcal{T}$. Specifically, we evaluate three modified mel-filter configurations: First, we adopt the HTK strategy for spectral partition. Second, we use $L_{1}$ and $L_{2}$, as well as the case with no normalization. Third, based on the default mel setup, we contaminate the mel filter with Gaussian noise of varying strength, defined as, $\tilde{\mathcal{A}}=\mathcal{A} + \sigma\mathbf{N}$, where $\mathbf{N}\sim\mathcal{N}\left(0, \mathbf{I}\right)$ and $\sigma$ denotes the noise strength, ranging from $10^{-7}$ to $10^{-2}$.}

\lad{Fig.~{\ref{fig:appendix_sensitivity}} showcases spectral visualizations under varying levels of filter mismatch, with MCDA-3 (illustrated in Sec.~V-E of the main text) used for inference. Several key observations can be made: First, even when trained exclusively with the Slaney strategy, our vocoder can effectively adapt to the unseen HTK configuration, as shown in Figs.~{\ref{fig:appendix_sensitivity}}(b) and (f)). Second, different filter normalization schemes impact reconstruction quality to varying degrees. For instance, in Figs.~{\ref{fig:appendix_sensitivity}}(c)–(e), the reconstructed spectrum shows minimal distortion under $L_1$ normalization, whereas severe distortion is observed for unnormalized filters and $L_2$ normalization, as shown in green boxed regions. A similar trend holds for the HTK scenario (see Figs.~{\ref{fig:appendix_sensitivity}}(g)–(i)). Third, our method demonstrates robust performance for Gaussian noise levels with $\sigma \leq 10^{-5}$, but suffers from significant noise-induced distortion when $\sigma$ increases to $10^{-3}$, as shown in Fig.~{\ref{fig:appendix_sensitivity_performance}}. This is because relatively large Gaussian noise can severely corrupt the predicted subspace, leading to substantial mismatch with the subspace distribution in the pretrained sampling pool $\mathcal{T}$.}
\vspace{-0.35cm}
\subsection{Illustrations Toward the Vocoder Network}
\label{sec:illustrations-toward-vocoder-network}
\lad{For better demonstration and reproducibility of the proposed vocoder network, we provide detailed network hyper-parameter setups and feature size after each layer, which are summarized in Table~{\ref{tbl:hyper-parameter}}.}

\subsection{Objective Performance on More Datasets}\label{sec:comparison-more-benchmark}
\lad{To further evaluate the generalizability of the proposed approach, we conduct experiments on two additional evaluation test sets. The first is a sound effect evaluation set collected in~{\cite{liu2024rfwave}}, which consists of 300 audio clips. The second is from  AISHELL3~{\cite{shi2020aishell}}, a Mandarin speech dataset, from which we randomly select 200 clips for evaluation. Table~{\ref{tbl:objective-metric-sound-effect-aishell3}} reports the objective evaluation results on two out-of-distribution datasets. Evidently, RNDVoC-shared achieves overall competitive performance against BigVGAN and PeriodWave, and outperforms these baselines on the sound effect scenario, which further validating the robustness across diverse audio types.}

\subsection{Parameter Configurations under Different Number of Subbands}\label{sec:config-subband}
\lad{Recall that in Sec.~IV-D of the main text, we propose a novel scaling scheme called subband-scaling, where consistently improved performance can be achieved by adopting more fine-grained subband-division strategy. In Table~{\ref{tbl:subband-division-other}}, we present detailed parameter configurations for different number of subbands.} 

\subsection{Quantitative Results of MCDA Variants}\label{sec:quantitative-mcda}
\lad{Recall that in Sec.~IV-E of the main text, we compare the performance between ``w/o MCDA'' and three MCDA variants. In Table~{\ref{tbl:objective-mcda-Fm}} and Table~{\ref{tbl:objective-mcda-fmax}}, we list the corresponding quantitative results. In Table~{\ref{tbl:objective-mcda-Fm}} we fix $f_{max}=12$ kHz and vary the values of $F_{m}$, whereas in Table~{\ref{tbl:objective-mcda-fmax}}, we fix $F_{m} = 100$ and vary the values of $f_{max}$.}

\subsection{Application to Speech Enhancement}\label{sec:application-to-speech-enhancement}
In recent years, neural vocoders have been utilized for speech enhancement (SE) task, where a SE network is first utilized for mel denoising, and a pretrained or jointly-trained neural vocoder is followed for waveform reconstruction~{\cite{liu2021voicefixer}}. Following this idea, we briefly investigate the performance of different vocoders when applied to SE task. Two types of representative SE networks are adopted, which belong to discriminative and generative streams, respectively. For the former, the network structure is similar to RNDVoC, where six DPBs are stacked to estimate the target mel-spectrogram. For the latter, following~{\cite{chen2023schrodinger}}, we build a Schr\" odinger bridge between the noisy and target mel-spectrograms, and a 1D-UNet serves as the score estimation function. For reverse process, the number of function evaluations (NFE) is set to 50. 

In Fig.~{\ref{fig:appendix_discriminative_generative}}, we present the framework of the utilized discriminative and generative methods for SE task, where for the former, a dual-path network is devised to estimate the target mel-spectrogram, and for the latter, a Schrodinger bridge is established to shift the feature domain from target to noisy mel-spectrograms in the forward process and the noise is gradually removed in the reverse process. Detailed structure of the adopted discriminative SE network is shown in Fig.~{\ref{fig:appendix_discriminative}}. Specifically, given the input noisy spectrogram $\mathbf{Y}^{mel}\in\mathbb{R}^{1\times F_{m}\times T}$, it first passes a Conv2d layer to obtain a shallow feature map $\mathbf{F}_{0}\in\mathbb{R}^{C\times F_{m}\times T}$, where $C$ is the channel size and is set to 256 herein. Later, $M=6$ dual-path blocks are employed for modeling along the time and frequency axes, and the network structures are similar to that of RNDVoC. Finally, another Conv2d layer is utilized to directly estimate the target mel-spectrogram $\mathbf{\tilde{X}}^{mel}$. For the training loss, both $\mathcal{L}_{1}$ and $\mathcal{L}_{2}$ are adopted, given as:
\begin{align}
	\label{eqn:62}
	\mathcal{L}_{dis} = \frac{1}{FT}\sum_{f,t}\left(\left\|\mathbf{\tilde{X}}^{mel}_{f,t} - \mathbf{X}^{mel}_{f,t}\right\|_{1} + \left\|\mathbf{\tilde{X}}^{mel}_{f,t} - \mathbf{X}^{mel}_{f,t}\right\|_{2}\right).
\end{align}

For the generative method, following~{\cite{chen2023schrodinger,jukic2024schr}}, we establish the Schr\" odinger bridge (SB) between the noisy and target mel-spectrograms. Specifically, a Schr\"odinger bridge is defined as minimization of the Kullback-Leibler divergence $D_{KL}$ between a path measure $p$ and a reference path measure $p_{ref}$, subject to the boundary conditions:
\begin{align}
	\label{eqn:63}
	\min_{p \in \mathcal{P}_{[0,T]}} D_{\text{KL}}(p,p_{\text{ref}}) \quad \text{s.t.} \quad p_0 = p_x, \quad p_T = p_y,
\end{align}
where $\mathcal{P}_{[0, T]}$ is the space of path measure on $\left[0, T\right]$. Given a forward stochastic differential equation (SDE):
\begin{align}
	\label{eqn:64}
	d\textbf{x}_{t} = \mathbf{f}\left(\mathbf{x}_{t}, t\right)dt + g\left(t\right)d\textbf{w}_{t},\quad \textbf{x}_{0} = \textbf{x},
\end{align}
where $t\in\left[0, T\right]$ denotes the current time for the process, $\textbf{x}_{t}\in\mathbb{R}^{D}$ denotes the state of the $t$ time, $f\left(\textbf{x}_{t}, t\right)$ is the dift term, $g\left(t\right)$ is the diffusion coefficient, and $\textbf{w}_{t}$ is a standard Wiener process. If $p_{ref}$ is defined by Eq.~({\ref{eqn:63}}), then the SB can be equivalent to a pair of forward-backward SDEs, given by:
\begin{align}
	\label{eqn:65}
	d\textbf{x}_{t} = [\textbf{f} + g^2(t) \nabla \log \Psi_{t}] dt + g(t) d\textbf{w}_t, \quad \textbf{x}_0 \sim p_x,
\end{align}
\begin{align}
	\label{eqn:66}
	d\textbf{x}_{t} = [\textbf{f} - g^2(t) \nabla \log \overline{\Psi}_{t}] dt + g(t) d\overline{\textbf{w}}_t, \quad \textbf{x}_{T} \sim p_x,
\end{align}
where $\Psi$ and $\overline{\Psi}$ are the score functions of the optimal forward and backward process, respectively. Following the conclusion in~{\cite{jukic2024schr}}, we adopt the variance-exploding (VE) noise schedule, which exhibits relatively better performance over the variance-preservation (VP) counterpart. To be specific, for the marginal Gaussian distribution $p_{t} = \overline{\Psi}_{t}\Psi_{t}\sim\mathcal{N}_{\mathbb{C}}\left(\boldsymbol{\mu}_{x}\left(t\right), \sigma_{x}^{2}\left(t\right)\right)$, the mean $\boldsymbol{\mu}_{x}\left(t\right)$ and the variance $\sigma_{x}^{2}\left(t\right)$ can be expressed as:
\begin{equation}
	\label{eqn:67}
	\boldsymbol{\mu}_{x}\left(t\right) = w_{x}\left(t\right)\mathbf{Y}^{mel} + w_{y}\left(t\right)\mathbf{X}^{mel},\quad \sigma_{x}^{2} = \frac{\alpha_{t}^{2}\overline{\sigma}_{t}^{2}\sigma_{t}^{2}}{\sigma_{T}^{2}},
\end{equation}
where $w_{x}\left(t\right) = \frac{\alpha_{t}\overline{\sigma}_{t}^{2}}{\sigma_{T}^{2}}$, $w_{y}\left(t\right) = \frac{\overline{\alpha}_{t}\sigma_{t}^{2}}{\sigma_{T}^{2}}$, and $\alpha_{t} = \exp^{\int_{0}^{t}f\left(\tau\right)d\tau}$, $\sigma_{t}^{2} = \int_{0}^{t}\frac{g^{2}\left(\tau\right)}{\alpha_{\tau}^{2}}d\tau$, $\overline{\alpha}_{t}=\alpha_{t}\alpha_{T}^{-1}$, and $\overline{\sigma}_{t}^{2} = \alpha_{T}^{2} - \alpha_{t}^{2}$. For VE case, both the drift term $f\left(t\right)$ and diffusion coefficient $g\left(t\right)$ are scalars and set as
\begin{align}
	\label{eqn:68}
	f\left(t\right) = 0,\quad g^{2}\left(t\right) = ck^{2t},
\end{align}
where the parameters $\left\{c, k\right\}$ are empirically set to $\left\{0.4, 2.6\right\}$. For training, the 1D-UNet is adopted, where $\textbf{x}_{t} = \boldsymbol{\mu}_{x}\left(t\right) + \sigma_{x}\left(t\right)\textbf{z}$ is the network input and $\mathbf{Y}^{mel}$ is the condition term. $\textbf{z}\sim\mathcal{N}_{\mathbb{C}}\left(0, \mathbf{I}\right)$ is a standard Gaussian noise. We adopt a data matching objective for training, shown as:
\begin{align}
	\label{eqn:69}
	\min_{\theta} \mathcal{E}_{(\mathbf{X}^{mel},\mathbf{Y}^{mel}),t,\mathbf{z}} \left\| s_{\theta}(\mathbf{x}_{t}, \mathbf{Y}^{mel}, t) - \mathbf{X}^{mel} \right\|_{2}^{2},
\end{align}
where $s_{\theta}\left(\cdot\right)$ is the score function. For reverse process, the first-order SDE sampler is employed, given by:
\begin{align}
	\label{eqn:70}
	\mathbf{x}_t = \frac{\alpha_t \sigma_t^2}{\alpha_{\tau} \sigma_{\tau}^2} \mathbf{x}_{\tau} + \alpha_t \left(1 - \frac{\sigma_t^2}{\sigma_{\tau}^2}\right) \mathbf{\hat{x}}_{\theta}(\tau) + \alpha_t \sigma_t \sqrt{1 - \frac{\sigma_t^2}{\sigma_{\tau}^2}}\mathbf{z}.
\end{align}

The environmental noise is taken from DNS-Challenge noise set{\footnote{https://github.com/microsoft/DNS-Challenge}}, and the clean speech is from LibriTTS dataset. During training, we mix clean speech and noise clips in a on-the-fly manner, and the signal-to-noise ratio (SNR) is randomly sampled from -5 dB to 15 dB. All SE models are trained for 500K steps. For model evaluations, speech clips are from \textit{dev-clean+dev-other} and noise clips are from MUSAN noise set~{\cite{snyder2015musan}}. The testing SNRs is chosen from $\left\{-5, -2, 0, 2, 5, 10\right\}$ dB to cover both low and high SNRs.

Table~{\ref{tbl:denoising-performance}} presents the metric performance of different neural vocoders when leveraged for SE task, where all the neural vocoders are pretrained on the LibriTTS benchmark, and kept frozen for waveform generation. Except for PESQ and VISQOL, extended short-time objective intelligibility (eSTOI)~{\cite{jensen2016algorithm}} and DNSMOS with P.808 criterion~{\cite{reddy2021dnsmos}} are also included, where the former is to evaluate the speech intelligibility and latter is a non-intrusive metric to simulate the human rating process with a pretrained NN model. One can observe that all the vocoders can effectively improve the speech quality, and the proposed RNDVoc-shared overall outperforms other methods except in DNSMOS and VISQOL. We notice that although the scores of the intrusive metrics (PESQ, eSTOI, VISQOL) for generative method is inferior to the discriminative counterpart, the former exhibits higher scores in the non-intrusive DNSMOS. We attribute the reason as the over-smoothing effect of discriminative methods, which can result in the audible electric-like artifacts and notably decrease subjective perception. Besides, one can see that the overall performance gap is not as prominent as in the clean cases, \emph{i.e.}, Tables~{4} and {5} in the main text, indicating that the performance bottleneck mainly lies in the enhancement of target mel-spectrograms.

\begin{figure*}[h]
	\centering
	\vspace{0pt}
	\includegraphics[width=0.82\textwidth]{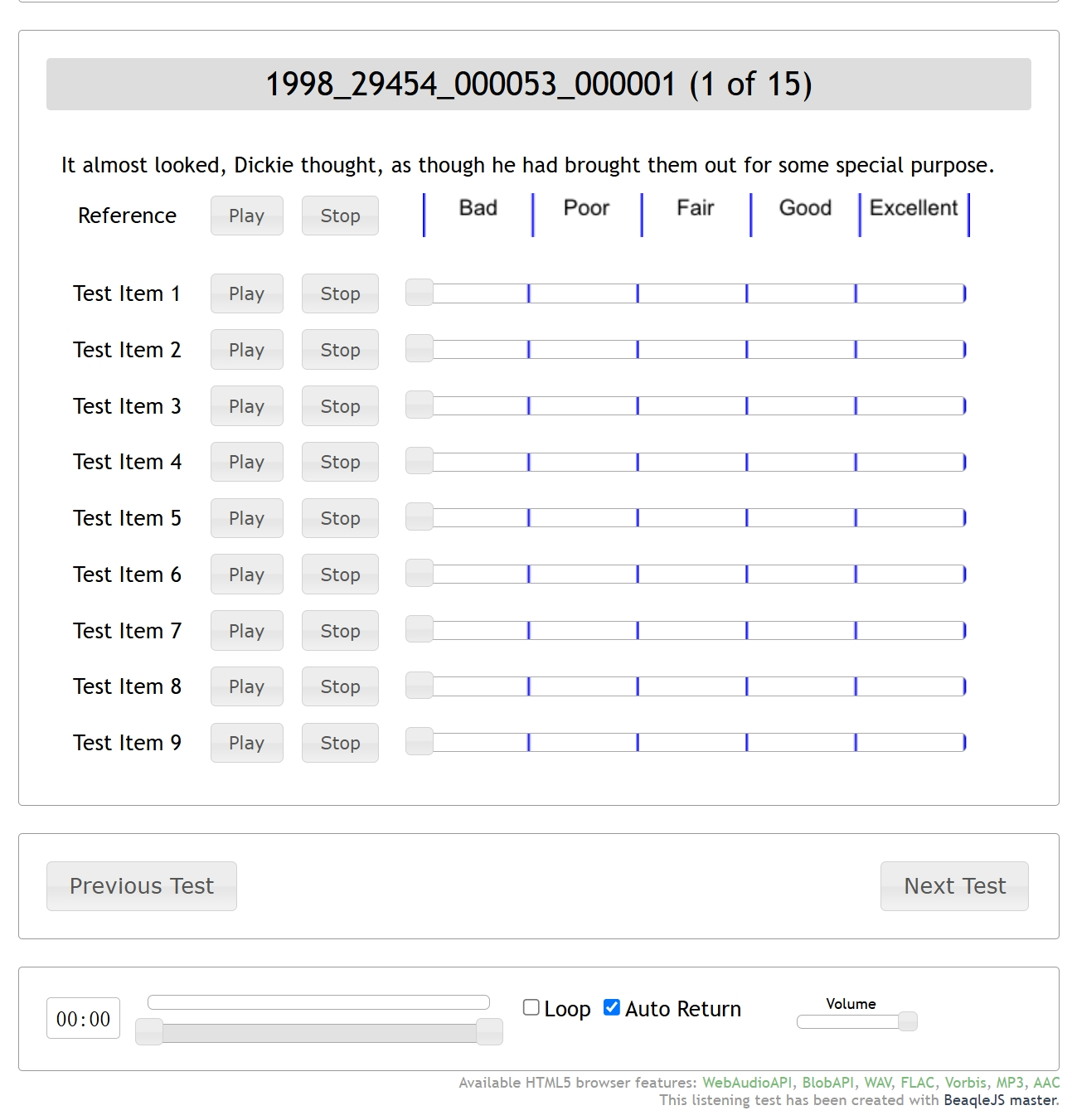}
	\vspace{-2pt}
	\caption{Screenshot of speech similarity for MUSRA testing.}  
	\label{fig:appendix_mushra}
	\vspace{-1pt}
\end{figure*}
\begin{figure*}[h]
	\centering
	\vspace{0pt}
	\includegraphics[width=0.86\textwidth]{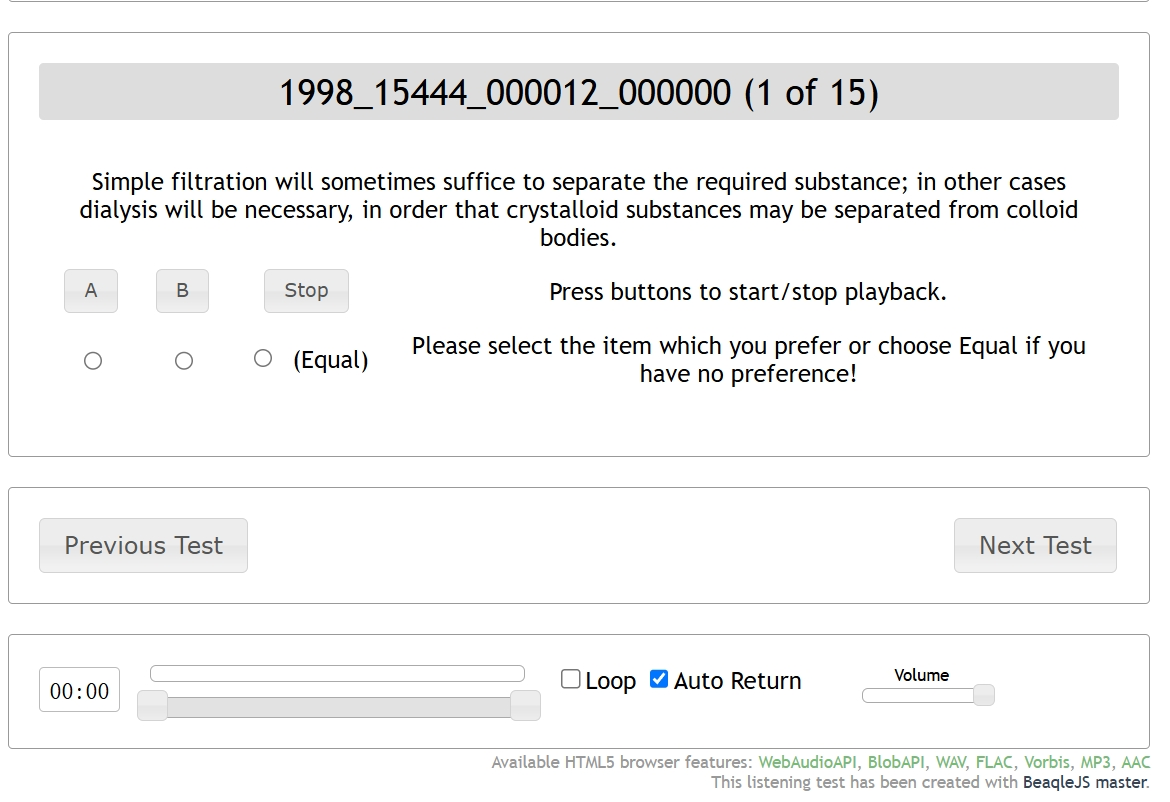}
	\vspace{-2pt}
	\caption{Screenshot of ABXTest.}  
	\label{fig:appendix_abtest}
	\vspace{-1pt}
\end{figure*}	

\subsection{More spectral visualizations}\label{sec:spectral-visualization}
\lad{From Fig.~{\ref{fig:spectral-visualization-1}} to Fig.~{\ref{fig:spectral-visualization-4}}, we present more spectral visualizations generated by different neural vocoders, whose clips are from vocal type of the MUSDB18 test set. Evidently, compared to existing baselines, RNDVoC can better restore detailed harmonic components, validating its potential to unseen audio types.}

\subsection{Listening Test with MUSHRA}\label{sec:listen-test-mushra}
We totally hired forty-three participants who are major in audio/music signal processing, and thirty-five people finished the test. During each testing, the utterance and algorithm orders are randomly shuffled and kept blind for participants. Figs.~{\ref{fig:appendix_mushra}}-{\ref{fig:appendix_abtest}} present screenshots of the webpages for MUSHRA and ABX test on the BeaqleJS platform.

{\small
	
	\bibliography{refs}
}

\end{document}